\documentclass[useAMS,usenatbib]{mnras}
\usepackage{multirow}
\usepackage{graphicx}	
\usepackage{rotate}
\usepackage{amsmath}
\usepackage{amssymb}
\usepackage{wasysym}
\usepackage{color}

\newcommand{\ms}{\mathcal{M}_{\star}}
\newcommand{\solarm}{\mathcal{M}_{\odot}}

\title[Global and local stellar mass assembly of galaxies from the MaNGA survey]{SDSS IV MaNGA: The global and local stellar mass assemby histories of galaxies}
\author[Ibarra-Medel et al.]{Hector J. Ibarra-Medel$^{1}$\thanks{E-mail: hibarram@astro.unam.mx},  
Sebasti\'an F. S\'anchez$^{1}$, 
Vladimir Avila-Reese$^{1}$, \newauthor 
H\'ector M. Hern\'andez-Toledo$^{1}$, 
J. Jes\'us Gonz\'alez$^{1}$, 
Daniel Thomas$^{6}$,
Niv Drory$^{4}$,\newauthor
Kevin Bundy$^{7}$,
Mariana Cano-D\'iaz$^{1}$, 
Alexandre Roman-Lopes$^{2}$,\newauthor 
Dmitry Bizyaev$^{3,4}$,
Elena Malanushenko$^{3}$,
Kaike Pan$^{3}$.
\\$^{1}$Instituto de Astronom\'ia, Universidad Nancional Aut\'onoma de M{\'e}xico, Box 70-264, M{\'exico} City, M{\'e}xico
\\$^{2}$Departamento de F\'isica, Facultad de Ciencias, Universidad de La Serena, Cisternas 1200, La Serena, Chile
\\$^{3}$Apache Point Observatory and New Mexico State University, P.O. Box 59, Sunspot, NM, 88349-0059, USA
\\$^{4}$Sternberg Astronomical Institute, Moscow State University, Moscow
\\$^{5}$McDonald Observatory, The University of Texas at Austin, 1 University Station, Austin, TX 78712, USA
\\$^{6}$Institute of Cosmology \& Gravitation, University of Portsmouth, Dennis Sciama Building, Portsmouth, PO1 3FX, UK
\\$^{7}$Kavli Institute for the Physics and Mathematics of the Universe, Todai Institutes for Advanced Study, the University of Tokyo,\\ Kashiwa, Japan 277- 858}

\begin{document}
\date{Accepted 2016 August 22. Received 2016 May 10; in original form 2016 September 2}

\pagerange{\pageref{firstpage}--\pageref{lastpage}} \pubyear{2014}

\maketitle


\begin{abstract} 
By means of the fossil record method implemented through Pipe3D, we reconstruct the global and radial stellar mass growth histories (MGHs) of an unprecedentedly large sample of galaxies, ranging from dwarf to giant objects, from the ``Mapping Nearby Galaxies at the Apache Point Observatory" survey.  We confirm that the main driver of the global MGHs is mass, with more massive galaxies assembling their masses earlier (downsizing), though for a given mass, the global MGHs segregate by color, specific star formation rate (sSFR), and morphological type. From the inferred radial mean MGHs, we find that at the late evolutionary stages (or for fractions of assembled mass larger than $\sim 80\%$), the innermost regions formed stars on average earlier than the outermost ones (inside-out). At earlier epochs, when the age resolution of the method becomes poor, the mass assembly seems to be spatially homogeneous or even in the outside-in mode, specially for the red/quiescent/early-type galaxies. The innermost MGHs are in general more regular (less scatter around the mean) than the outermost ones. For dwarf and low-mass galaxies, we do not find evidence of an outside-in formation mode; instead their radial MGHs are very diverse most of the time, with periods of outside- in and inside-out modes (or strong radial migration), suggesting this an episodic SF history. Blue/star-forming/late-type galaxies present on average a significantly more pronounced inside-out formation mode than red/quiescent/early-type galaxies, independently of mass. We discuss our results in the light of the processes of galaxy formation, quenching, and radial migration. We discus also on the uncertainties and biases of the fossil record method and how they could affect our results. 
\end{abstract}


\section{Introduction}

The study of how galaxies did assemble spatially their stellar masses is of paramount relevance for understanding the overall picture of galaxy evolution.  {A first inference on this can be provided by the mean mass- or luminosity-weighted age gradients obtained from line-strength indices and photometric studies of local galaxies. A more complete analysis is provided by the study of the color-magnitude diagram from resolved stars in nearby galaxies of by the fossil record method using Integral Field Spectroscopy observations (IFS; see below).  Several studies have shown that a significant fraction of the local galaxies present negative mean age gradients \citep[for recent works see e.g.,][]{Wang:2011aa,Lin:2013aa,Li:2015aa,Dale+2016}, which can be interpreted as earlier star formation (SF) of the inner regions with respect to the outer ones; this could imply that these galaxies assembled their stellar masses from inside to out.} Other kind of studies based on look back observations, suggest rather an uniform radial mass growth at early/intermediate redshifts \citep[][]{vanDokkum+2013,Patel+2013}. More recently, \citealp[][]{Perez+2013} have reported that the differences in the inner-to-outer stellar mass assembly histories of local galaxies studied with IFS depends on the total galaxy stellar mass; for their less massive galaxies, the trend even inverts suggesting an outside-in stellar formation mode. Some studies of dwarf galaxies report indeed positive age gradients for them \citep{Bernard+2007,Gallart:2008aa,Zhang:2012aa}. The radial way in which galaxies assemble their mass (or quench their SF) seems to depend also on their morphological type. For instance, for a small sample of early-type galaxies, \citet[][]{Sanchez-Blazquez:2007aa} have concluded that a solely inside-out or outside-in scenario is ruled out for these galaxies \citep[see also ][]{Mehlert:2003aa,Kuntschner:2010aa}. 
Summarizing, it is not yet well established how did galaxies on average assemble radially their stellar masses or quench their SF. The aim of this work is to provide a significant contribution on this question. 

According to the current cosmological paradigm, galaxies form from gas that cools and falls into the center of dark matter halos. Therefore, the galaxy gas accretion and the consequent SFR rate are expected to be associated to the cosmological dark matter accretion rate \citep[see e.g.,][]{vandenBosch2002,Faucher-Giguere+2011,Rodriguez-Puebla+2015}.  On the other hand, the disks formed inside growing dark matter halos are predicted to assemble from the inside out \citep[][and more references therein]{Gunn1982,Silk1987,Mo+1998,Avila-Reese+2000,Roskar+2008,Mo+2010}. The formation of spheroids is believed to occur mostly from the morphological transformation of disks, either by mergers or by internal dynamical processes \citep[see for reviews e.g.,][]{Mo+2010,Brooks+2016}. During early gas-rich mergers, strong bursts of SF {that consume most of the gas are expected to happen, resembling this the so-called ``monolithic collapse'' model \citep{Eggen+1962}; late major mergers, responsible also of spheroid formation, add an spatially extended population of \textit{ex situ} (likely old) stars \citep[e.g.,][]{Hopkins+2009,Rodriguez-Gomez+2016}, and produce a significant stellar radial mixing in the primary galaxy}. Therefore, the spatial mass assembly of spheroid-dominated galaxies is expected to depart from the inside-out mode of disk galaxies.  Moreover, galaxies along their evolution may suffer several phases of disk destruction and rebuilding, leaving this a complex imprint in their { present-day stellar population spatial distributions}. The other aspects that may play a relevant role in assessing the observed  { stellar population spatial distribution} of galaxies are when, where, and why the SF is shut off in galaxies. This shut off process is commonly called in the literature as "quenching".

According to the literature, there is not a general mechanism that explains the quenching of galaxies; rather exist multiple hypotheses that try to explain the shut off as a function of mass and environment \citep[e.g.,][]{Lin:1983aa,Dekel:1986aa,Efstathiou:1992aa,Birnboim:2003aa,Di-Matteo:2005aa,Slater:2014aa,Tal:2014aa}. Ram-pressure, strangulation, and harassment are some of the mechanisms related to environmental conditions that can explain the shutoff of SF in satellite galaxies \citep[e.g.,][]{Gunn:1972aa,Larson:1980aa,Farouki:1981aa,Hidalgo+2003,Tal:2014aa}. Mass dependent processes related to the heating up of the intrahalo gas can help to stop the SF of central galaxies \citep[e.g.,][]{Guo:2014aa,Tal:2014aa}; for example, the virial shock heating in massive halos \citep[the halo quenching model,][]{Birnboim:2003aa,Dekel:2006aa} and the AGN feedback \citep[e.g.,][]{Kauffmann:2004aa,Bower:2006aa,Guo:2014aa}, in particular the powerful outflows in quasars that deplete the gas content in massive galaxies \citep{Sanders:1988aa,Di-Matteo:2005aa}. On the other hand, since the SFR of galaxies is expected to primarily depend on the halo dark matter accretion rate (see above), then if the latter significantly drops, then the former shoould drop too \citep[cosmological quenching; e.g.,][]{vandenBosch2002,Feldmann:2015aa}. Whatever the dominant process is, the form of how the SF can be deactivated correlates with the galaxy mass \citep[e.g.,][]{Woo:2013aa}. 

In order to constrain the scenarios and processes of galaxy evolution above described, it is crucial to obtain information about how the stellar mass grows during the evolution of the galaxy as a function of radius. Considering that the light of a galaxy is an assembly of multiple star contributions, it is possible to infer how these stars assembled the observed galaxy light through its life. This method uses stellar evolutionary models to get the energy spectra distribution (SED) of single stellar populations (SSP) at different ages and metallicities. Thus, the problem is reduced to look for the best combination of SSPs that mimics the observed galaxy SED. Then, it is possible to obtain information of the stellar mass per SSP at different time steps. As a result, it is feasible to reconstruct the SF or stellar mass growth history (SFH or MGH) of an observed galaxy. This approach based in stellar population synthesis (SPS) is known in the literature as the \textit{fossil record method} \citep[e.g.,][]{Tinsley:1980aa,Buzzoni:1989aa,Bruzual:2003aa,Walcher:2011aa}. The fossil record method has been widely used to recover the global SFH and stellar masses by using spectral and photometric data of galaxies in large surveys like the Sloan Digital Sky Survey (SDSS) one \citep[e.g.,][]{Kauffmann:2003aa,Kauffmann:2003ab,Cid-Fernandes:2005aa,Gallazzi:2005aa,Tojeiro+2007}.

With multi-band and spectroscopic data of good spatial resolution, it is possible to resolve the SFH (or MGH) in space \citep[e.g.,][]{Brinchmann:2000aa,Kong:2000aa,Perez-Gonzalez:2008aa,Lin:2013aa}.  These studies used to be limited by the size of the slit or fiber (for the spectroscopy data) or by the number of photometric filters that are used to adjust the SED. These limitations can retrieve important bias in the interpretation of the data. A fixed fiber (or slit) size cannot locally resolve the galaxy properties because the observed galaxy spectrum integrates all or a substantial fraction of the light within the aperture. For the case of photometric studies, it is required a large enough number of photometric bands to reproduce the galaxy spectrum \citep[e.g.,][]{Benitez:2009aa}. It is important to correct the observed photometric fluxes by emission lines to accurate fit the observed galaxy SED. However, with the advent of the IFS, 
 it is possible to access to spectral data resolved through different galaxy regions at the same time and obtain the spectral information of the local galaxy properties \citep[e.g.,][]{Bacon:2001aa,Cappellari+2011,Croom:2012aa,Sanchez:2012aa,Cid-Fernandes:2013aa,Bundy:2015aa}. With IFS data and SSP decomposition, it is feasible to obtain information about how galaxies assembled globally and locally their stellar masses. 
 
\citet{Perez+2013} used the advantages of the IFS to resolve the radial gradient of the stellar mass assembly. They analyzed $105$ galaxies from the Calar Alto Legacy Integral Field Area survey \citep[CALIFA,][]{Sanchez:2012aa} and explored the mass growth gradient from $10^{9.8}$ to $10^{11.26}\ \solarm$. With the MaNGA survey \citep[Mapping Nearby Galaxies at the Apache Point Observatory,][]{Bundy:2015aa}, it is possible to perform a spatially resolved study of the MGHs for an unprecedentedly large galaxy sample both in number and in mass range. The MaNGA survey is one of the three core programs of the fourth generation of the SDSS, and  plans to get IFS observations for $10000$ galaxies (selected by stellar masses) within a redshift range of  $0.01<z<0.18$ \citep{Bundy:2015aa,Li:2015aa,Law:2015aa}. MaNGA performs dithered observations with integral field units (IFU) that cover a spectroscopic range of $3600$ to $10300\ \AA$, with a resolution of $R\sim2000$. MaNGA provides $17$ fiber-bundle IFUs with a field of view that covers from $12''$ (19-fiber IFU) to $32''$ (137-fiber IFU). For a more detailed description of the MaNGA survey and its instrumentation, 
see \citet{Gunn:2006aa}, \citet{Bundy:2015aa}, \citet{Drory:2015aa}, and \citet{Yan:2016aa}. Some Early results with MaNGA are the study of \citet{Li:2015aa} that maps the gradients of stellar population and star formation indicators D4000 and Halpha covering a large wavelength range.\citet{Wilkinson:2015aa} maps the gradients of stellar population parameters and dust attenuation, discussion of impact of different fitting techniques and IFU size. Finally, \citet{Belfiore:2015aa} performs an spatially resolved study of emission line diagnostic diagrams, maps of gas metallicities and element ratios.

{In this paper we present the global and spatially-resolved normalized MGHs of the so far observed galaxies in the MaNGA survey, namely those to be reported in the SDSS Data Release 13. We focus our analysis on galaxies from the Primary sample, covering a mass range from dwarf to giant galaxies, and with numbers that overcome those of the few previous related works. Our goal is to explore how galaxies did assemble globally and locally their stellar masses as a function of mass. The large number of galaxies allows us to study this question also as a function of color, specific SFR (sSFR), and morphology, by separating the galaxies in each mass bin at least in two groups according to these properties. It should be said also that our analysis differs in several aspects to those presented in previous works, for instance, those of \citet{Perez+2013} and \citep[][]{Gonzalez-Delgado+2016}.}

The outline of the paper is as follows. In Section \ref{analysys}, we explain the sample selection, its analysis and the methodology. In Section \ref{results-meanMGHs}, we discuss both the global and radially-resolved mean stellar MGHs as a function of the final galaxy mass, color, SF activity, and morphology. In Section \ref{inMGH}, we recover the individual radial mass assembly gradient for the final sample of galaxies at the times when the $90\%$, $70\%$ and $50\%$ of their total masses were assembled. In Section \ref{discussion}, we compare our results with previous works, we discuss the possible implications of our inferences for the galaxy evolution paradigm, and comment on the caveats of these inferences. Finally, in Section \ref{conclusions}, we present the summary of the paper and our main conclusions.

\section{The Sample and the Methodology} \label{analysys}

{ We study 
the 1260 galaxies from the MaNGA product launch 4 (MPL-4, which is part of the SDSS Data Release 13) that were observed from March 2014 to June 2015. With this sample we are able to explore a large mass range, from dwarf to giant galaxies.} 
The structural parameters, such as S\'ersic half-light radii ($R_{s,50}$), Petrosian radii ($R_{p,50},R_{p,90}$), position angles (PA) and axis ratios ($b/a$), and the absolute magnitudes in different bands, were taken initially from the NASA-Sloan Atlas\footnote{\url{http://www.nsatlas.org/}} (NSA), that also contains data from the SDSS III \citep{Eisenstein:2011aa}.

The $R_{s,50}$ radius, defined as the radius where half of the light for a S\'ersic profile is contained, is calculated from a two-dimensional fit to such a profile. This fit assumes that all galaxies are well described by the same surface brightness profile, independently of the morphological type. The $R_{p,50}$ (and $R_{p,90}$) radius is calculated given the measured Petrosian magnitude, where such magnitude was thought to include most of the object's flux and independent of distance \citep{Petrosian:1976aa}. However, different fractions of the light of each galaxy are missed depending on the surface brightness profile. This introduces a systematical uncertainty in the Petrosian half-light radii, which can depend, for instance, on the galaxy concentration $C\equiv R_{p,90}/R_{p,50}$. In an effort to correct for this effect, \citet[][]{Graham:2005aa} (see equation 6), assuming a S\'ersic profile, provides a procedure that makes use of $C$. Thus, having the reported $R_{p,50}$ and $C$ for each galaxy, the corrected half-light radius,  $R_{50}$, can be calculated. We adopted the corrected radius in this work to characterize the size of the galaxies.

We have performed a visual morphological classification of the 1260 MPL-4 MaNGA galaxies to be presented elsewhere (Hernandez-Toledo et.al 2016, in prep). In this sample we have found 112 galaxies (9\% out of the total sample) with strong signatures of interaction or in the process of merging. We exclude these galaxies from our analysis, leaving then 1149 galaxies. We select then only galaxies that are in the MaNGA Primary sub-sample (those with a radial coverage that reaches 1.5 effective radii) and exclude those from the ``Color-Enhanced'' sub-sample (they were specially added to balance the color distribution at a fixed stellar mass); for details on the MaNGA sub-samples, see \citet[][]{Bundy:2015aa}. {\bf After this, remain 601 galaxies.} We have selected only the Primary sub-sample to avoid any aperture bias in our study, and the Color-Enhanced sub-sample was excluded to avoid biases with respect to the real color distributions in our MGHs averaged in mass bins. 
{Finally, in our main analysis we will not take into account galaxies more massive than $10^{11.2} \solarm$ (see below for our definition of total stellar mass). We have found that for some reason, the currently observed (MPL-4) most massive MaNGA galaxies in the Primary sub-sample are biased to an excess of blue and star-forming (by $H_{\alpha}$ emission) objects.  } { We are left then with 533 galaxies. }

\begin{figure}
\begin{center}
\includegraphics[width=0.9\linewidth]{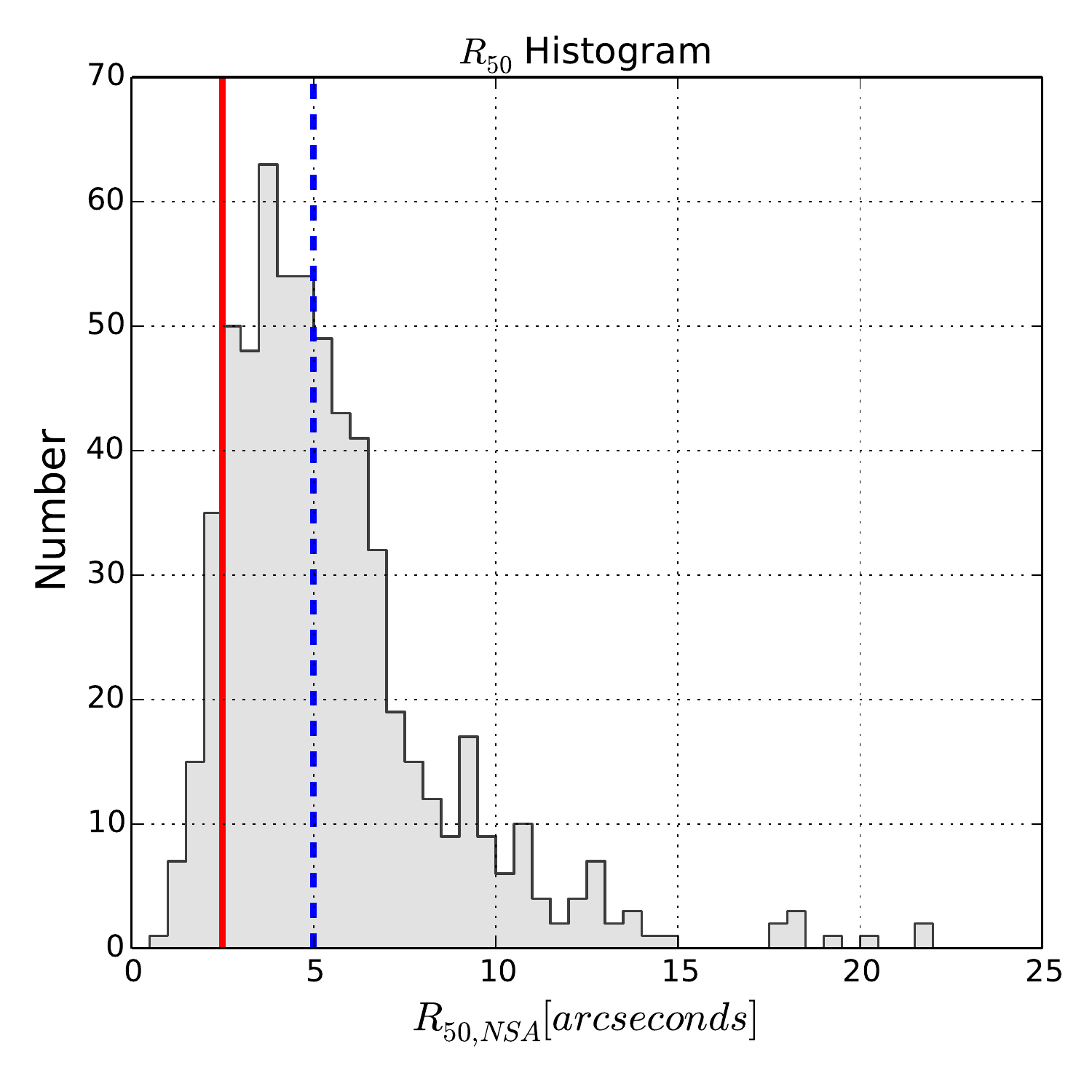}
\includegraphics[width=0.9\linewidth]{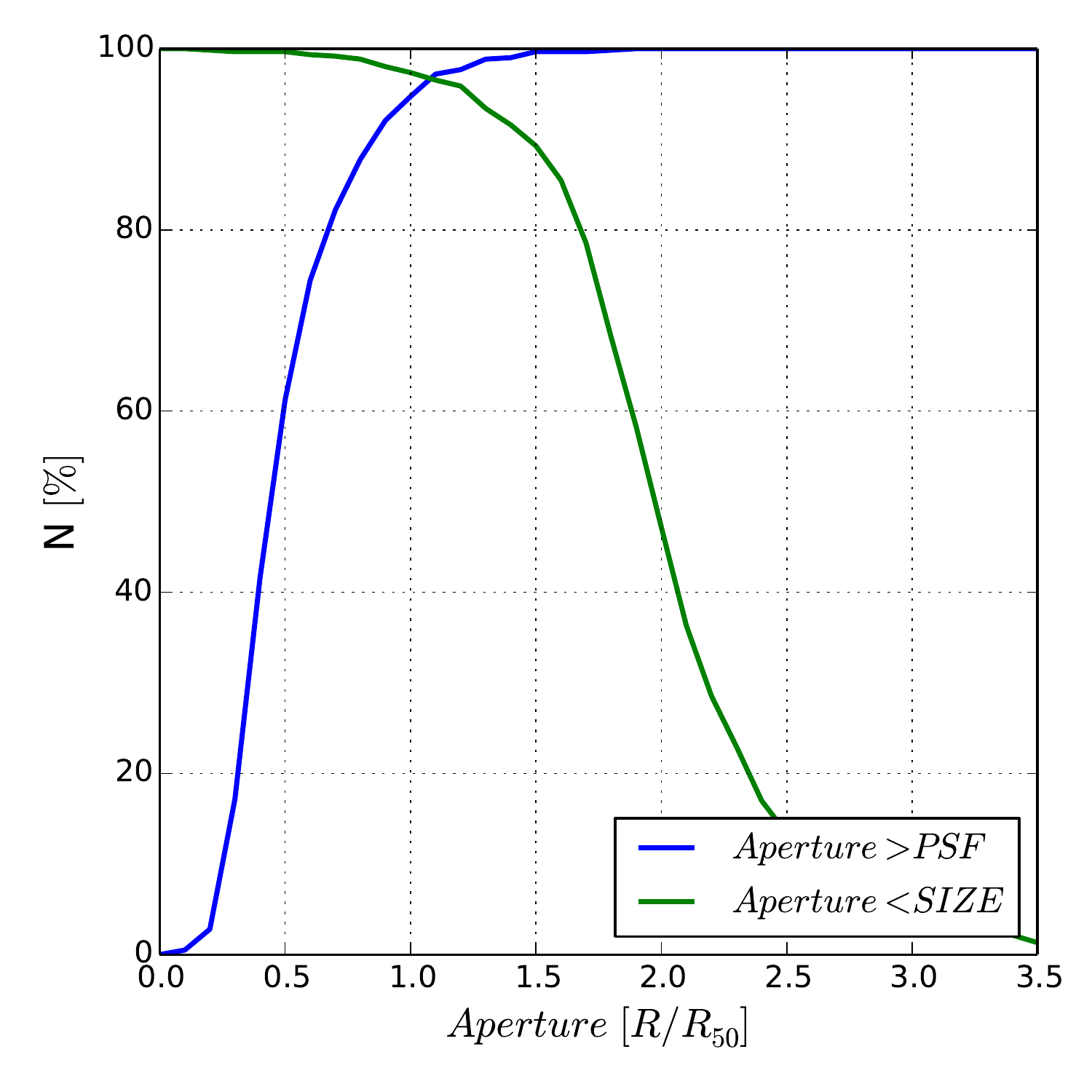}
\end{center}
\caption{{\it Upper:} Histogram of effective radii ($R_{50}$) for 533 MaNGA galaxies compiled from the NASA-Sloan Atlas (NSA). The red line represents the reconstructed PSF of the IFU datacubes; whereas the blue dashed line represents the limit when $0.5R_{50}=PSF$. {\it Lower:} Fraction of galaxies that have a radius larger or equal to the IFU PSF (blue line). The green line represents the fraction of galaxies that have a radius less or equal to the IFU field of view. }\label{fig00}
\end{figure}
\begin{figure*}
\begin{center}
\includegraphics[width=\linewidth]{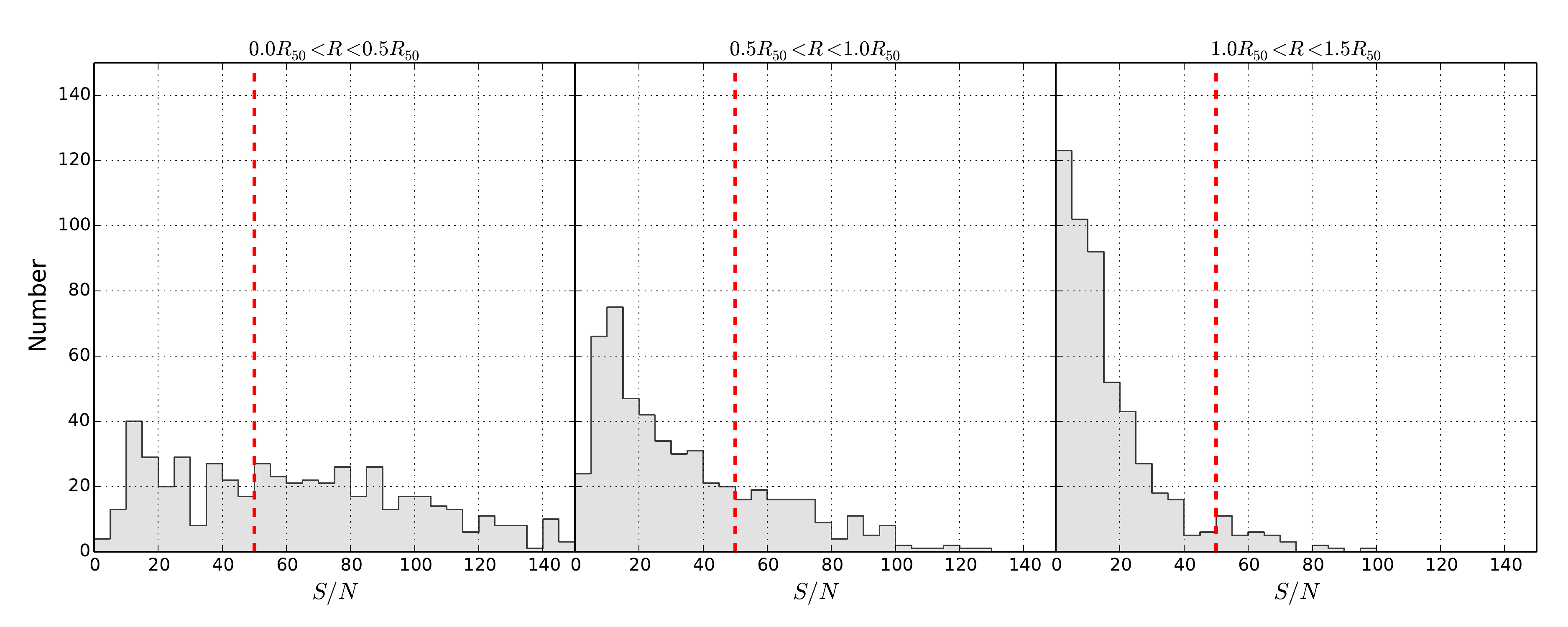}
\end{center}
\caption{Distribution of the S/N ratio within the three radial regions: $R<0.5R_{50}$, $0.5R_{50}<R<R_{50}$, and $R_{50}<R<1.5R_{50}$. The red dashed lines represent the S/N threshold of 50 that is used by Pipe3D.}\label{figSN}
\end{figure*}

For our spatially-resolved analysis, we will consider three radial galaxy regions: $0<R<0.5R_{50}$, $0.5R_{50}<R<R_{50}$ and $R_{50}<R<1.5R_{50}$. We estimate the radial distances from the galactic center by de-projecting the distances with the use of the galaxy axis ratio $a/b$ and the PA reported from the NASA-Sloan Atlas\footnote{\url{http://www.nsatlas.org/}} (NSA). These regions consider the resolution effects from the reconstructed point spread function (PSF) in the MaNGA datacubes, and the IFU bundle size. Since the MaNGA FWHM PSF is $2.5''$ \citep{Bundy:2015aa}, the minimum aperture that we should integrate without suffering the effects of the PSF is $0.5R_{50}$. In Figure~\ref{fig00}, we show the histogram of $R_{50}$ in arcseconds for all the sample. The thick red line represents the MaNGA FWHM PSF. The blue dashed line represents an aperture two times the value of the PSF; this line lies in the mode of the $R_{50}$ distribution. The bottom panel of Figure~\ref{fig00} shows the fraction of galaxies that have a radius (as a fraction of $R_{50}$) larger than the MaNGA FWHM PSF (blue line) and have an IFU size larger than that extension (green line). We find that $60\%$ of all the galaxies have a $0.5R_{50}$ radius larger than or equal to the value of the reconstructed PSF. For scales smaller than $0.5R_{50}$, the fraction of galaxies where the PSF is smaller than the optical extension drastically drops to zero; i.e., they are not resolved.  The distribution of $R_{50}$ indicates that $95\%$ of the sample have an $R_{50}$ larger than the PSF, whereas the fraction increments to $99\%$ for $1.5R_{50}$. {Therefore, the  great majority of the galaxies are well resolved, and we will use the whole sample for our analysis.}

We observe that almost all the sample contains the $0.5R_{50}$ ($99\%$), $1R_{50}$ ($95\%$), and $1.5R_{50}$ ($90\%$) regions within the IFU field of view. For apertures larger than $1.5R_{50}$, the fraction of galaxies that have a size within the IFU field of view decreases to $45\%$ at $2R_{50}$ and $2\%$ at $3.5R_{50}$. Hence, we set the maximum aperture to $1.5R_{50}$ to trace the galaxy outskirts without being affected by the IFU size. The first region ($R<0.5R_{50}$) maps the inner galaxy parts without being severely affected by the effects of the PSF. For the second region ($0.5R_{50}<R<R_{50}$), the width of the ring is large enough to not mix different regions by the convolution effects of the PSF. This region maps the intermediate region between the galaxy core and the galaxy outskirts, with one radial bin in terms of the PSF size. For the last radial region ($R_{50}<R<1.5R_{50}$), the maximum aperture traces the periphery for all our galaxy sample and assures that $90\%$ of the sample is observed with an IFU size of at least $1.5R_{50}$.

\begin{figure*}
\begin{center}
\includegraphics[width=0.3\linewidth]{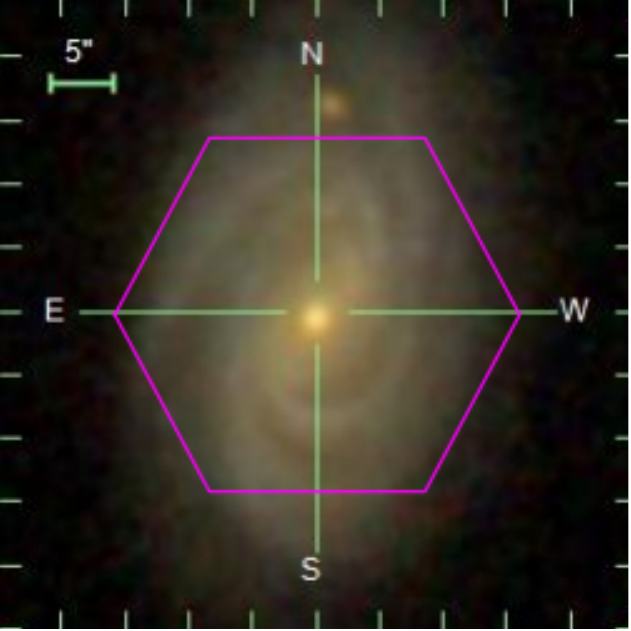}
\includegraphics[width=0.67\linewidth]{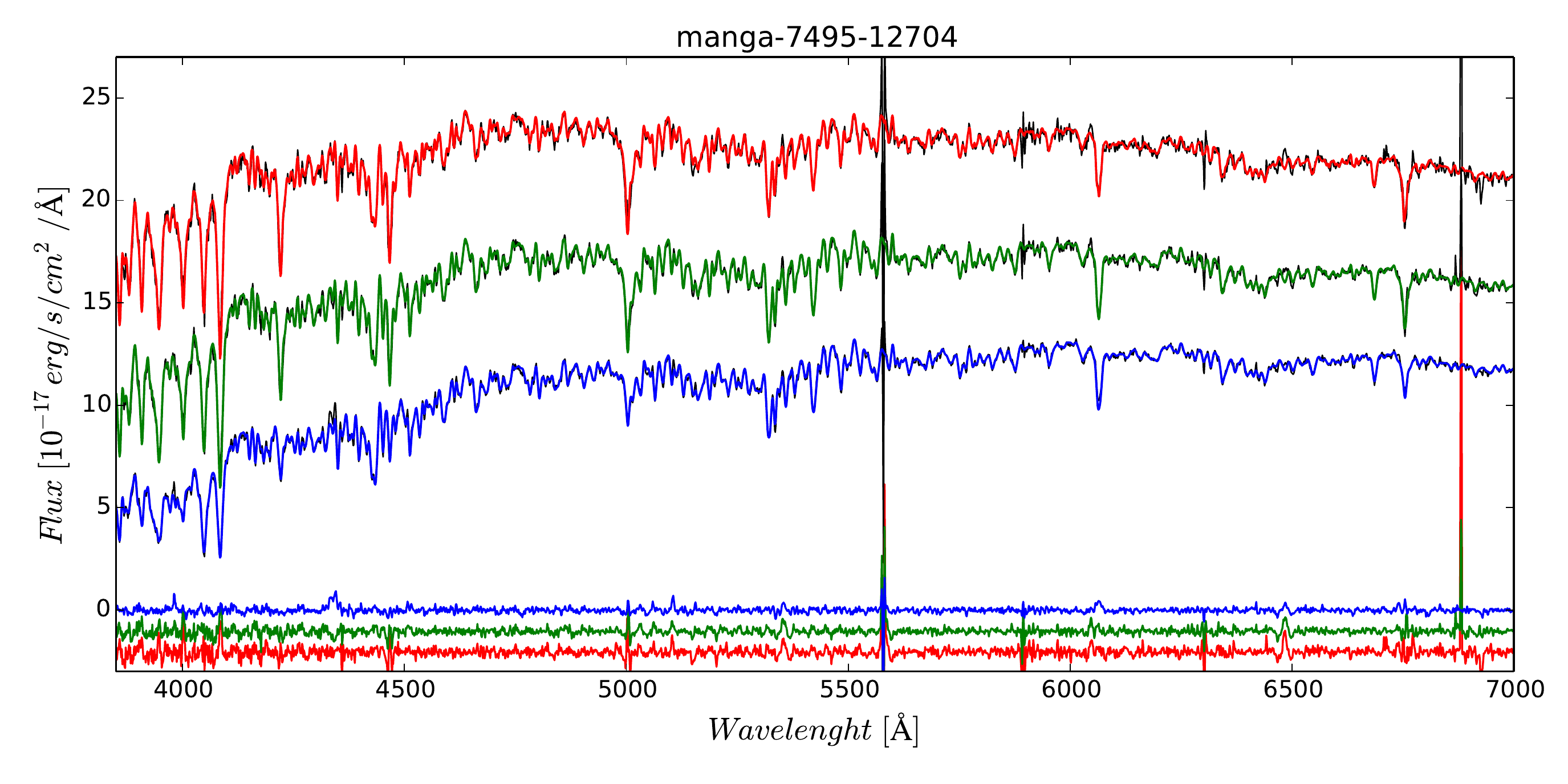}
\includegraphics[width=0.33\linewidth]{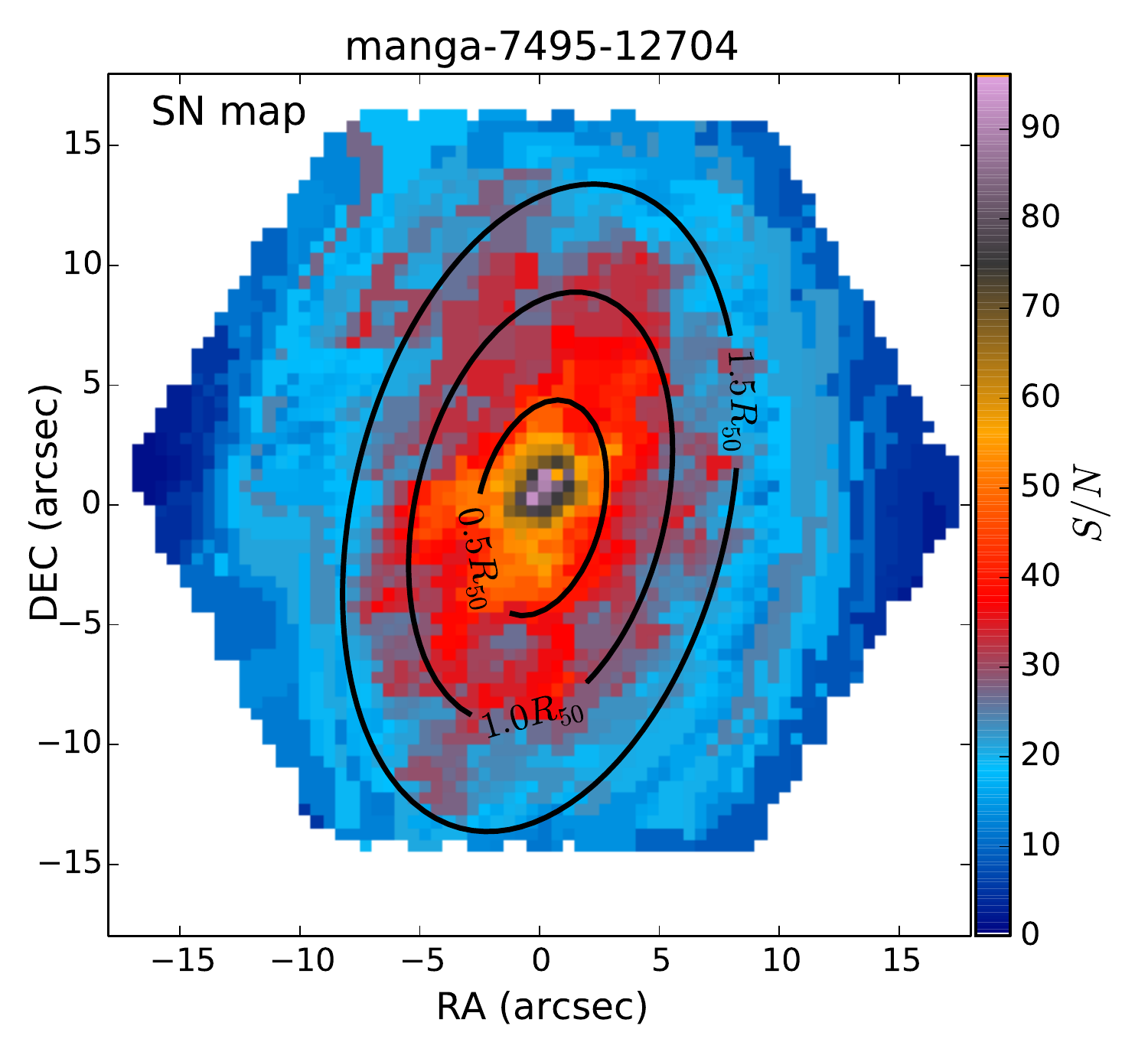}
\includegraphics[width=0.33\linewidth]{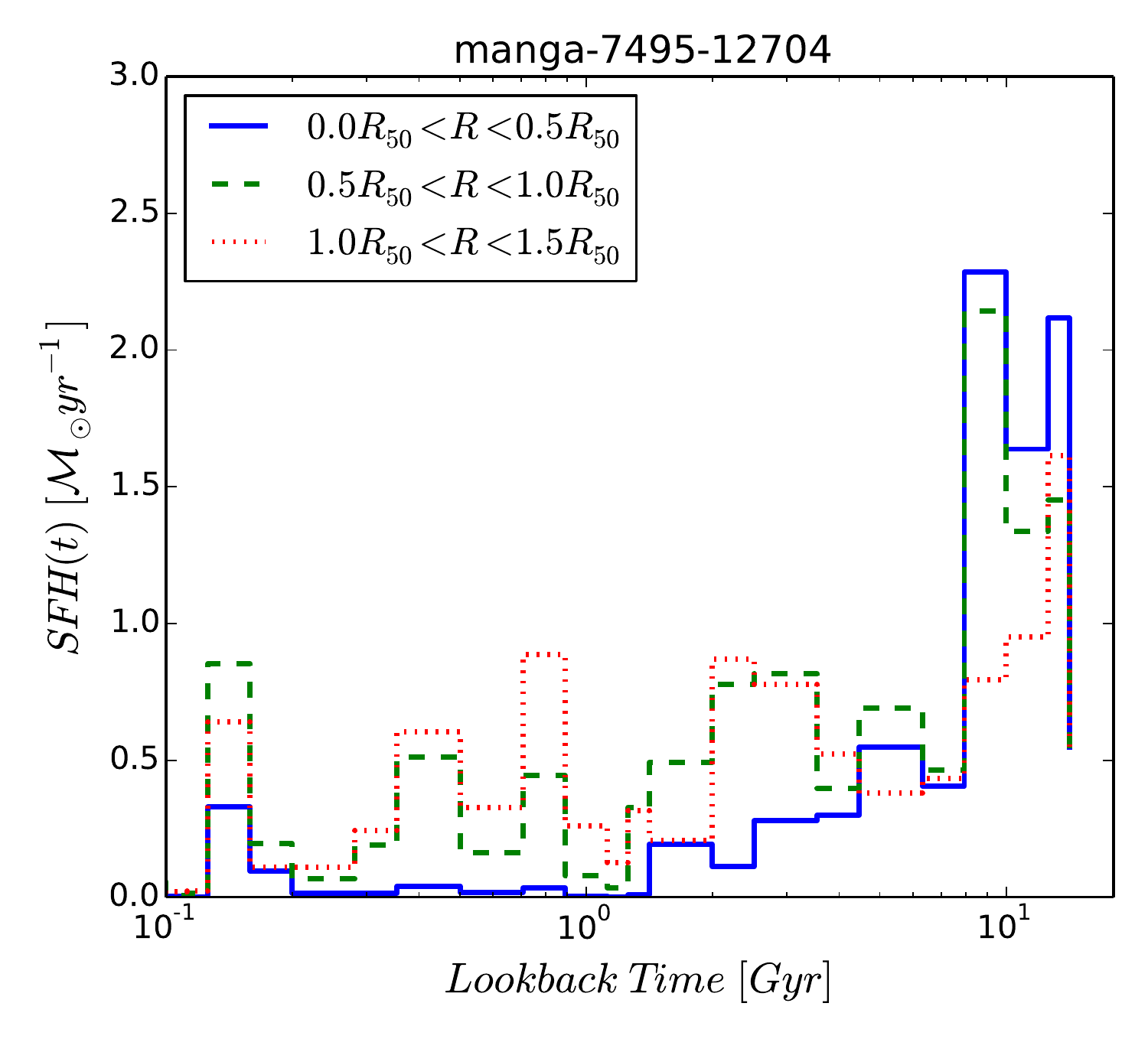}
\includegraphics[width=0.33\linewidth]{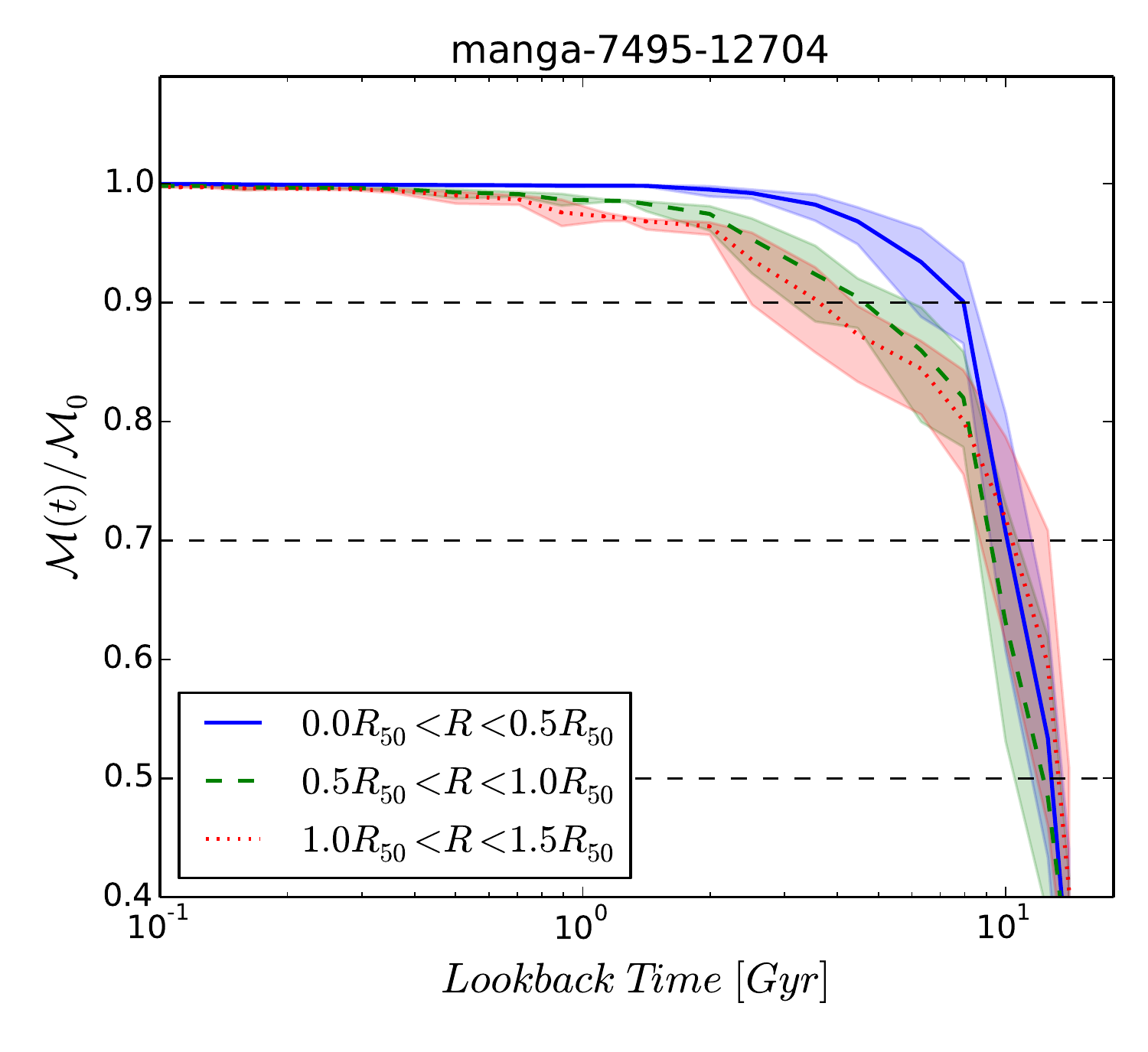}
\end{center}
\caption{Example of our analysis for the galaxy SDSS J134145.21+270016.9 (MaNGA ID 12-129618) observed with the 127 fiber IFU configuration. This is a Milky Way-sized galaxy, $\mathcal{M}(\leq1.5R_{50}) = 3.1\times10^{10}\solarm$. The upper left panel presents the optical image ($g,r,i$) and the observed field (ciyan lines) of the 127 fiber IFU bundle.  The upper right figure presents the integrated synthetic stellar (without gas) spectrum fit from Pipe3D at three radial regions:  $R<0.5R_{50}$ (blue),  $0.5R_{50}<R<R_{50}$ (green),  and $R_{50}<R<1.5R_{50}$ (red). The corresponding integrated observed spectra are shown with black lines. Each spectrum  (synthetic and observed) is shifted in flux amplitude to avoid overlapping. The residuals of the fits are plotted below with the same color code and shifted in flux to avoid overlapping. The lower left panel presents the S/N map of the IFS spectra. We over-plotted the three radial regions as contours. The lower middle panel shows the reconstructed SFH at the three different radial regions. Finally, the lower right panel shows the reconstructed MGHs at the three different radial regions. The shaded color areas within the MGHs are the $1\sigma$ error estimates for each MGH. The color code of the lines for the lower middle and right panels are the same as the upper right panel.}\label{test1}
\end{figure*}

\begin{figure*}
\begin{center}
\includegraphics[width=0.3\linewidth]{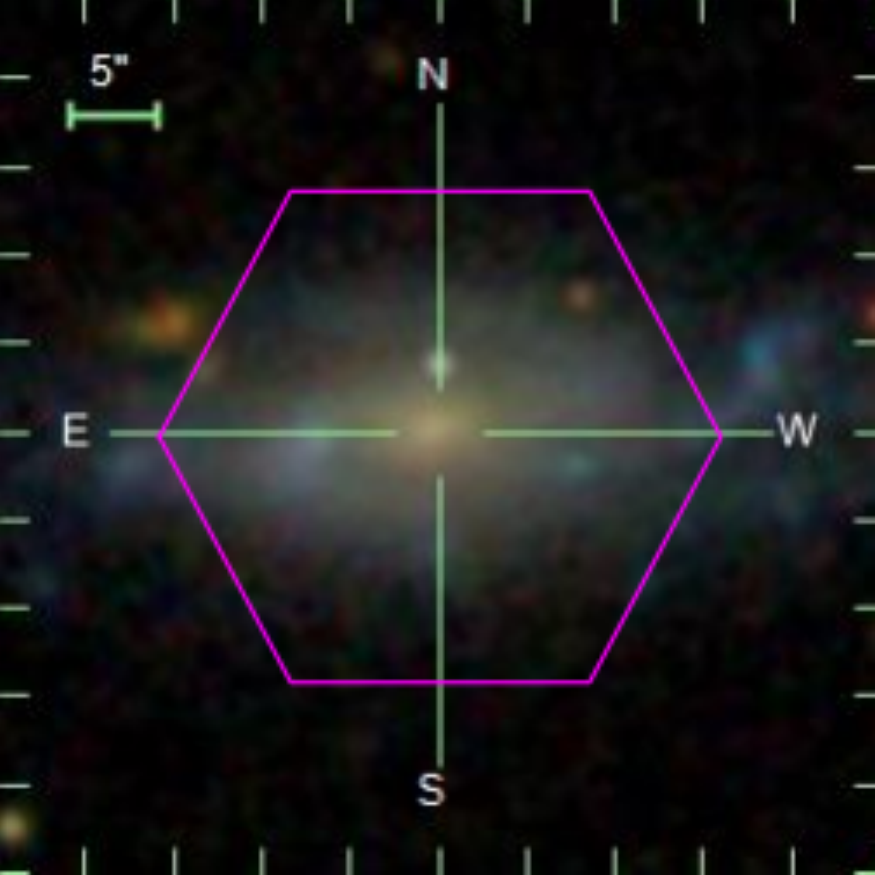}
\includegraphics[width=0.67\linewidth]{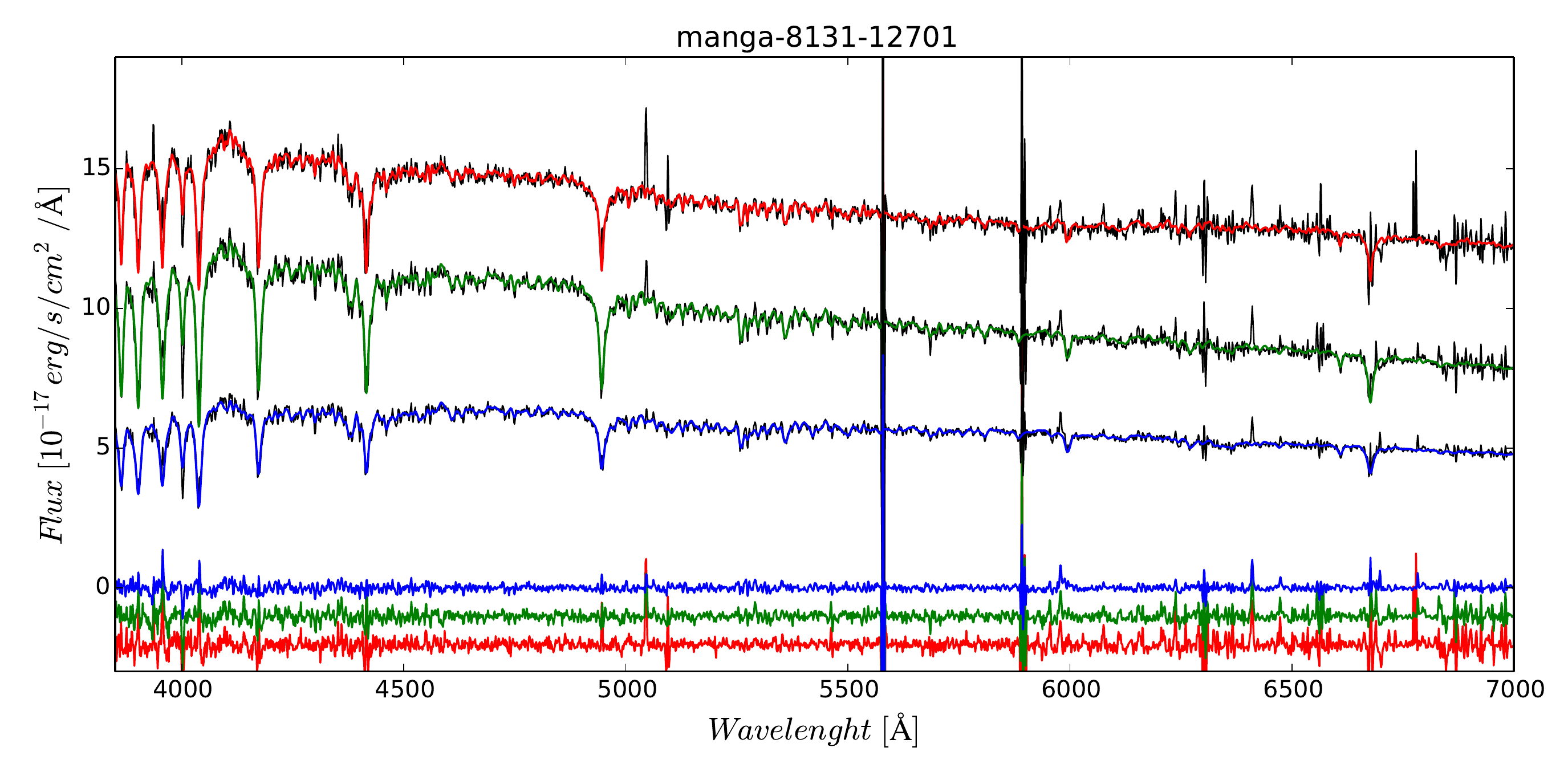}
\includegraphics[width=0.33\linewidth]{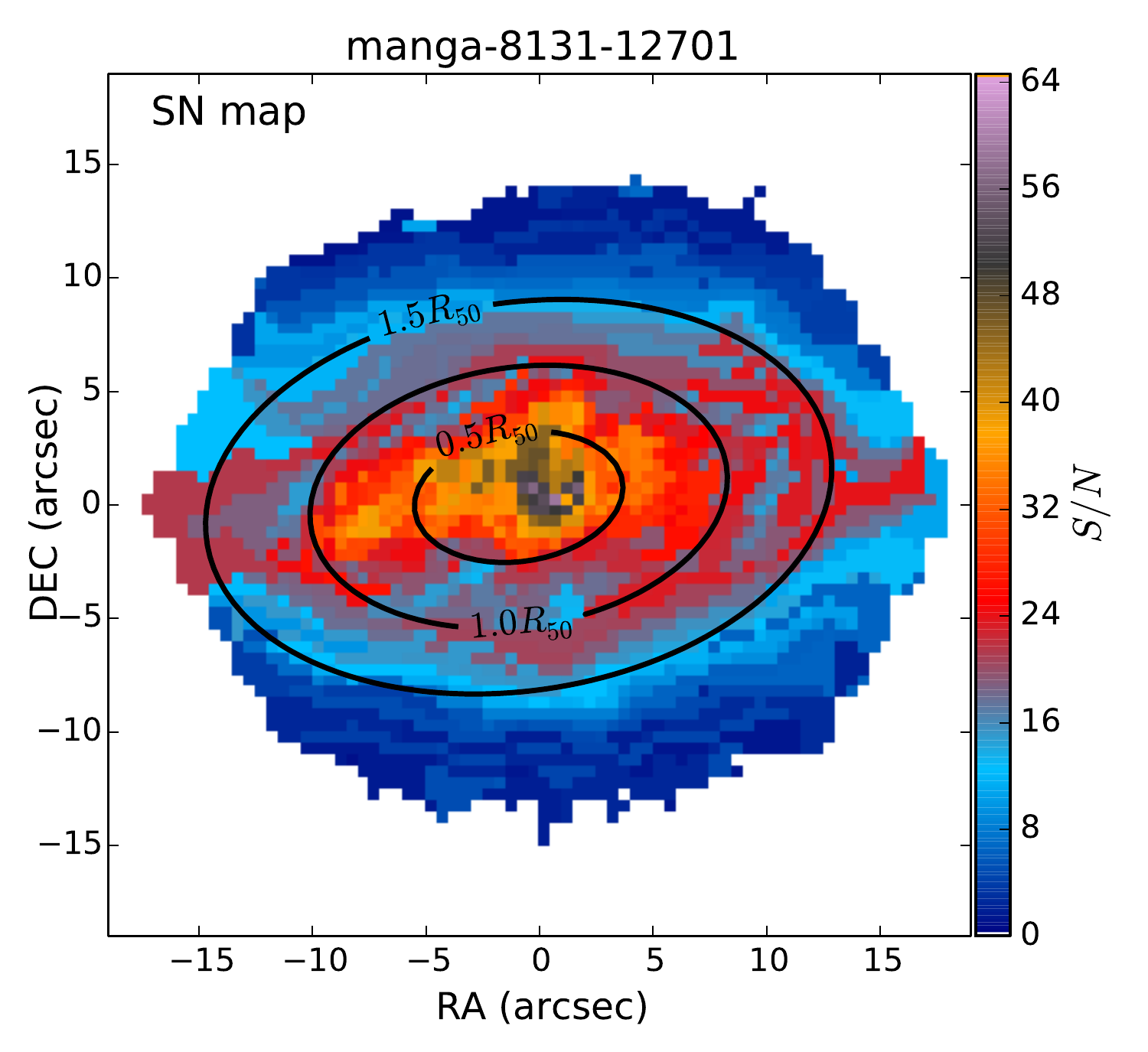}
\includegraphics[width=0.33\linewidth]{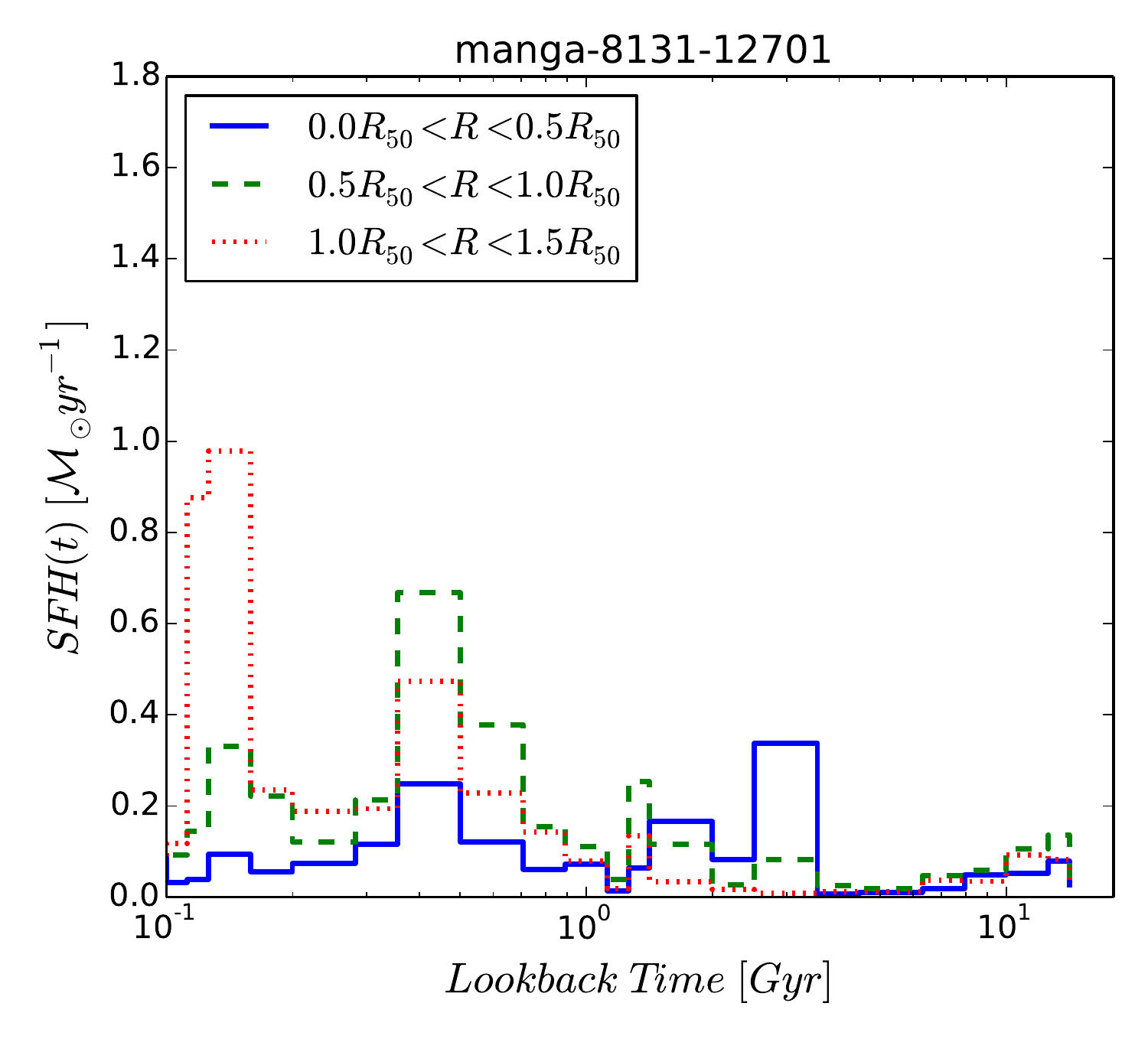}
\includegraphics[width=0.33\linewidth]{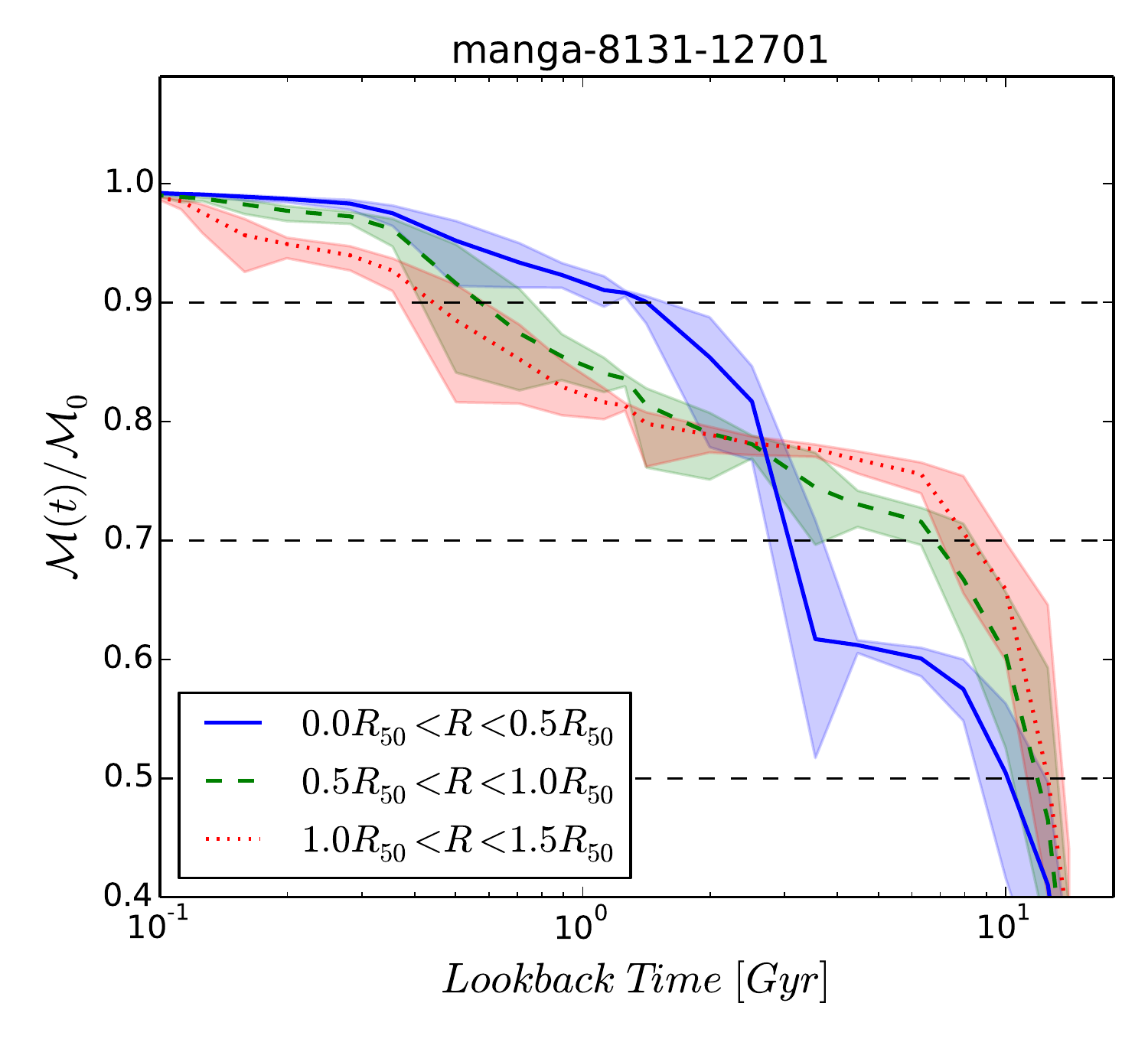}
\end{center}
\caption{Same as Figure \ref{test1} but for the galaxy SDSS J072425.92+385643.0 (MaNGA ID 1-584473) observed with the 127 fiber IFU configuration. This is a
low-mass galaxy, $\mathcal{M}(\leq1.5R_{50})=2.3\times10^{9}\solarm$.}\label{test2}
\end{figure*}

\subsection{Stellar Population Analysis}

We perform the stellar population synthesis of the IFU datacubes as provided in the MaNGA site by using Pipe3D. Following, we briefly summarize the procedure used to analyze the underlying stellar population \citep[for further details, refer to][]{Sanchez:2016aa,Sanchez:2016ab}. For each datacube and individual spaxel, a signal-to-noise (S/N) analysis is performed. Afterwards, a spatial binning is applied in order to achieve a homogeneous S/N threshold level of 50. It is important to say that this S/N threshold is not necessary achieved at each spatial bin, but it determines the resolution of the spatial binning according with the intensity of the flux at each spaxel, see \citet{Sanchez:2015ab} and \citet{Cid-Fernandes:2014aa} for more details. Therefore, for each galaxy we do not necessary achieve a S/N of 50. In Figure \ref{figSN} we show the S/N distribution {\bf of the binned spectra} for all galaxies in our initial sample at the three radial regions. 
{For the innermost radial region, we obtain that the averaged S/N ratio is higher or equal to 50 in most of the galaxies, whereas for the outermost radial bin, the S/N ratio peaks in a value around 5.}
The minimum S/N that we get is about 2. According to Table 1 of \citet{Sanchez:2016ab}, the precision of Pipe3D as a function of the S/N for the case of the stellar age is on the order of 0.14 dex for a S/N of 63.6, 0.13 dex for a S/N of 32.2, 0.1 dex for a S/N of 6.9, and 0.07 dex for a S/N of 3.4. Nevertheless, the precision is on the order of $\approx0.1$ dex, that are similar to the reported accuracy of other procedures \citep[e.g., Starlight,][]{Cid-Fernandes:2014aa}.
Examples of the obtained S/N maps are shown in the left lower panels of Figures \ref{test1} and \ref{test2} for a Milky Way-sized and a low-mass galaxy, respectively. 

The next step is to fit each spectrum {binned to reach a S/N of 50} to a reduced stellar template to derive the kinematics properties and the dust attenuation, adopting the \citet{Cardelli:1989aa} extinction law. Then, the stronger emission lines are fit using a set of Gaussian functions to remove them from the binned datacubes. Finally, the continuum is fitted with a linear combination of SSP templates (dust attenuated), convolved and shifted to take into account the derived kinematics,  by using the values provided by the previous analysis.\footnote{ We note that our procedure assumes that all components (or SSP ages) at every point in the galaxy share the same kinematics. This assumption is strong and by relaxing it might have some effect on the results. This problem has not been yet discussed in detail in the literature.} 

The adopted stellar library is described in detail by \citet{Cid-Fernandes:2013aa}. It comprises 156 templates that cover 39 stellar ages (1 Myr to 14.2 Gyr), and 4 metallicities ($Z/Z_{\odot}=$ 0.2, 0.4, 1, and 1.5). These templates were extracted from a combination of the synthetic stellar spectra from the GRANADA library \citep{Martins:2005aa} and the SSP library provided by the MILES project \citep{Sanchez-Blazquez:2006aa, Vazdekis:2010aa, Falcon-Barroso:2011aa}, {\bf using a Salpeter IMF. } This library has been extensively used within the CALIFA collaboration in different studies \citep[e.g.][]{Perez+2013, Cid-Fernandes:2013aa, Gonzalez-Delgado:2014aa}. We present two examples of the stellar spectra fits in the upper right panels of Figures~\ref{test1} and ~\ref{test2}. Then, by taking into account the mass-to-light ratio of each individual template within the library ($M/L_{i,j,k}$), the stellar mass-loss ($f_{i,j,k}$), the weight in light of each component derived from the fitting ($c_{i,j,k}$), and the internal dust corrected surface brightness corresponding to each individual Voxel ($I_i$), it is possible to derive the stellar mass density at each particular look back time using the formula $\mu_{i,j}=\sum_{k=0}^{met}f_{i,j,k}c_{i,j,k}M/L_{i,j,k}I_i.$ The index $i$ denotes the index of the Voxels, $j$ indicates the age of the SSP template and $k$ identifies the metallicity of the SSP.

\subsection{The Mass Growth Histories}
\label{Method_MGHs}

We define the normalized stellar MGH as $\mathcal{M}(t)/\mathcal{M}_0$, where $\mathcal{M}(t)$ is the cumulative stellar mass (taking into account the stellar mass loss) at a given epoch $t$, and $\mathcal{M}_0$ is this mass at the final epoch corresponding in principle to the observed redshift $z$;\footnote{The $\mathcal{M}_0$ masses are slightly different to the NSA masses because the latter are calculated (1) all at the same redshift ($z=0.1$), (2) with a different method than us and using the SDSS data rather than the IFS SDSS-IV data, and (3) using the Chabrier IMF instead of the Salpeter one. In spite of these many differences, the NSA masses are shifted at all masses on average by only $\approx -0.14$ dex with a small scatter and no dependence on mass. The main reasons of this difference is due to the different IMFs used in both cases.} the cumulative SFH is related to the MGH but without taking into account the stellar mass loss. From the SPS technique in Pipe3D, we recover the cumulative SFH resolved at $39$ stellar ages for each Voxel. To study the radial differences in the mass assembly of galaxies, we calculate the normalized MGHs within a given area $A$ (a circle or an annulus) as follows: \begin{equation}\left(\frac{\mathcal{M}_j}{\mathcal{M}_0}\right)_{A}=\frac{\sum_{A}\mu_{i,j}A_{}+\mathcal{M}_{j-1}}{\sum_{j=0}^{age} \sum_{A(R)}\mu_{i,j}A_{}}, \label{MGH} \end{equation} where $\mathcal{M}_j$ is the accumulated stellar mass within the area $A$ at the time $j$, and the sum over $A$ is for all the $i$ Voxels within this area. In the lower medium and right panels of Figures~\ref{test1} and ~\ref{test2}, we show examples of the recovered SFHs and MGHs, respectively, for the already mentioned massive and low-mass galaxies.  The shaded color areas in the MGH panels are the error estimates for each radial MGH. The error comes from the given errors on the stellar mass maps from Pipe3D (per template age) plus the age error estimated as the time step between the SSP templates. We define this time step as the time at which the archaeological method can resolve temporary the MGHs, see \S\S \ref{age-uncertainties} for more details.

We apply eq. (\ref{MGH}) to infer the MGHs within the three radial galaxy regions mentioned above: $0<R<0.5R_{50}$, $0.5R_{50}<R<R_{50}$, and $R_{50}<R<1.5R_{50}$. The total galaxy MGH is calculated within $R \leq 1.5R_{50}$.
The MaNGA galaxies are within a redshift range of $z=0.007$ to $z=0.148$. This redshift range corresponds to a look-back time (LBT) interval of $\approx 2$ Gyr. Therefore, the fossil record method recovers the MGHs and final stellar masses through different initial LBTs for each galaxy. In order to compare the normalized MGHs among all galaxies in the sample, the same initial LBT (where the final stellar mass $\mathcal{M}_0$ is defined) should be used. For an approximation to this, we select only galaxies with $z<z_{\rm lim}$ and assign them the same initial LBT, equal to the time corresponding to $z_{\rm lim}$ (LBT$_{\rm lim}$). Then, we generate a grid of ages from LBT$_{\rm lim}$ to the maximum resolved SSP age ($14.12$ Gyr) and calculate by interpolation the values of the normalized masses at these ages.  We redefine the final stellar mass as the accumulated stellar mass at $z_{\rm lim}$ within $1.5R_{50}$. Hereafter, we refer to this approach as the \textit{truncation} method.

We explore four redshift limits: $z_{\rm lim}=0.037$, $z_{\rm lim}=0.074$, $z_{\rm lim}=0.111$, and $z_{\rm lim}=0.148$. These limits cover the range of MaNGA redshifts in four uniform bins. With the largest $z_{\rm lim}$, of course, we recover the largest number of galaxies, however significant late MGH information gets lost for those galaxies that have observed redshifts much lower than $z_{\rm lim}$. The latter is a problem for galaxies with active late SF-driven mass growth; this is the case of most of the low-mass and dwarf galaxies. In the lowest redshift sub-sample ($z\leq0.037$) there are $454$ galaxies with a corresponding LBT$_{\rm lim}$ of $0.5$ Gyr. In the highest redshift sample ($z\leq0.148$), all the $533$ galaxies are included but with a LBT$_{\rm lim}$ of $2$ Gyr; that is, all the MGHs are interpolated to start from a time 2 Gyr ago, which correspond typically to yet active epochs of mass growth for low-mass/dwarf galaxies.    

With the truncation method we attempt to homogenize the MGHs to the same initial LBT. However, some biases on the MGHs can be introduced in this operation, given that each galaxy has actually a different initial LBT (initial $z$). This is why 
we will use also a sub-sample of galaxies in a narrow redshift sample, $z=0.037\pm0.005$. Hereafter, we refer to this approach as the \textit{redshift selection} method. This selection ensures that the initial LBT for the selected galaxies is almost the same for all the galaxies in the sub-sample. In this case, the interpolation for reconstructing the MGHs at the same redshift cut is not required.
The obvious problem is that in such a narrow redshift range, there are not too many galaxies (only 192). The mean MGHs calculated for the redshift selection sub-sample will be used as a kind of control sample to check for potential biases in the mean MGHs calculated from the more numerous redshift truncation sub-samples.

\subsubsection{Mean MGHs in Stellar Mass Bins}
\label{meanMGHs}

One of our aims is to explore the spatially-resolved MGHs as a function of mass. For this, we will calculate the mean of the MGHs in four different stellar mass bins:  $10^{8.5}<\mathcal{M}_0/\solarm<10^{9.3}$ (dwarf galaxies), $10^{9.3}<\mathcal{M}_0/\solarm<10^{10}$  (low mass galaxies), $10^{10}<\mathcal{M}_0/\solarm<10^{10.7}$ (intermediate mass galaxies), and $10^{10.7}<\mathcal{M}_0/\solarm<10^{11.2}$ (high mass galaxies). As mentioned above, we do not take into account galaxies more massive than $10^{11.2} \solarm$.
%

The galaxies in MaNGA are selected to have a roughly flat stellar mass distribution, that is, an equal number of galaxies per mas bin. This selection criteria returns a much higher number of massive galaxies with respect to less massive ones from what we expect from the local galaxy stellar mass function. Since massive galaxies are rare, such galaxies should be selected from larger volumes and therefore, larger redshifts. The shortcoming of such a selection for our study is that more massive galaxies tend to have larger initial LBTs than less massive ones, and since we need to cut the LBT at the same epoch for all galaxies, this forces us to make the cut at the higher redshifts of the observed massive galaxies. Therefore, we lose the late evolution of the less massive galaxies that are observed at lower redshifts (see above). {Then, we need to find a compromise between not losing many (mostly massive) galaxies and not losing the latest evolutionary stages of (mostly low-mass) galaxies. } We explore the differences in the stellar mass histograms between the lowest and highest $z_{\rm lim}$ sub-samples. For the former ($z\leq0.037$; 454 galaxies), there are $63$ galaxies within the $10^{8.5} < \mathcal{M}_0/\solarm < 10^{9.3}$ mass bin, and the maximum number ($200$ galaxies) is attained in the $10^{10}< \mathcal{M}_0/\solarm <10^{10.7}$  mass bin. In comparison, for the latter sub-sample ($z\leq0.148$; 533 galaxies), we obtain $66$ galaxies in the $10^{8.5} <\mathcal{M}_0/ \solarm < 10^{9.3}$ mass bin whereas the maximum number ($202$ galaxies) is attained also in the $10^{10}<\mathcal{M}_0/\solarm <10^{10.7}$ mass bin. Thus, for masses $<10^{10.7}\ \solarm$ both sub-samples have similar number distributions. The strong difference is only in the most massive bin ($10^{10.7}<\mathcal{M}/\solarm <10^{11.2}$), where there are $63$ galaxies at $z\leq0.037$ and $139$ at $z\leq0.148$. Therefore, for the highest mass bin, $\approx 55\%$ of the galaxies observed at $z\leq0.148$ are lost in the $z\leq0.037$ sub-sample. 

As mentioned above, for dwarf and low-mass galaxies, late evolutionary times are relevant, and since the number of them remains roughly the same in the sub-samples at higher $z_{\rm lim}$, we consider the lowest redshift sub-sample ($z\leq0.037$) as the optimal for carrying out our study. In this way we guarantee reliable MGHs for dwarf and low-mass galaxies from $\approx 500$ Myr, though in demerit of a larger statistics for the most massive galaxies (by lowering more $z_{\rm lim}$, we can follow the MGHs from even more recent epochs but the loss in number of galaxies at all mass bins become already significant). In any case, we will present also the mean MGHs by using the full galaxy sample (up to the highest $z_{\rm lim}$), but with the initial LBT of all galaxies re-normalized to $z_{\rm lim}=0.148$, that is from $\approx 2$ Gyr.

Once the galaxy sub-sample is selected, within each mass bin we stack the radially-resolved MGHs and calculate the mean MGHs for the three radial regions, as well as the corresponding variances and errors of the mean. For each of the three radial regions, the MGH variance in the mass bin $k$ is defined as \begin{equation}\sigma(t)^2_k=\frac{\sum_{g}^{N_k}\left(\left\langle\frac{\mathcal{M}(t)}{\mathcal{M}_0}\right\rangle_k-\left(\frac{\mathcal{M}(t)}{\mathcal{M}_0}\right)_g\right)^{2}}{N_k-1},\label{variance}\end{equation} where $N_k$ is the number of galaxies within the mass bin $k$, $g$ stands for each galaxy in this bin, and $\left\langle\frac{\mathcal{M}(t)}{\mathcal{M}_0}\right\rangle$ is the average spatially-resolved MGH in the mass bin. The variance quantifies the population scatter. The standard error of the mean, defined as $\Delta \langle\frac{\mathcal{M}(t)}{\mathcal{M}_0}\rangle_k=\sigma(t)_k/\sqrt{N_k}$, quantifies how precise is the inferred true average of the given population. 

Since the normalized MGHs at different radii are expected to display a large variety of shapes, it is important to have an estimate of the statistical significance of the corresponding stacked MGHs as well as of the distribution of the individual MGHs around the mean.  For this, we perform a $\chi^2$ analysis. For each galaxy $g$ within the mass bin $k$ and for a given radial region, we define the $\chi^2$ of its MGH as:\begin{equation}\chi^2_g=\sum_t^{N_t}\frac{\left(\left\langle\frac{\mathcal{M}(t)}{\mathcal{M}_0}\right\rangle_k-\left(\frac{\mathcal{M}(t)}{\mathcal{M}_0}\right)_g\right)^2}{\sigma(t)_k^2 (N_t-1)}, \label{chi2}\end{equation} where $N_t$ is the number of stellar ages larger than LBT$_{\rm lim}$ used by Pipe3D. Hence, we obtain $N_k$ values of $\chi^2_g$ per mass bin and for each radial region. If the population of the respective MGHs follows a normal distribution, then the mean MGH and variance ($\langle\frac{\mathcal{M}(t)}{\mathcal{M}_0}\rangle_k$, $\sigma(t)_k^2$) are statistically representative of the global behavior of the population. A normal distribution of the MGHs implies that the distribution of $\chi^k_g$ follows a reduced $\chi^2$ distribution with $N_t-1$ degrees of freedom. We will apply this test to the obtained MGHs and their means and variances.

\section{The mean mass growth histories} 
\label{results-meanMGHs}

{
Following, we present our results on the mean global (\S\S \ref{globalMGHs}) and radial (\S\S \ref{r-meanMGHs}) MGHs for galaxies in the sub-sample truncated at $z_{\rm lim} = 0.037$; as discussed in \S\S \ref{Method_MGHs}, this sub-sample seems to be the most adequate for studying the MGHs, as a compromise in number and late-time evolution recovery. In \S\S \ref{age-uncertainties}, we estimate the age uncertainties on the calculated MGHs. In \S\S \ref{tests}, we test the results obtained with the truncated method against possible biases introduced in the normalized MGHs due to forcing them to an equal initial LBT for all galaxies, and compare these results with those obtained for the full sample ($z_{\rm lim} = 0.148$), where the MGHs are truncated at a larger LBT than in the $z_{\rm lim} = 0.037$ sub-sample. Finally, in \S\S \ref{MGH-properties}, the global and radial mean MGHs are presented for the populations of blue/red, star-forming/quiescent, and late-/early-type galaxies in our four mass bins. 
}

\begin{figure}
\begin{center}
\includegraphics[width=\linewidth]{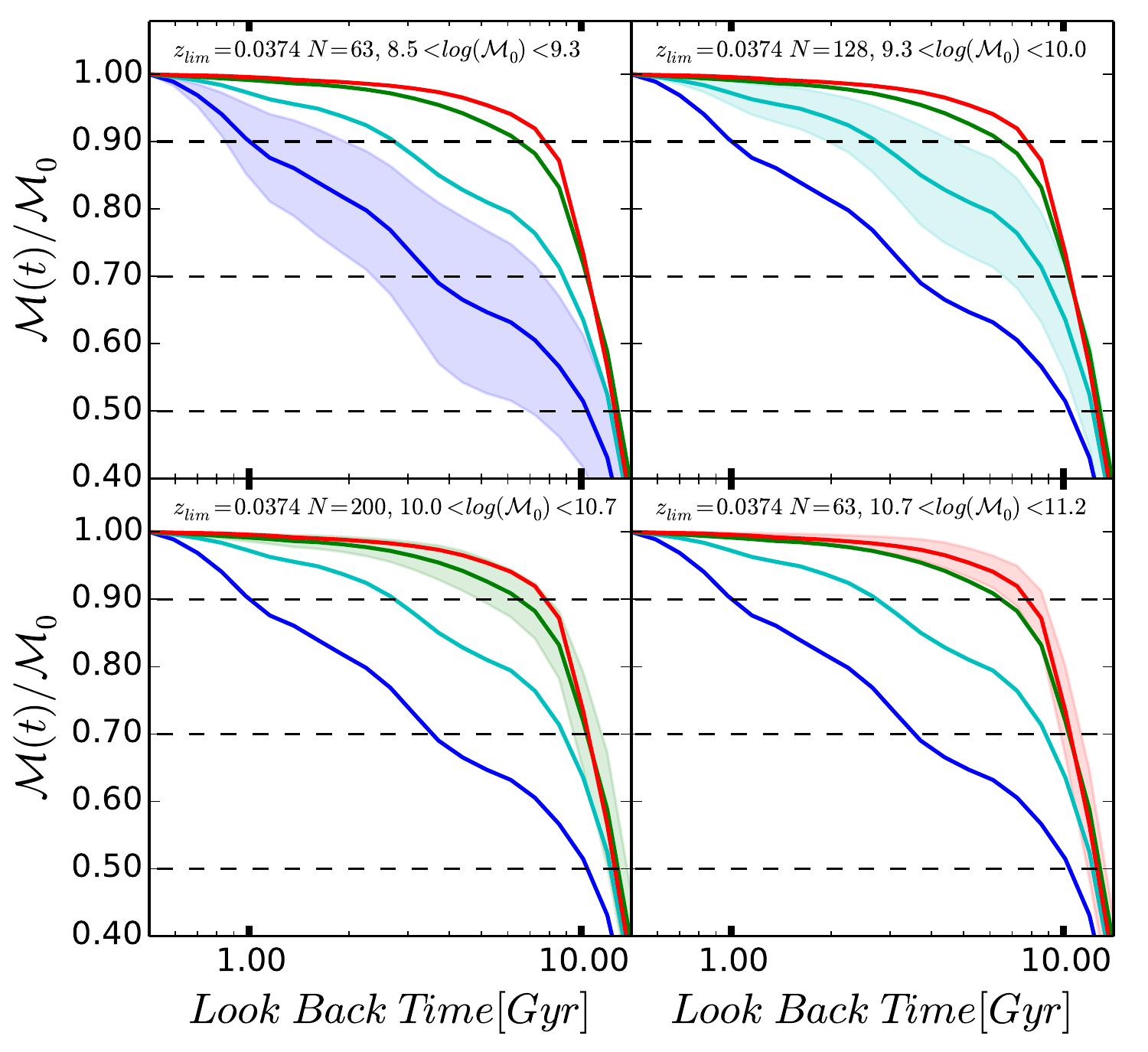}
\end{center}
\caption{Global mean MGHs integrated within $R<1.5R_{50}$ obtained from the truncated method with $z_{\rm lim}=0.037$. Each line (repeated in the four panels) corresponds to the mean in one of our four mass bins:  $10^{8.5} < \mathcal{M}_0/\solarm < 10^{9.3}$ (blue), $10^{9.3}< \mathcal{M}_0/\solarm <10^{10.0}$ (cyan), $10^{10.0}<\mathcal{M}_0/\solarm <10^{10.7}$ (green), and $10^{10.7}<\mathcal{M}/\solarm <10^{11.2}$ (red). The respective population standard deviations are shown in each panel with the shaded regions. }
\label{fig02a}
\end{figure}

\subsection{Global mean MGHs: downsizing}
\label{globalMGHs}

The global mean MGHs (within $1.5R_{50}$) in our four mass bins are plotted in Figure \ref{fig02a}. In each panel we show the number of galaxies and the population standard deviation (shaded area) of the corresponding mass bin. The more massive the galaxies are, the earlier they assembled their total stellar masses on average. This trend, called archaeological downsizing, is well known from previous fossil record inferences, look back empirical studies, and semi-empirical approaches (see \S\S \ref{comparison} below for references). According to our results, dwarf galaxies ($\mathcal{M}_0\lesssim10^{9.3}\solarm$) formed, on average, the 50\%, 70\%, and 90\% of their masses at $\approx 10$, $3$, and $1$ Gyrs ago, respectively, though the scatter among the individual MGHs is large. These galaxies have delayed SFHs/MGHs with respect to the more massive ones, and present a large diversity of histories, evidencing a quite episodic SFH. Galaxies in the more massive bins show that a significant fraction of their final masses (defined at LBT$_{\rm lim}=0.5$ Gyr) were assembled long ago. For example, in the $10^{9.3}<\mathcal{M}_0/\solarm < 10^{10}$ bin, the 70\% of the final mass was assembled on average at LBTs of $\approx 8.3$ Gyr; for more massive galaxies, these LBTs are even larger. We see also a trend of less scatter (diversity) in the global MGHs as more massive are the galaxies. {However, this can be partially due to the low age resolution of the fossil record method at large LBTs (see \S\S \ref{age-uncertainties} below), when the massive galaxies were in their active phases of mass assembly.  }

\begin{figure*}
\begin{center}
\includegraphics[width=0.95\linewidth]{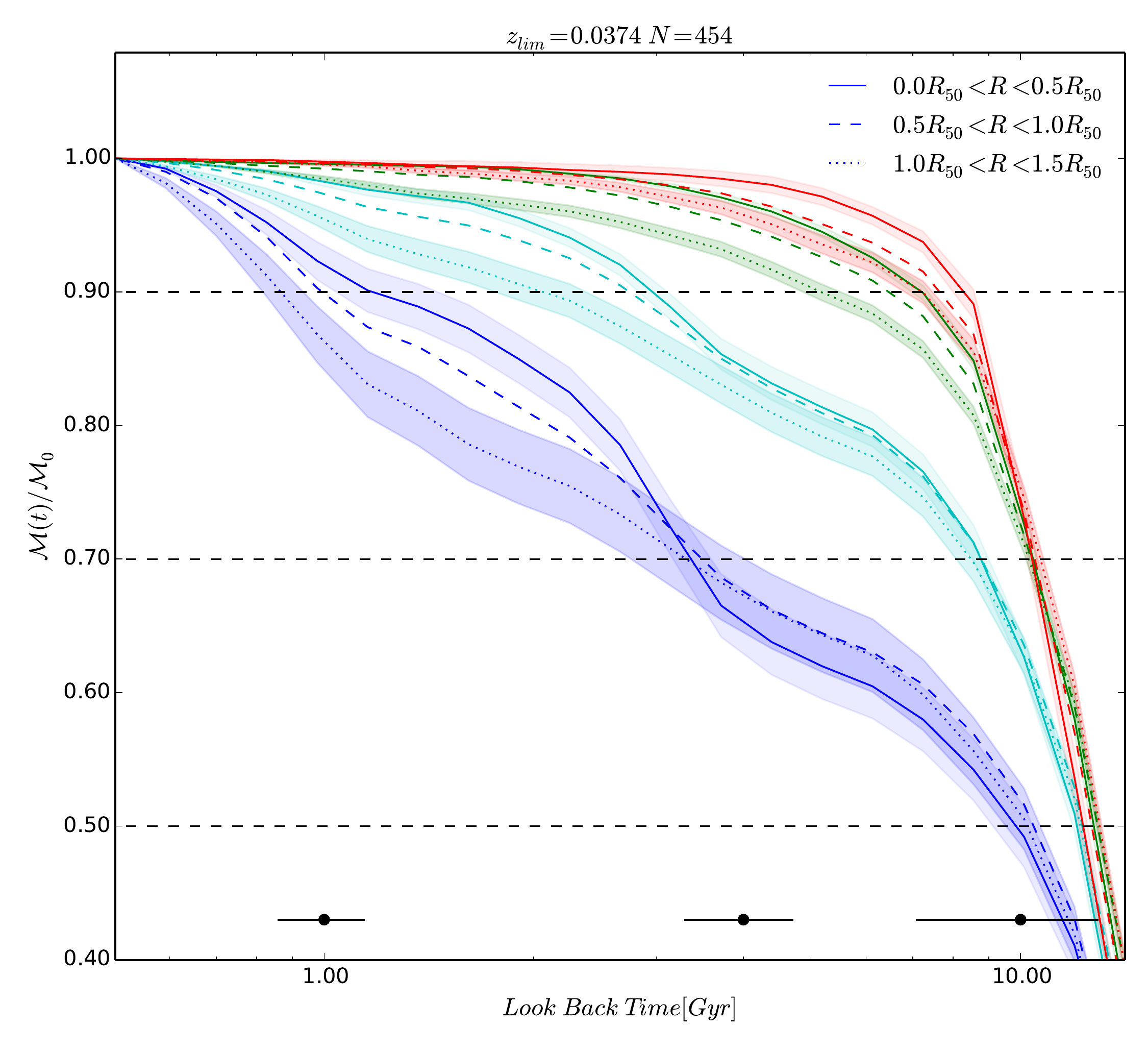}
\end{center}
\caption{Radial mean stellar MGHs obtained from the truncated method with $z_{\rm lim}=0.037$ for our four mass bins (the color code is the same as in Fig. \ref{fig02a}). The radial stacking was performed in three regions:  $R<0.5R_{50}$ (solid lines),  $0.5R_{50}<R<R_{50}$ (dashed lines),  and $R_{50}<R<1.5R_{50}$ (doted lines). The shaded areas represent the errors on the mean (not shown for the intermediate radial region). The horizontal error bars are the estimated age uncertainties in the recovery of the MGHs at three LBTs: 1, 4 and 10 Gyr (see \S\S  \ref{age-uncertainties}). 
}\label{fig02}
\end{figure*}
\subsection{Mean MGHs at different radial regions}
\label{r-meanMGHs}

For all the mass bins, the mean MGHs at different radial regions show that at a 90\% of the assembled mass in these regions, the mass growth is proceeding on average from the inside out (Fig. \ref{fig02}).\footnote{Our working hypothesis is that stars were formed within the observed (wide) regions in such a way that their formation times correspond to the mass assembly time of these regions. This hypothesis is not valid under strong radial migration and significant accretion of stars formed ex-situ; see subsection \ref{migration} below for a discussion on this question.} The differences between the times that the innermost (solid lines) and outermost (dotted lines) regions assembled 90\% of their masses are larger than the standard error of the means (shaded regions), excepting for the $10^{9.3}<\mathcal{M}_0/\solarm<10^{10}$ mass bin (estimates of the age uncertainty in the mean MGHs at different ages are shown with the horizontal error bars in Fig. \ref{fig02}; see \S\S \ref{age-uncertainties} below for details). 

For dwarf galaxies ($\mathcal{M}_0\leq10^{9.3}\solarm$), the inside-out mode remains until $\approx 3.3$ Gyr ago, when the fraction of both inner and outer mass assembled is $75-80\%$. For earlier times and lower mass fractions, the differences among the mean MGHs for each radial region are smaller than their error on the mean. Individually, we observe inner and outer MGHs that cross each other one or more times, suggesting periods of outside-in an inside-out formation (or strong radial migrations). Thus, our results suggest for dwarfs galaxies, a stellar mass growth from a not well defined gradient at early epochs to an inside-out formation at late epochs. This trend is similar for the low-mass galaxies ($10^{9.3}<\mathcal{M}_0/\solarm < 10^{10}$) but the inner and outer mean MGHs are closer than in the case of dwarfs. For $\mathcal{M}_0>10^{10}\solarm$, a trend of inside-out formation mode is seen since $\approx 9-10$ Gyr ago (when 80-85\% of the inner and outer region stellar masses were formed). At earlier epochs, the differences in the mean radial MGHs become negligible. 

\begin{figure}
\begin{center}
\includegraphics[width=\linewidth]{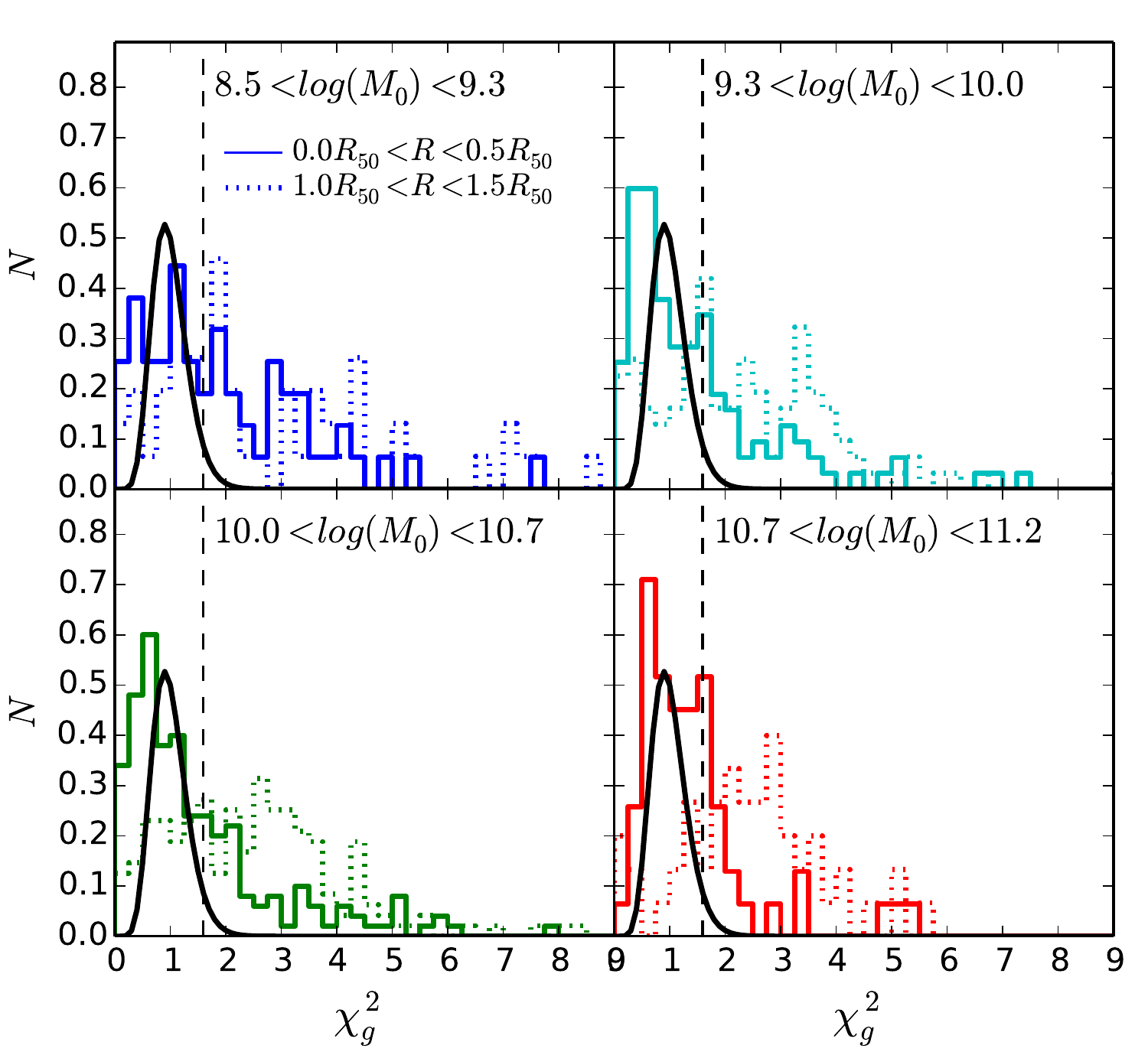}
\end{center}
\caption{Distributions of the individual $\chi^2_g$ values in each mass bin for the innermost (solid lines) and outermost (dotted lines) stellar MHGs. The black solid line represents $\chi^2$ distribution with $N_t-1$ degrees of freedom. The black dashed lines represent the $\chi^2$ value at which the ideal distribution returns a $2\sigma$ confidence limit. }\label{fig07}
\end{figure}

The general trends presented in Fig. \ref{fig02} refer to stacked (mean) MGHs in different mass bins. How representative are these stacked MGHs of the individual MGHs? To answer this question we study the distributions of the $\chi^2_g$ estimator of each individual spatially-resolved MGH presented in subsection \ref{meanMGHs} (see eq. \ref{chi2}). Figure \ref{fig07} shows these distributions for the innermost (solid lines) and outermost (dotted lines) regions. The black solid lines correspond to a reduced $\chi^2$ distribution with $N_t-1$ degrees of freedom, where $N_t$ is the number of time points used to interpolate the MGHs. The vertical dashed lines show the $2\sigma$ confidence level; MGHs with $\chi^2_g$ values larger than this level, strongly depart from the corresponding mean MGH.

For dwarf galaxies ($10^{8.5}\solarm<\mathcal{M}_0<10^{9.3}\solarm$), only $\approx 49\%$ and $30\%$ of the MGHs (for the innermost and outermost radial region respectively) are within the $2\sigma$ confidence level. For low-mass galaxies  ($10^{9.3}\solarm<\mathcal{M}_0<10^{10}\solarm$), $62\%$ of the innermost MGHs ($R<0.5R_{50}$) are within the $2\sigma$ level, while in the case of the outermost MGHs ($R_{50}<R<1.5R_{50}$) this fraction drops to $34\%$. For the intermediate-mass galaxies ($10^{10}<\mathcal{M}_0/ \solarm<10^{10.7}$), we observe a similar trend: $63\%$ of the innermost MGHs remain within the $2\sigma$ level, whereas only $29\%$ do it for the outermost MGHs. Note that the $\chi_g^2$  distribution of the innermost MGHs approaches well the reduced $\chi^2$ distribution. For the massive galaxies ($10^{10.7}<\mathcal{M}_0/ \solarm<10^{11.2}$), $68\%$ of the innermost MGHs remain within the $2\sigma$ level whereas only $20\%$ do it for the outermost MGHs. Therefore, for galaxies more massive than $\sim 10^{9.3} \solarm$, more than $50\%$ of the individual innermost MGHs are well represented by the corresponding mean MGHs plotted in Figure~\ref{fig02}. For the outermost regions, the fraction drops to $20-30\%$. These results suggest that the innermost regions of these galaxies assembled typically more regularly than the outermost regions. For dwarf galaxies,  both the inner and outer regions seem to have assembled with a large diversity of histories in most of the cases; due to this diversity, the mean inner and outer MGHs remain only as a very rough description of most of the respective individual MGHs.

\subsection{Age uncertainties in the MGHs}
\label{age-uncertainties}

In \S\S\ref{caveats} we will discuss on the intrinsic limitation of the fossil record method regarding the age determination of the oldest SSP contributions to the observed spectrum. Such a limitation introduces a large uncertainty in the earliest phases of the recovered MGHs. {This is why only a few stellar spectrum templates separated by large time periods are used at large LBTs: the spectral features of old SSPs barely change with age. On the other hand, the spectral features of young SSPs change significantly with age, and therefore, a large number of spectrum templates separated by short time periods should be used.  We estimate the uncertainties in age for our mean MGHs based on the age separations between the stellar spectrum templates since these separations correspond namely to the age ``resolution'' of the fossil record method. At LBTs of 10 Gyr we obtain an uncertainty of $\approx \pm 3$ Gyr, while at LBTs of 1 Gyr we obtain an uncertainty of $\approx \pm 150$ Myr (see \S\S \ref{caveats} for a discussion). In Figure~\ref{fig02}, the estimated age uncertainties at three representative LBTs are shown with horizontal error bars. The uncertainty is large for large LBTs but significantly reduces at low LBTs. Thus, the mean MGHs at early epochs, when $\sim 50-70\%$ of the final mass was assembled, are highly uncertain.  Since all the galaxies and radial bins are analyzed with the same fossil record method, in spite of the large uncertainties at the level of individual determinations at these stages, the differences in the mean MGHs among different mass bins, and even among different radial bins, are probably systematic. 
}


\subsection{Testing the results from the $z_{\rm lim}=0.037$ sub-sample}
\label{tests}

\begin{figure}
\begin{center}
\includegraphics[width=\linewidth]{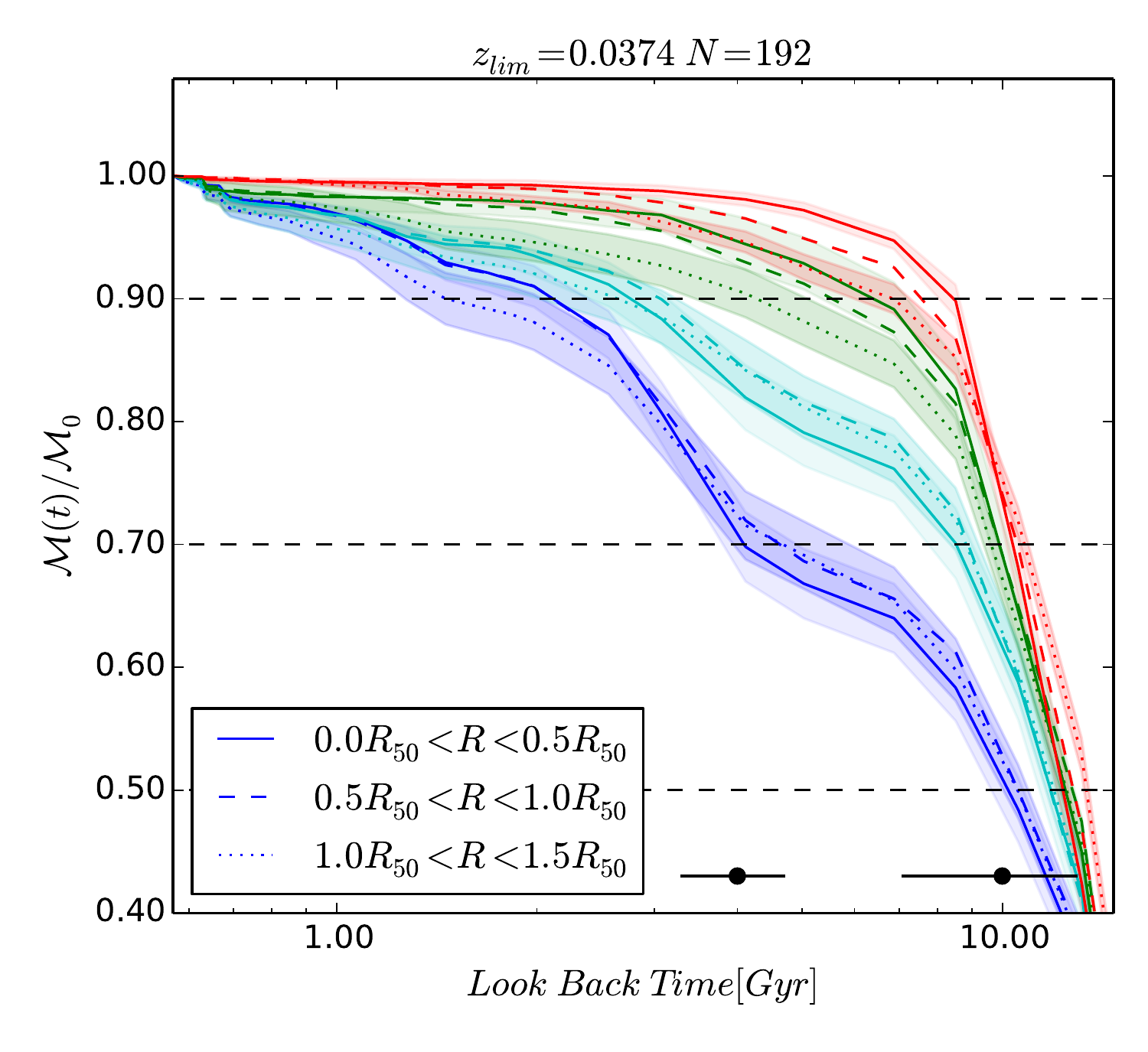}
\end{center}
\caption{Mean radial stellar MGHs as in Figure~\ref{fig02}, but for galaxies only in the narrow redshift bin $z=0.037\pm 0.005$ (redshift selection method). 
The line and color codes are the same as in Figure~\ref{fig02}.}\label{fig03}
\end{figure}

The truncation method at a given $z_{\rm lim}$ (for example 0.037) used above could introduce some bias in the normalized MGHs because all galaxies with lower redshifts than $z_{\rm lim}$ are forced to have the same initial LBT, equal to the epoch corresponding to $z_{\rm lim}$. To test against this possible bias we use the sub-sample selected in the narrow redshift range $z=0.037\pm0.005$. The mean normalized MGHs in the three radial regions (solid, dashed, and dotted lines) and for each of the four mass bins (red, green, cyan, and blue lines) obtained using this sample are depicted in Figure~\ref{fig03}, to be compared with Figure~\ref{fig02}. We implement a Kolmogorov-Smirnov (K-S) test to compare whether the reconstructed spatially-resolved MGHs with the truncation and redshift selection methods are compatible. At high masses ($10^{10.7}< \mathcal{M}/\solarm <10^{11.2}$) the two methods recover the same distribution within a $95\%$ confidence level. At the $10^{9.3}<\mathcal{M}/\solarm < 10^{10}$ and $10^{10} < \mathcal{M}/\solarm < 10^{10.7}$ mass bins, the two distributions are compatible with each other at the limit of the critical value criterion of the K-S test. At the lowest mass range ($10^{8.5} < \mathcal{M}/\solarm < 10^{9.3}$), we can say that the two reconstructed MGHs at the different radii are similar. The fact that the mean MGHs obtained from the $z_{\rm lim} = 0.037$ sub-sample are not significantly different from those obtained from the $z=0.037\pm0.005$ sub-sample, shows that the truncation of the MGHs at the same initial LBT (LBT$_{\rm lim} =0.5$ Gyr) does not introduce a bias in the main results. Therefore, we feel confident in using the mean normalized MGHs estimated with the \textit{truncation} method for $z_{\rm lim} = 0.037$ (Figure~\ref{fig02}). For the subsamples with higher values of $z_{\rm lim}$, the K-S test show similar or worse results than for the $z_{\rm lim} = 0.037$ subsample.

\begin{figure*}
\begin{center}
\includegraphics[width=0.9\linewidth]{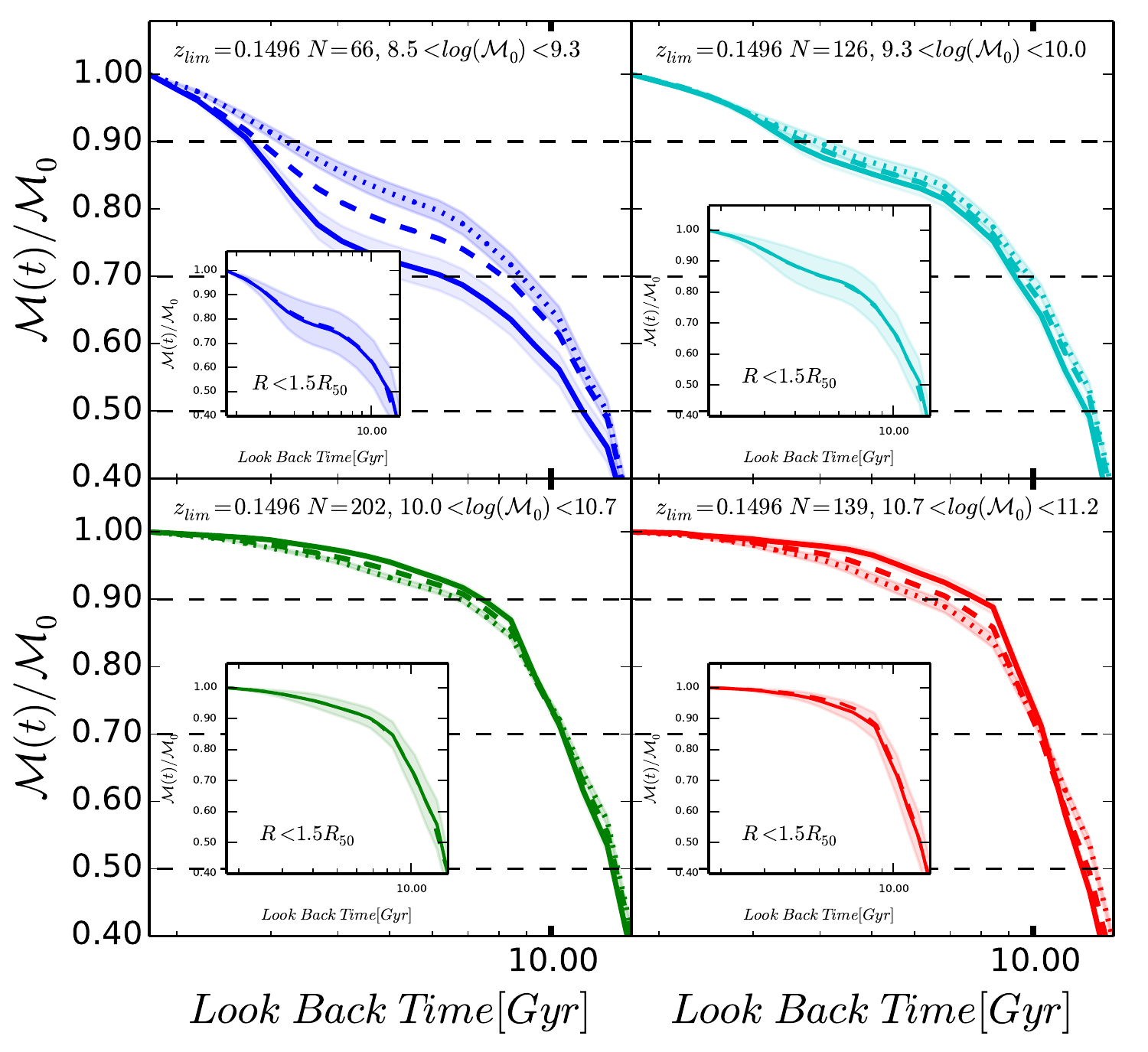}
\end{center}
\caption{Mean MGHs for the highest redshift sample ($z_{\rm lim}=0.148$) at three galaxy radial regions:  $R<0.5R_{50}$ (solid lines), $0.5R_{50}<R<R_{50}$ (dashed lines), and $R_{50}<R<1.5R_{50}$ (doted lines), and for the fourth mass bins: $10^{8.5} < \mathcal{M}_0/\solarm < 10^{9.3}$ (upper left panel), $10^{9.3}< \mathcal{M}_0/\solarm <10^{10.0}$ (upper right panel), $10^{10.0}<\mathcal{M}_0/\solarm <10^{10.7}$ (lower left panel), and $10^{10.7}<\mathcal{M}/\solarm <10^{11.2}$ (lower right panel). The MGHs start at $\approx 2$ Gyr. The color code is the same as in Figure~\ref{fig02}. The shaded areas represent the errors of the mean. The inset panels show the global mean MGHs integrated within $R<1.5R_{50}$ and the population standard deviations (same as in Figure~\ref{fig02a}). The dashed lines within the insets are the mean MGHs integrated within $R<1.5R_{50}$ from the $z_{\rm lim}=0.037$ sub-sample and re-normalized to LBT$_{\rm lim}\approx2$ Gyr corresponding to the highest redshift limit, $z_{\rm lim}=0.148$.}\label{test3}
\end{figure*}

In the previous Section, we have explained why the full galaxy sample (up to $z_{\rm lim}=0.148$) was not used for our analysis of the mean MGHs, but instead a sub-sample truncated at $z_{\rm lim}=0.037$. For completeness, we present now the mean normalized MGHs obtained using the full galaxy sample. As mentioned above, the difference in number between the $z\leq0.037$ and $z\leq0.148$ samples is mostly in the most massive bin ($10^{10.7}-11^{11.2}$ $\solarm$); in this bin, there are 76 objects more ($55\%$) in the latter sample than in the former one. The normalized MGHs calculated from the $z_{\rm lim}=0.148$ sample start from the corresponding limit LBT, $\approx 2$ Gyr, {which is larger than the 0.5 Gyr limit LBT from the $z_{\rm lim}=0.037$ sample. Therefore,  to compare the normalized MGHs of both samples we need to cut the MGHs from the latter at 2 Gyr, and re-normalize the MGH to the final mass $\mathcal{M}_0$ calculated now at this time. Thus we lose the last 1.5 Gyr of mass growth in the galaxies from the $z_{\rm lim}=0.037$ sample.}

In Figure~\ref{test3} we present the global and spatially-resolved mean MGHs from the $z_{\rm lim}=0.148$ sample for each mass bin. {Note that at difference of Fig.~ \ref{fig02}, the MGHs start now from $\approx 2$ Gyr. } For the case of the global MGHs, the corresponding ``clipped" mean MGHs from the $z_{\rm lim}=0.037$ sample are also plotted (dashed lines). In fact, they are indistinguishable from those of the $z_{\rm lim}=0.148$ sample. If any, the mean MGH of the most massive bin from the $z_{\rm lim}=0.148$ sample is slightly shifted towards later epochs than the one from the $z_{\rm lim}=0.037$ sample, but well within the standard deviation. The comparison obtained for the global MGHs are very similar for the radial MGHs. {This comparison shows us that the used galaxy sub-sample with $z_{\rm lim}=0.037$, while loses many massive galaxies with respect to the full sample ($z_{\rm lim}=0.148$), it recuperates the same mean normalized MGH of massive galaxies from the full sample. However, the advantage of the $z_{\rm lim}=0.037$ sample is that allows to follow the MGHs to later evolutionary stages (since LBT$\approx 500$ Myr), which is relevant for the less massive galaxies. 

We have also calculated the normalized MGHs for galaxies more massive than $10^{11.2} \solarm$ from the full sample. The mean MGH of them is significantly shifted towards late epochs with respect to the $10^{10.7}<\mathcal{M}_0/ \solarm<10^{11.2}$ mass bin, resembling rather the mean MGH of low-mass galaxies. As mentioned above, the most massive galaxies in the currently observed (MPL-4) Primary MaNGA galaxies are biased to be blue and star-forming (by $H_{\alpha}$ emission), being this the reason of their late stellar mass assembly. This is why we preferred do not include in our analysis galaxies more massive than $10^{11.2} \solarm$. 
}

\begin{figure*}
\begin{center}
\includegraphics[width=0.95\linewidth]{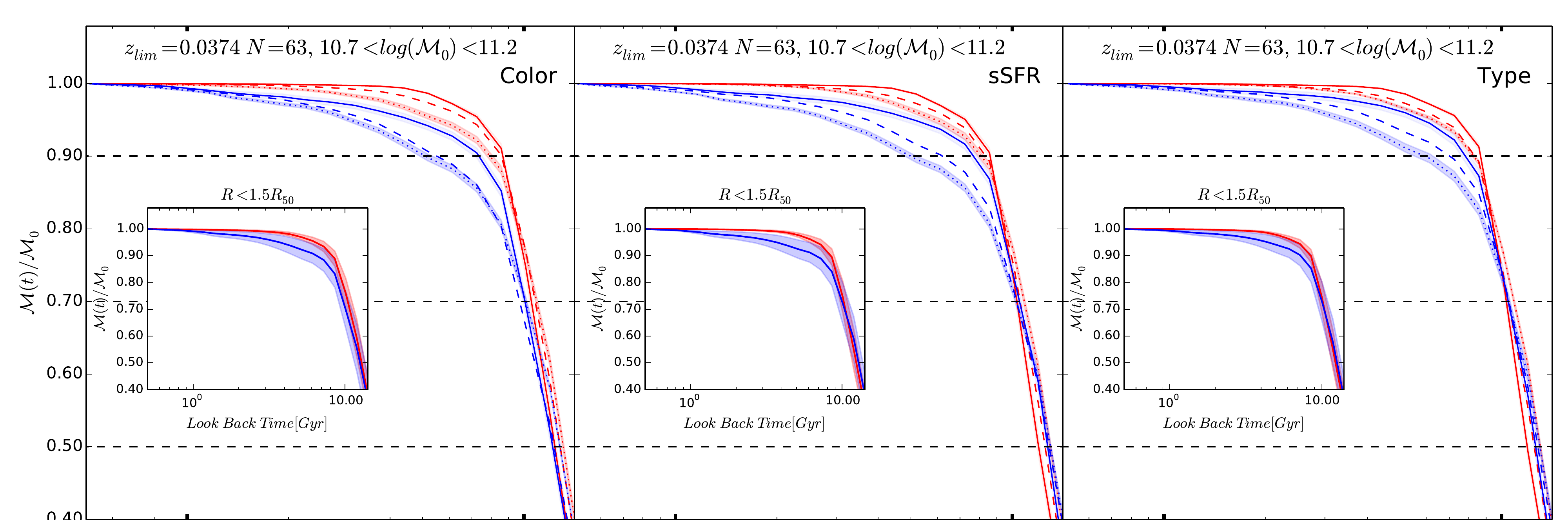}
\includegraphics[width=0.95\linewidth]{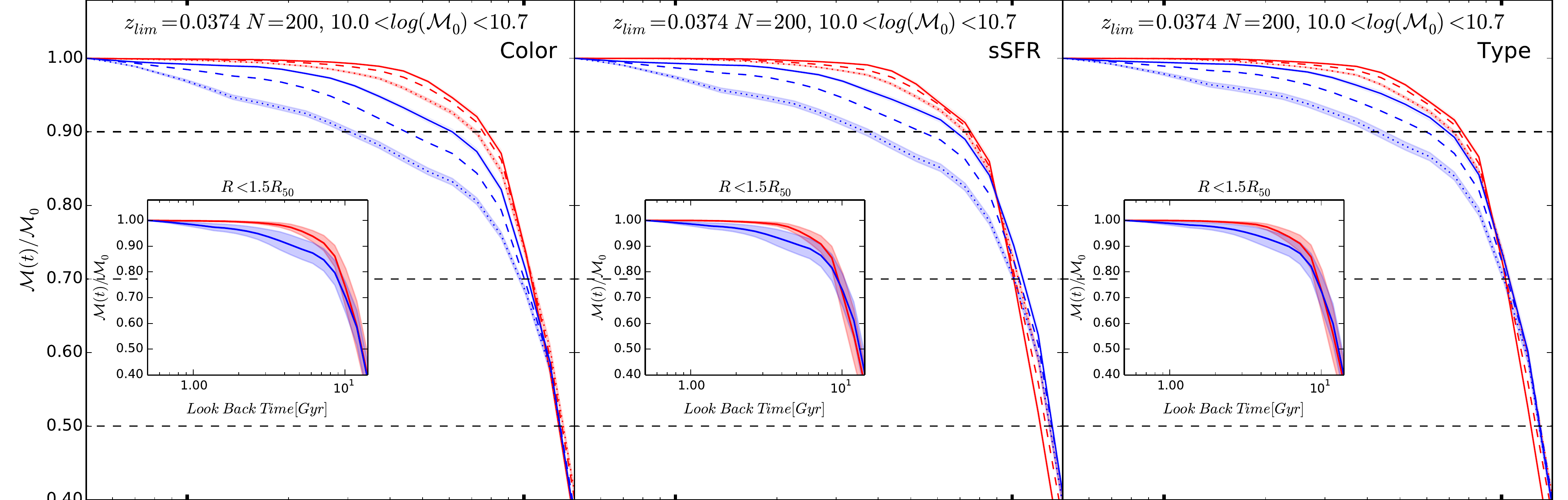}
\includegraphics[width=0.95\linewidth]{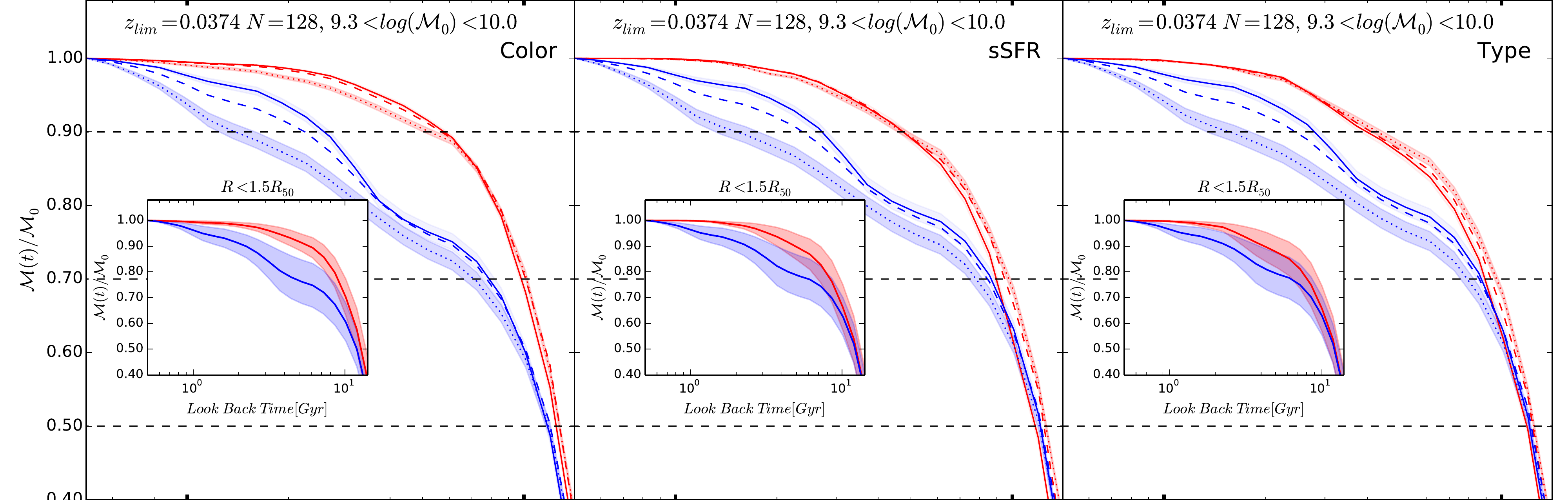}
\includegraphics[width=0.95\linewidth]{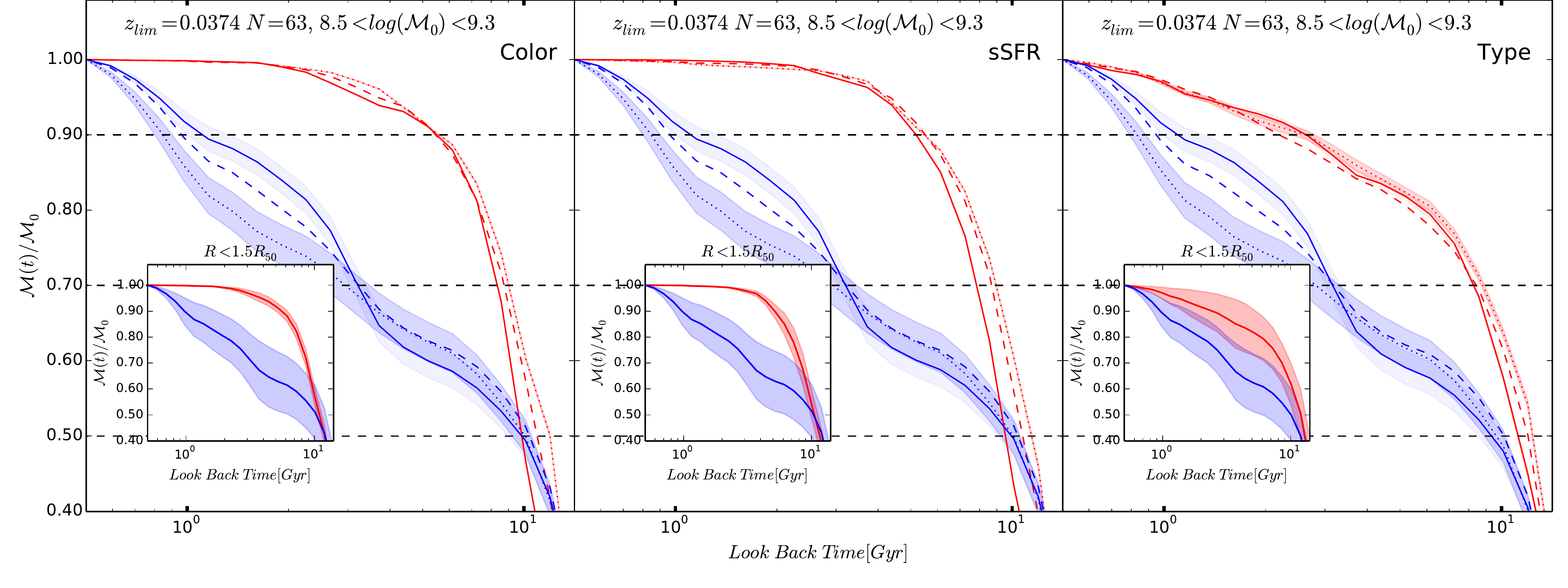}
\end{center}
\caption{Mean radial MGHs segregated by color, sSFR, and morphology; each row is for a mass bin. The line code is the same as in Fig.~\ref{fig02}. Red lines are for red/quiescent/early-type galaxies and blue lines for blue/star-forming/late-type galaxies. The mean global MGHs are shown in the insets.}
\label{fig08}
\end{figure*}

\subsection{Dependence of the MGHs on galaxy properties} 
\label{MGH-properties}

{We have reported above that the global and spatially-resolved MGHs of galaxies depend on average on their masses. However, galaxies of the same mass can have very different properties and it is of interest to explore whether the stellar MGHs depend on these properties. Given the large number of galaxies in our sample, we can separate them at least in two general populations (according to the chosen property) at each mass bin. The main properties of galaxies are their morphological type, color, and sSFR. It is well known that these properties correlate with mass; as galaxies are more massive than $\sim 10^{10.3} \solarm$, they tend to be of earlier morphological types, redder, and more quiescent, while less massive galaxies, tend to be of later morphological types, bluer and more star forming \citep[see for a review][]{Blanton+2009}. }
From the NSA, we have the global $g-r$ colors and from the outputs from Pipe3D we obtain the specific sSFRs of our MaNGA galaxy sample, and from our own visual morphological classification, we have the galaxy types. We divide the sample only into two broad populations considering each one of these properties. The criteria reported in \citet{Lacerna+2014} for the DR7 SDSS are applied to separate galaxies into blue and red, $(g-i) = 0.16 [\log(\ms/\solarm) -10.31] + 1.05$,  and into star forming and quiescent, $\log$(sSFR) $= -0.65[\log(\ms/\solarm) - 10.31] -10.87$. Regarding morphology, we define as early-type galaxies those classified as ellipticals and lenticulars, including those classified as S0a;  objects latter than these types are defined as late-type galaxies. Recall that we have excluded from our analysis strongly interacting and merging galaxies. 

{In Fig. ~\ref{fig08} we present the global (insets) and radial mean MGHs for the blue/red (first column), star forming/quiescent (second column), and late-/early-type (third column) galaxies in our four mass bins, from less to more massive ones as the lines go from top to bottom. The numbers of blue/red,  star-forming/quiescent, and late-/early-type galaxies in the $10^{10}-10^{10.7} \solarm$ bin are: 90/110, 117/83,  and 143/57, respectively. In the largest mass bin, as expected, dominate the red/quiescent galaxies with almost the same number of late-/early-type galaxies. The corresponding numbers are 22/41, 24/39 and 29/34, while in the lower mass bins, dominate the blue/star forming/late-type galaxies (59/4, 59/4 and 55/8). In the $10^{9.3}-10^{10} \solarm$ bin, the corresponding numbers are 86/42, 97/31 and 98/30.

At all masses, the global mean MGHs clearly segregate by color, sSFR, and morphological type (see the insets). Red/quiescent/early-type galaxies assemble earlier on average their masses than blue/star-forming/late-type galaxies, at least since $\sim 10$ Gyr ago. Moreover, the less massive are the galaxies, the larger are the differences, mainly because the blue/star-forming/late-type delay more and more their mass assembly as the mass is smaller, while the red/quiescent/early-type galaxies assemble most of their masses always relatively early.  For the $10^{10}-10^{10.7}$ $\solarm$ mass bin, where both populations are roughly equally represented in the bimodal distribution, the differences in the mean global MGHs between the two populations are much smaller than the differences seen in the mean global MGHs between small and giant galaxies (Figure \ref{fig02a}). 

Regarding the radial MGHs, in the $10^{10}-10^{10.7}\ \solarm$ mass bin, both galaxy populations show an inside-out formation mode since high/intermediate LBTs. However, this trend is clearly more pronounced for blue/star-forming/late-type galaxies. These galaxies form their outermost regions significantly later than in the case of red/quiescent/early-type galaxies. The differences in the radial mass growth mode between the two populations of galaxies are larger than those seen as a function of mass (Figure \ref{fig02}). The $\chi^2$ analysis of the individual MGHs show that the innermost/outermost MGHs of blue galaxies are within the $2\sigma$ deviation from their respective mean MGHs in 61\%/36\% of the cases. For red galaxies, these fractions are not too different: 64\%/27\%.

For the dwarf and low-mass galaxies, the radial mean MGHs of blue/star-forming/late-type galaxies are actually similar to those shown in Fig. \ref{fig02} (without any separation into two populations). This is because at low masses, these galaxies dominate in number. However, the mean MGHs of the few red/quiescent/early-type galaxies in these mass bins depart strongly from the corresponding mean MGHs seen in Fig. \ref{fig02}: their radial MGHs are much earlier and similar among them than the corresponding mean MGHs, which are dominated by blue/star-forming/late-type galaxies.  For the most massive bin, $10^{10.7} < \mathcal{M}_0>/\solarm <10^{11.2} $, the radial mean MGHs of red/quiescent/early-type galaxies are actually similar to those shown in Fig. \ref{fig02} since this population dominates in number at large masses. The few blue/star forming/late-type galaxies in this mass bin depart slightly from the corresponding mean MGHs seen in Fig. \ref{fig02} in the direction of a later and more pronounced inside-out mass assembly growth.  

In general, blue/star-forming/late-type galaxies assemble radially on average following the inside-out mode, being this behavior more pronounced at all epochs for the galaxies more massive that $\sim 10^{10.3} \solarm$. Instead, the red/quiescent/early-type galaxies have on average a more coeval radial mass assembly (flat gradient) at all masses; there is even a weak hint that as less massive they are, the radial growth turns on more to the outside-in mode. 
}

\begin{figure} 
\begin{center}
\includegraphics[width=\linewidth]{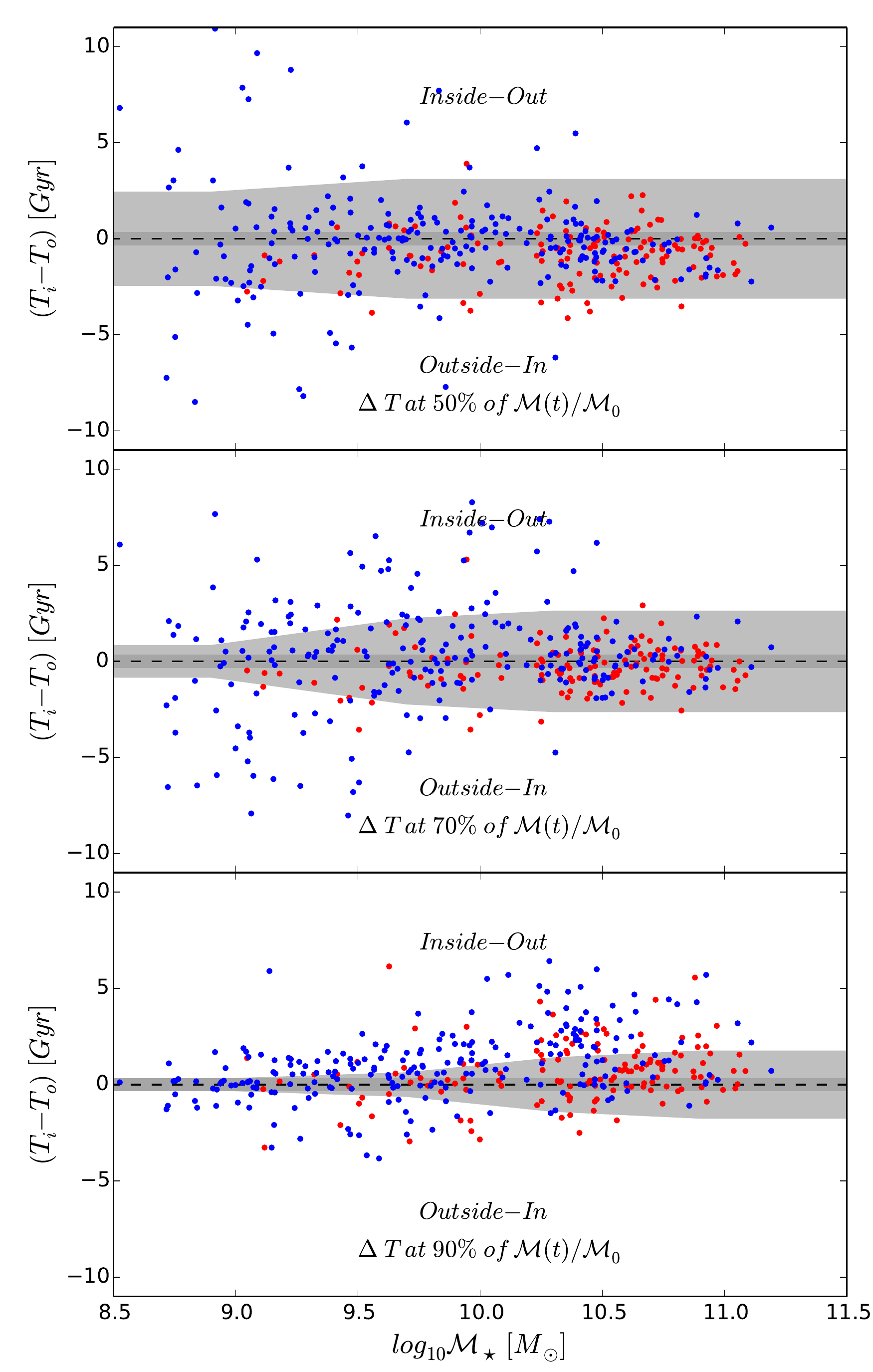}
\end{center}
\caption{Difference on the formation times for the inner and outer regions ($\Delta T_{\rm i-o}=T_i-T_o$) when the MGHs reaches their $50\%$ (upper panel), $75\%$ (middle panel) and $90\%$ (lower panel) of their total masses. For both panels, the blue points represent the selected blue galaxy sample whereas the red points represent the red galaxy sample (see text). The light grey area represents the not resolved time region of the archaeological method. The dark grey region represents the galaxy mixed time (300 Myr).}
\label{fig09}
\end{figure}

\section{Individual mass assembly gradients}\label{inMGH}
 
The results presented above refer to the average behavior inferred from the stacked spatially-resolved MGHs. As mentioned in \S\S \ref{r-meanMGHs} and \ref{MGH-properties}, an important fraction of the individual MGHs, specially those of the outermost regions, deviate strongly from their corresponding mean MGHs. So, it is of interest to study also the MGHs individually and look for some population trends as a function of mass. In order to quantify the individual MGHs with only one parameter, we calculate the difference of the LBTs between the innermost ($<0.5R_{50}$; $T_i$) and outermost ($1R_{50}<R<1.5R_{50}$; $T_o$) regions at which each MGH attains a specific percentage of these region's total masses. If $\Delta T_{\rm i-o}\equiv T_i - T_0$ is positive, then the stars in the inner region were assembled first in comparison with those in the outer region; this should correspond to an inside-out formation mode. On the contrary, a negative value of $\Delta T_{\rm i-o}$ should correspond to an outside-in mode. We consider that the inside-out or outside-in modes are undetermined for $\left|\Delta T_{\rm i-o}\right|<t_{\rm dyn}$; in this case, we can say that the SF gradient is flat. The galaxy dynamical time, $t_{\rm dyn}$, is thought as the typical timescale in which global dynamical processes occur. A conservative upper value for $t_{\rm dyn}$ in local galaxies is 300 Myr; there is a trend to lower values of $t_{\rm dyn}$ as smaller are the galaxies.

In Figure~\ref{fig09}, we depict $\Delta T_{\rm i-o}$ at the $50\%$, $70\%$, and $90\%$ mass fractions as a function of the total mass for galaxies in our $z_{\rm lim}=0.370$ subsample. For completeness, blue and red galaxies are plotted with blue and red colors, respectively.  The dark gray bands show $\pm 300$ Myr around $\Delta T_{\rm i-o}=0$ (see above). The light gray bands are estimates of the time resolution allowed by the fossil record method at different epochs and masses. 
{According to the mean MGHs (see Fig. \ref{fig02}), the LBTs when the radial MGHs reach $50\%$ of their masses are $> 10$ Gyr. At these epochs, the uncertainty in the age determination is of 2-4 Gyr. On the other hand, the LBTs when the radial MGHs reach $90\%$ of their masses are $\sim 1$ Gyr. At $1$ Gyr, the age uncertainty is $\sim \pm150$ Myr. Formally, each $\Delta T_{\rm i-o}$ value inside the light gray regions has an uncertainty due the lack of temporal resolution that does now allow to say whether the radial growth at that epoch is inside-out or inside-in. 
At the 50\% mass fractions, the majority of the galaxies have $\Delta T_{\rm i-o}$ values lower than the age uncertainty. Only a few massive (red) galaxies show evidence of a clear outside-in mode, and many of the low-mass/dwarf galaxies show clear evidence of either inside-out or outside-in mode.  At the 70\% mass fractions, most of galaxies more massive than $\sim 5\times 10^{9} \solarm$ have values of $\vert{\Delta T_{\rm i-o}\vert}$ smaller than the age uncertainty ($\approx 2.5$ Gyr), so that it is not possible to say with precision whether they are in the inside-out or outside-in mode. Less massive galaxies continue presenting a large scatter, with many of them being in a clear outside-in mode and many others being in a clear inside-out mode. At the 90\% of the mass fractions, when the age uncertainties are smaller, most of  galaxies more massive than $\sim 10^{10} \solarm$ show clear evidence of inside-out mass assembly, while for less massive galaxies, both radial growth modes are possible, with a small preference to the inside-out mode. 

In summary, due to the lack of age resolution for old stellar populations, the age differences $\Delta T_{\rm i-o}$ at epochs when $\sim 50\%$ of the innermost/outermost masses were formed are very uncertain. The situation improves for later epochs, when larger mass fractions are formed. Our results are consistent with an scenario, where the most common evolutionary trend for galaxies more massive than $\sim 5\times 10^{9} \solarm$ is that of transiting from a flat/outside-in gradient mode to an inside-out mode, being more frequent this trend as galaxies are more massive. Galaxies less massive than $\sim 5\times 10^{9} \solarm$ present large differences between the formation epochs of the innermost and outermost regions at early epochs, both in direction of the inside-out or outside-in mode; these differences decrease at later epochs.
}

\section{Discussion} 
\label{discussion}

\subsection{Comparison with previous works}
\label{comparison}

We have analyzed the galaxies from the Primary sample of the the ongoing MaNGA survey (excluding strongly interacting/merging galaxies) and inferred their global stellar MGHs. Our results, in particular for the $z_{\rm lim} = 0.037$ sub-sample (454 galaxies; see \S\S \ref{globalMGHs}), confirm for a large stellar mass range, the strong downsizing trend previously found through fossil record inferences of SDSS galaxies samples, based on observations with only one central fiber \citep[e.g.,][]{Thomas:2005aa,Panter+2007,Cid-Fernandes+2007,Gallazzi+2008,Tojeiro+2009,Thomas:2010aa}. The MaNGA galaxies are covered with many fibers across them;  in particular, for 99\% of the galaxies studied here, this coverage attains at least 1.5 $R_{50}$ (Fig. \ref{fig00}). 

A direct comparison with previous works is not possible, due to the different assumptions for the epoch at which the final mass of the galaxies is defined (this mass is relevant for normalizing the MGHs), the different mass binning and sample selection used in each work, etc. For example, in \citet{Cid-Fernandes+2007}, the reported normalized histories start from an age of $10^6$ yrs and they consider that this is equal to the initial LBT, i.e., they assume that all galaxies are observed at $z=0$ and their final masses correspond to this epoch. Moreover, these authors report results only for star forming galaxies and their histories refer to those of the mass converted into stars (cumulative SFH) instead of the histories of mass locked in stars,  i.e. discounting the mass returned to the interstellar medium as we do. A rough comparison of our global MGHs for star forming galaxies in different mass bins with those of \citet{Cid-Fernandes+2007} shows qualitatively similar behaviors. However, their histories attain the 70-90\% of the final mass later (lower LBTs) than ours. We have also calculated the global cumulative SFHs. As expected, when the stellar mass loss is not taken into account, the histories tend to be shifted to later formation epochs (smaller LBTs). However, the shifts are small and do not explain the apparent differences with the results from \citet{Cid-Fernandes+2007}. 

\citet{Leitner:2012aa} plots the archaeological cumulative SFHs in several mass bins from SDSS galaxies obtained from the ``versatile spectral analysis'' (VESPA; \citealp{Tojeiro+2007,Tojeiro+2009}). He presents both cases of SPS models used in VESPA: \citet{Bruzual:2003aa} and \citet{Maraston2005}; for intermediate ages (around 2-10 Gyrs), a given fraction of the cumulative SFH is attained earlier when using the \citet{Maraston2005} model. A rough comparison with our results, shows that in our case a given fraction of the cumulative SFH is attained even earlier than the VESPA inferences with the \citet{Maraston2005} SSPs.

In conclusion, the global (inside 1.5$R_{50}$) MGHs or cumulative SFHs inferred here for the MaNGA galaxies as a function of mass, thought qualitatively similar to those inferred previously from the SDSS galaxies with available spectral information, seem to imply an earlier mass assembly of the first 50-70\% of the final mass than in these previous studies (see below for a discussion on possible systematic differences among the fossil record inferences).

{ More recently, \citet{McDermid+2015},  by means of the fossil record method, have presented the cumulative SFHs integrated up to one effective radius for a sample of early-type galaxies from the Atlas$^{\rm 3D}$ survey \citep{Cappellari+2011}. Their results are consistent with those obtained here for early-type galaxies (Fig. \ref{fig08}). For instance, \citet{McDermid+2015} find that 50\% of the stellar mass of their sample (dominated by intermediate/massive galaxies) has been formed on average within the 2 first Gyr after the big bang (i.e, at LBTs larger than $\approx 11.7$ Gyr), in good agreement with the MGHs of our early-type galaxies.  They also find a clear downsizing trend, as in our case. However, for small masses, their cumulative SFHs imply a later mass assembly than in our case. We should bear in mind that both analysis are intrinsically different in many aspects. For example, the wavelength range covered by Atlas$^{\rm 3D}$ does not include the Balmer break and the very important absorption lines bluer than this break. Therefore, it is expected to be less sensitive to stellar populations younger than 1 Gyr, and more sensitive to recent SF traced by H$\beta$; this may produce a bias specially for galaxies with late MGHs.  Also, although the dust attenuation in early-type galaxies is not very high \citep[$A_V\lesssim 0.2$ mag;][]{Gonzalez-Delgado+2016}, it may have an effect in particular in the derivation of the young stellar components in the stellar fitting. The dust attenuation was not taken into account by \citet{McDermid+2015}, since the wavelength range preclude from including it.
}

The downsizing trend of stellar mass assembly of galaxies has also been determined from multi-wavelength observational studies of galaxy populations at different redshifts, from which the galaxy stellar mass function and/or the SFR--$\ms$ relation at different epochs are built \citep[e.g.,][see for a review \citealp{Fontanot:2009aa} and more references therein]{Cimatti+2006,Bell+2007,Drory+2008,Perez-Gonzalez:2008aa,Pozzetti+2010,Leitner:2012aa,Bauer+2013,Munoz+2015,Tomczak+2016}. The first time the concept of  downsizing was introduced comes just from one of such studies \citep[][]{Cowie+1996}. Note that with these look back studies, the MGH of an individual galaxy can not formally be followed as in the case of the fossil record method, though average histories can be constructed statistically. Semi-empirical individual stellar MGHs can be constructed also by connecting at each redshift the observed galaxy populations with the dark matter halo populations predicted in cosmological $N$-body simulations; by means of this galaxy-halo connection at different epochs, the MGHs of the halos are used to trace the stellar MGHs of the linked galaxies \citep[][]{Conroy+2009,Firmani+2010,Behroozi+2010,Moster+2010,Behroozi+2013,Moster+2013}.  

Again, the downsizing trend of our global mean MGHs is in qualitatively agreement with most of the inferences obtained with the look back and semi-empirical methods mentioned above. However, our inferences at all scales are in the extreme of early stellar mass formation. The larger differences are for the earliest stages of mass assembly, for instance, at epochs when $\sim 50$\% of the mass was assembled. As will be discussed is \S\S \ref{caveats}, the recovery precision of the fossil record method becomes very poor for the old stellar populations that compose a spectrum in such a way that the early phases of the MGHs result very uncertain.

Regarding the spatially-resolved stellar MGHs, we may compare our results with those from the 105 CALIFA galaxies presented in \citet{Perez+2013}.  In general, their MGHs at all radii attain a given fraction of the final mass later than ours. \citet{Perez+2013} conclude that massive galaxies show a clear inside-out formation mode since early epochs (e.g., when $\approx 50\%$ of the corresponding final masses were assembled), while  for the least massive galaxies in their sample ($\mathcal{M}_0 \approx 3-5 \times 10^9 \solarm$), the outer regions formed stars earlier than the inner ones (outside-in formation mode), at least until $\sim 80\%$ of their final masses were assembled (\citealp{Pan:2015aa} arrived to a similar conclusion but using NUV$-r$ color gradients and the central $D_n$4000 index for SDSS galaxies along the color-$\mathcal{M}_0$ diagram). We should note that \citet{Perez+2013} integrated the light beyond $1.5R_{50}$. For galaxies more massive than $\sim 10^{10} \solarm$, our results are in qualitative agreement with those of \citet{Perez+2013}, although in our case the inside-out behavior reveals on average at larger mass fractions ($>70$\%)/later epochs and the age differences between inner and outer regions are not too large such as in their case.\footnote{We note that \citet{Perez+2013} report the MGHs in a hybrid way: for radii smaller than $1R_{50}$, the MGHs are for the regions contained inside a given radius ($<0.1, 0.5$ and $1R_{50}$), and for radii larger than $1R_{50}$, the MGH is for the regions larger than this radius and not for those contained inside a given radius.} For low-mass galaxies, we do not confirm the outside-in formation claimed by \citet[][]{Perez+2013} and \citet{Pan:2015aa}. Instead, we find that at these and lower masses, the spatially-resolved MGHs are very diverse: along their evolution $\approx 22$\% of them keep the inside-out formation trend, $\approx 15$\% keep the outside-in formation trend, $\approx40$\% transit from outside-in in the past to inside-out more recently, $5-20$\% transit from inside-out to outside-in, and the rest do not show evidence of significant radial gradients in their MGHs (see Figs. \ref{fig09}).

In the literature, there are also studies based on long-slit spectroscopy of few targeted galaxies, where their spatially-resolved SF histories are inferred. For example,  \citet{MacArthur+2009} and \citet[][]{Sanchez-Blazquez+2011}, from full spectrum fits to high S/N data in central/intermediate regions of nearby spiral galaxies find that the majority of the stellar mass is composed of old ($>10$ Gyr) stars, in agreement with our results of very early stellar mass assembly. These authors find also that a larger fraction of young stars are present in the external parts of the disk compared with the inner parts, generally consistent with a moderate inside-out formation mode.

\subsection{Implications for galaxy evolution}

Our analysis confirms that the way galaxies assemble {\it globally} their stellar mass (from an `archaeological' point of view)  depends strongly on their scale (Fig. \ref{fig02a}): the global MGHs are significantly delayed in time as less massive the  galaxies are; this downsizing trend agrees well with many previous determinations (see \S\S \ref{comparison}). A main result reported here is that the global MGHs also segregate on average according to their color, sSFR, and morphological type (insets in Fig. \ref{fig08}), but this segregation, in a given mass bin, is less significant than the one seen across all the masses. Instead, our analysis suggests that the spatially-resolved mean MGHs depend more on galaxy color, sSFR or type than on mass: blue/star-forming/late-type galaxies had on average a more pronounced inside-out formation than red/quiescent/early-type galaxies (Fig. \ref{fig08}). For the latter galaxies, the age differences between their inner and outer MGHs are small, implying nearly flat formation gradients on average. 

According to our results, the MGH shapes of the innermost ($R<0.5R_{50}$) and outermost ($1R_{50}<R<1.5R_{50}$) regions  imply a clear inside-out formation for most of galaxies (at all scales) {\it only} at epochs when $\sim 80-90\%$ of the final masses of these regions were reached, that is, only during the late evolutionary stages of galaxies. We find that 70\% of the galaxies show an inside-out gradient at the 90\% of the corresponding inner and outer final masses. From figure \ref{fig02}  we observe that, for galaxies more massive than $10^{10} \solarm$, the differences in time between the formation of 90\% of the mass of the innermost and outermost regions are around 1.5--2 Gyr (for red/quiescent/early-type galaxies the differences are significantly smaller on average than for blue galaxies),  while for smaller masses, these difference are around 0.5 Gyr.   

At epochs when $70-50\%$ of the inner/outer regions were formed, their corresponding mean MGHs tend to converge to similar tracks. In fact, there are significant variations in these early MGHs on a galaxy-by-galaxy basis: one finds galaxies with inside-out gradients as well as with outside-in and nearly flat gradients (Fig. \ref{fig09}); therefore, the mean MGHs may not be a fair representation of the individual MGHs. The largest variations are for galaxies less massive than $10^{10} \solarm$, evidencing this very irregular paths of early radial mass growth for these galaxies, specially the dwarf ones. Models and numerical simulations show indeed that the mass assembly of galaxies in shallow gravitational potentials is strongly affected by the SF-driven feedback, mainly by the supernova outflows, which redistribute spatially the cold gas and the stars forming from this gas, and promote highly pseudo-stochastic SFR histories \citep[e.g.,][]{Stinson+2007,Avila-Reese+2011,Brook+2011,Gonzalez-Samaniego:2014aa,Governato+2015,El-Badry:2016aa}. On the other hand, when small galaxies become satellites, the environmental effects (see references in the Introduction) can drive to gas removal from the satellite, first from its outskirts, causing this a quenching of SF from outside to inside. It is interesting that despite the early assembly of our analyzed low-mass and dwarf galaxies, it is very irregular spatially, at late evolutionary stages (for instance, when 80-90\% of the inner/outer final masses are attained), there is a trend to present regular inside-out gradients as seen both in the mean MGHs of Fig. \ref{fig02} and in most of the individual inner-to-outer assembly time differences in Fig. \ref{fig09}. This suggests that the radial mass assembly of these galaxies tends to get more regular as their masses grow. 

Regarding galaxies more massive than $10^{10} \solarm$, for most of them it seems to be a trend of transiting from a nearly flat gradient at early epochs to a clear inside-out gradient at late epochs (see Fig. \ref{fig09}), specially for the most massive ones. This trend is more noticeable for blue/star-forming/late-type galaxies. According to Fig. \ref{fig08}, at intermediate/late evolutionary stages, these galaxies present on average a larger difference in the inner-to-outer MGHs than red/quiescent/early-type galaxies. This difference is mainly due to the outermost regions ($>1R_{50}$), which for blue/star-forming/late-type galaxies grew significantly later than for red/quiescent/early-type galaxies. These results are in agreement at a qualitative level with the gradual inside-out mass build-up of massive disk galaxies predicted in the context of the hierarchical $\Lambda$ cold dark matter cosmology \citep[c.f.,][]{Firmani+2000, Avila-Reese+2000,Aumer+2014}, and agree with previous observational inferences using photometry and/or spectroscopy \citep[c.f.][]{Wang:2011aa,Lin:2013aa,Perez+2013,Gonzalez-Delgado+2014,Pan:2015aa,Li:2015aa,Pezzulli+2015,Dale+2016}.

The effects of SF quenching in the central regions (bulge) due to the AGN-driven feedback mentioned in the Introduction could work in the direction of increasing the differences between the inner and outer MGHs by quenching the SF in the inner regions and halting their growth, while the outer regions continue growing by gas accretion. On the other hand, a significant contribution from gas-rich minor mergers or misaligned gas infall work in the direction of reducing these differences, producing a more continuous mass assembly at all radii \citep{Aumer+2014}.

The AGN-driven quenching mechanism should play a more global role in the evolution of massive spheroid-dominated (early-type) galaxies, which do not form by gentle inside-out accretion of cold gas such as the disks. The large (classical) spheroids are thought to form mainly by early major mergers that induce strong bursts of SF uniformly across the whole galaxy and that promote the formation of a central supermassive black hole, with the consequent turn on of one or more phases of powerful AGNs. The powerful AGN can quench the SF in the inner regions, and later in the outer regions (see the Introduction for references), producing this some inside-out like gradient that could explain our inferences. However, this inside-out formation mode is on average small (see Fig. \ref{fig08}), with assembly differences between the innermost and outermost regions $<<1$ Gyr at mass fractions lower than 90\%. Therefore, the possible inside-out AGN quenching process is expected to act only on small time scales.  Massive early-type galaxies undergo also late dry mergers, which add (probably old) stellar populations formed \textit{ex-situ} \citep[see e.g. results from numerical simulations in][]{Hopkins+2009,Rodriguez-Gomez+2016} and produce radial redistribution of stars mainly in the outer regions. Then, the apparent flat or even outside-in mode at early epochs inferred from the early MGHs of the outermost regions of red/quiescent/early-type galaxies could not be a correct interpretation.  As discussed in \citet{Sanchez-Blazquez:2007aa}, the stellar mass assembly of early-type galaxies probably can not be explained solely by inside-out or outside-in scenarios. Our analysis shows indeed the lack of pronounced inside-out or outside-in formation mode for the red/quiescent/early-type galaxies, mainly due to the strong diversity in the outer MGHs.

\subsubsection{Radial migration and mergers}
\label{migration}
 
Our inferred stellar MGHs in different radial regions can be interpreted as the intrinsic ones only if we assume that the stellar populations remain in the same location where they formed in the galaxy. However, there are several internal dynamical effects across the evolution of galaxies that can trigger stellar radial migration. Moreover, during mergers, the accretion of stars formed ex-situ and some radial redistribution of the galaxy stars is expected to happen.

For galaxies less massive than $10^{10} \solarm$, the feedback-driven outflows can produce strong age-dependent stellar radial migration, with a cumulative effect such that the older stellar populations significantly migrate outwards. As the result, the true underlying age and metallicity gradients of galaxies are mixed or even inverted. The two mechanisms that dominate in this process are (1) the large-scale gas movements that drive fluctuations in the global gravitational potential, transferring energy to all stars, and (2) the outflowing/infalling gas that remains star-forming, producing young stars that migrate up to $\sim 1$ kpc  \citep[][]{El-Badry:2016aa}. The former mechanism acts in long time scales producing a systematic outward radial migration, which for old populations may account for distances as large as $1.5R_{50}$($z=0$). The latter mechanism lasts short time scales (up to 100 Myr), affecting only to young populations. The stellar outward migration and overall galaxy expansion of low-mass/dwarf galaxies have been observed in numerical simulations in both cold and warm dark matter cosmologies \citep{Governato+2015,El-Badry:2016aa,Gonzalez-Samaniego+2016}. 

The strong migration processes can explain partially the large diversity in the the spatially-resolved MGHs inferred here for galaxies less massive than $10^{10} \solarm$. On the other hand, since the migration is predicted to be mostly outwards, specially for the older stellar populations, we expect that the diverse MGH gradients inferred here for these galaxies correspond actually to underlying true stellar MGHs following a strong inside-out mode \citep[][show how strong can be this gradient change for their numerical simulations]{El-Badry:2016aa}.

For massive galaxies, the feedback-driven mechanisms of radial migration mentioned above are not efficient enough; they have deeper potential wells and lower gas fractions in central regions than less massive galaxies, so that they do not experience strong coherent fluctuations in stellar kinematics \citep{El-Badry:2016aa}. Radial migration in secularly evolving massive disk galaxies (without heating significantly the disk) has been proposed to be produced by dynamical mechanisms associated to spiral arms, bars, clumps, etc. \citep[e.g.,][see the latter paper for a review and more references therein]{Friedli+1994,Sellwood+2002,Roskar+2008,Minchev+2010,Minchev:2012aa,Di-Matteo:2014aa,Di-Matteo:2015aa,Sanchez-Blazquez+2014}. There are theoretical and observational pieces of evidence in favor of these mechanisms. However, it is a matter of discussion the level of net radial migration that they could produce in observed galaxies. {\bf For instance, \citet[][]{Schonrich+2009} and  \citet[][]{Roskar+2012} have suggested that the main mechanism of radial migration is basically a switch in the radial position of two stars in different circular orbits, something that does not produce a net radial flow of stars. However, it should be taken into account that the proportion of migrated stars depends on the density, which decreases with radius, leading this likely at the end to a net flow towards the external parts}.  Some numerical simulations show that the consequences of radial migration are only significant in the outer parts of galaxies, beyond 2-3 scale lengths \citep{Roskar+2008,Sanchez-Blazquez+2009,Di-Matteo:2014aa,Di-Matteo:2015aa}. From the observational side, studies of age and metallicity gradients show that there are not significant differences for galaxies with and without bar \citep{Sanchez-Blazquez+2014,Wilkinson:2015aa}, indicating that the spiral-bar mechanism proposed for radial migration \citep{Minchev+2010} is not as important as predicted. Furthermore, a direct and strong constrain for the radial migration is the observed abundance gradient \citep{Sanchez:2014aa,Sanchez:2015ab}; recent simulations and archaeological studies indicate that old stars present a strong abundance gradient (50\% of the mass in old stars have a gradient), which implies that these populations could not suffer significative radial mixing \citep[][S\'anchez-Bl\'azquez et al. 2016 in prep.]{Tissera:2013aa}.

Due to the radial migration discussed above, the true underlying radial MGHs of massive disk galaxies could be different to those inferred via the fossil record method. 
It could be that some outward migration is systematic in most of the massive disk galaxies in such a way the the outer regions of the observed galaxies have formed actually at inner radii. Therefore, if any, the underlying true MGHs could have a slightly more inside-out behavior than our inferences.  

For massive early-type galaxies, which are dominated by a dynamically hot spheroid, the effects of radial migration are not expected to be important. However, if late dry minor mergers happen, then it is possible that the outer regions will grow populated mainly with the old/intermedium age stars of the accreted satellites.  As discussed above, this could explain partially the apparent outside-in behavior of the inferred MGHs for our red/early-type massive galaxies at early epochs, being actually the intrinsic inner-to-outer MGH gradient nearly flat or even inside-out.

\subsection{Caveats of the method}
\label{caveats}

The fossil record method is a unique tool for inferring the {\it individual} evolution of observed galaxies \citep{Wilkinson:2015aa}. Nevertheless, we should be aware of its limitations and shortcomings in order to avoid an over-interpretation of the results. Following, we briefly discuss on some of these issues and how they could affect (or not) our main conclusions.

There is an intrinsic limit to the precision of the fossil record method in determining the age of the stellar populations that compose a spectrum because of the degeneracy in the typical features of similar age (this degeneracy increases for older stellar population spectra). As the result, the precision of age determinations in the spectrum inversion is relatively low. For high S/N spectra, \citet[][see also \citealp{Cid-Fernandes+2007}; \citealp{Walcher:2011aa}]{Ocvirk+2006} found that disentangling confounding degeneracies, coupled with meager optical SSP differences, lead to age resolution of (FWHM) $\Delta_{\rm age}=0.8$ dex (but see \citealp{Tojeiro+2007}, who refine the age uncertainty up to $\Delta_{\rm age}\approx 0.4$ dex. The rough estimates of the age uncertainty that we have presented for our mean MGHs in Figure \ref{fig02} are even lower,
between $\approx 0.2$ and $0.3$ dex \citep[see also][who study the uncertainties for the general case of Pipe3D]{Sanchez:2016aa}.
The logarithmic age resolution around any median log-age value is expected to be $\pm \Delta_{\rm age}/2$.  This uncertainty in stellar population ages drives the peak of the SFH to spread out, and given the logarithmic nature of variations in stellar population ages, the shape of the corresponding MGH is then modified systematically as explained below \citep[see][]{Leitner:2012aa}. 

On one side, the steep decreasing of the MGH at high LBTs becomes shallower because a fraction of the stellar mass formed around the peak of the SFH is spread into a tail of very old stellar populations; this bias makes the fossil record inferences to predict a fraction of stellar mass formed at too early times. In addition to this bias, when the spectrum-inversion method requires of old SSP contributions, since there only 3-4 of them above $\sim 7$ Gyr, the choice is for SSPs that could be actually older than the real ones; this biases again the early phases of the MGHs towards earlier epochs.  As the result of these biases, the (high) fractions of stellar mass assembled at high LBTs are likely overestimated in our MGHs. On the other side, because of the age uncertainty, a fraction of the stellar mass formed around the peak of the SFH is spread to lower LBTs shifting the intermediate parts of the MGHs to lower LBTs. This bias can be more significant for the low-mass/dwarf galaxies, which have the peak of their SFHs at intermediate/late epochs. According to \citet[][see his Fig. 6]{Leitner:2012aa}, the MGHs inferred from his look back approach can become similar to those obtained from the VESPA analysis (fossil record method) when these MGHs are smoothed with a Gaussian filter in log-age such that emulates the age uncertainty of the 
fossil record method. 
 
In addition to age resolution, there are substantial systematic uncertainties in the fossil record inferences related to the SPS and dust models, the IMF, etc. For instance, the cumulative SFHs inferred with the \citet{Bruzual:2003aa} SPS models are significantly more extended (later assembly of a given fraction) from the past until $\sim 2$ Gyr ago than those inferred with the \citet{Maraston2005} SPS models \citep[][e.g., see Fig. 6 of the latter author]{Sanchez:2016aa,Panter+2007,Tojeiro+2009,Leitner:2012aa}. The main difference between these two SPS models are in the treatment of thermally pulsating asymptotic giant branch stars. Thus, according to the SPS model used for the inversion, cumulative SFHs or MGHs different by several Gyrs can be obtained.

Regarding the extinction, the fossil record method constrains it by assuming that it is the same at all epochs. Otherwise, a self-consistent model for the evolution of dust in galaxies should be introduced. We know that on average the amount of dust was smaller at higher redshifts. Therefore, to assume that the dust fraction in galaxies is the same than today introduces a bias in the MGHs towards earlier mass assembly as older are the stellar population contributions. In order to evaluate the strength of this bias, we have a posteriori introduced a time-dependent extinction coefficient in the calculation of the surface brightness corresponding to each individual Voxel. A step function is used for the extinction coefficient: the value constrained in the fitting is used until a LBT of 2 Gyr and for larger LBTs, the value is set to 0. By comparing the mean MGHs obtained with this assumption for the extinction to those presented in Section \ref{results-meanMGHs} we find that the MGHs of low-mass/dwarf galaxies shift towards a slightly later mass assembly, while those of more massive galaxies remain almost the same or barely shift to an earlier mass assembly.  In any case, the differences are in general negligible.

We conclude that the statistical and systematic uncertainties in the fossil record method apparently work mostly in the direction of biasing the early mass assembly inferences to even earlier epochs (larger LBTs). {Therefore, the MGHs presented here for LBTs larger than $\sim $8 Gyr or mass fractions lower than 70\% should be taken with caution.}

\section{Conclusions}
\label{conclusions}

We have inferred the global and spatially-resolved stellar MGHs of the first set of galaxies observed in the MaNGA/SDSS-IV survey (to appear in the SDSS Data Release 13) by means of the fossil record method using Pipe3D.  We used only the Primary sample and excluded from our analysis the strongly interacting/merging galaxies as well as those from the Color-Enhanced MaNGA sample. Then, a sub-sample of 454 galaxies at redshifts lower than 0.037 was studied in particular in order to obtain normalized MGHs with the final mass $\mathcal{M}_0$ defined at the \textit{same} redshift for all galaxies ($z_{\rm lim}=0.037$ corresponding to an initial LBT of $\approx 500$ Myr). The sample was divided into four mass bins and, for each mass bin, the mean normalized MGHs at different radial regions (up to 1.5$R_{50}$) and their scatter were calculated. We have also calculated the mean normalized MGHs for galaxies separated into blue/red colors, star forming/quiescent, and late/early types. Moreover, the differences between the assembly times of a given mass fraction of the innermost and outermost regions, $\Delta T_{\rm i-o}$, was presented for each individual galaxy and for different values of the attained mass fraction. 

The main goal of the present work was to probe the general way galaxies assembled their stellar masses, both globally and at different radial regions, as a function of mass and other galaxy properties. Our study is aimed to explore only general trends; a more detailed separation of galaxies according to both their masses and properties, as well as their environment, will be presented in forthcoming papers. The main results from our analysis are as follow.

\begin{itemize}

\item The larger the final galaxy mass, the earlier on average was assembled most of this mass. For example, galaxies in the $\log$($\mathcal{M}_0/ \solarm$) = 8.5--9.3, 9.3--10.0, 10--10.7 and 10.71--11.2 bins assembled 70\% of their final masses at LBTs of 3.7, 10, 11, and 11 Gyr on average. Dwarf and low-mass galaxies present actually a large diversity of global stellar MGH shapes, showing that their stellar mass assembly is not only delayed with respect to more massive galaxies but also more episodic. On the other hand, in a given mass bin, red/quiescent/early-type galaxies assemble on average their masses earlier than blue/star forming/late-type ones. However, these differences are not so large as those among the different mass bins. 

\item {At late evolutionary stages (or high fractions of assembled mass), most of galaxies in all the mass bins show that the innermost regions formed stars earlier than the outermost ones (inside-out formation mode).} For instance, in the $\log$($\mathcal{M}_0/\solarm$) = 8.5--9.3, 9.3--10.0, 10--10.7, and 10.7--11.2 bins, the regions inside 0.5 $R_{50}$ attain 90\% of their final masses on average  0.5, 0.5, 1.5, and 1.0 Gyr before than regions outside 1 $R_{50}$ do it, respectively.  For galaxies more massive than $\sim 10^{10} \solarm$, the  inside-out trend in the mean radial MGHs remains until LBTs of $9-10$ Gyr, when 70--80\% of the inner and outer regions were in place. At earlier stages (lower assembled mass fractions), this trend tends to disappear; the inner and outer MGH differences become actually diverse {\bf and within the large age uncertainty of the method}. In the case of dwarf galaxies, the radial MGHs are very diverse most of the time, showing periods of outside-in and inside-out growth modes (or strong radial migration). 

\item {The outermost regions of galaxies present on average less regular MGHs (more scattered around the mean) than the innermost ones, being this trend more pronounced as more massive are the galaxies.  }

\item The way galaxies assemble their mass radially depends more on the galaxy color/sSFR/type than on its mass: blue/star-forming/late-type galaxies follow on average a significantly more pronounced inside-out formation mode than red/quiescent/early-type galaxies. {This is true specially for the late evolutionary stages; for the old stellar populations that are more common in the later galaxies, the divergent solutions of the fossil record methods at the earliest epochs of mass assembly, does not allow to establish a clear difference between the inner and outer MGHs of these galaxies}. For these galaxies, the outermost MGHs present also a large diversity of shapes, in many cases with signs of {older assembly than the innermost regions, specially for the small galaxies.}

\item {For galaxies more massive than $\sim 5\times 10^{9} \solarm$, the age differences $\Delta T_{\rm i-o}$ at epochs when $50\%$ of the innermost/outermost masses were formed are mostly below the age uncertainty. The situation improves at later evolutionary stages, when larger mass fractions form. Our results suggest a common trend of transiting from a flat/outside-in (or undetermined) gradient mode to an inside-out mode; this trend is more frequent for the more massive galaxies. Galaxies less massive than $\sim 5\times 10^{9} \solarm$ present large differences between the formation epochs of the innermost and outermost regions at early epochs, both in the direction of the inside-out or outside-in mode; these differences decrease at later epochs.
}


\end{itemize}

From the general trends obtained in our analysis, we can conclude that mass is the main driver of the \textit{global} stellar MGHs, while the way mass is \textit{radially} assembled is more related to the galaxy color/sSFR/type. The trend is that more massive galaxies form earlier than less massive galaxies (downsizing), and galaxies with radial MGHs more regular (less scattered around the mean) and with a stronger inside-out mode are more related to bluer/star-forming/later-type objects. The stellar mass assembly of galaxies less massive than $10^{10} \solarm$, specially the least massive ones, reveals itself as episodic and stochastic, both at the global and local level, with periods of outside-in and inside-out formation modes. However, as discussed in \S\S \ref{migration} this diversity of radial MGHs could be also due to the expected strong radial migration processes in these galaxies, where the older stellar populations suffer more net outward migration. {After taking into account such an outward migration, the intrinsic radial MGHs could follow instead an inside-out growth mode. }

For $\mathcal{M}_0>10^{10} \solarm$, the marked inside-out formation mode seen for most of the blue/star-forming/late-type galaxies, at least since $\sim 10$ Gyr ago, is in agreement with predictions of inside-out disk formation in the context of the hierarchical $\Lambda$ cold dark matter cosmology.  {If takes into account a possible net outward flow of stars due to migration mechanisms, a more pronounced intrinsic inside-out mode would be inferred.} 
In the case of red/quiescent/early-type galaxies,  {due to the poor age resolution when the ages are old, we can not say too much about the formation mode of these early-forming galaxies. However, at later epochs, when more than $70-80\%$ of the mass was assembled, they seem to follow on average a weak inside-out formation mode,} which could be produced by quenching of a powerful AGN, which shuts-off the SF first in the central regions and then in the outer ones; given the relatively small differences between the innermost and outermost MGHs (or relatively small values of $\Delta T_{\rm i-o}$), this effect is not expected to long not too much ($<<1$ Gyr). On the other hand, the weak hint of early outside-in formation in most of the red/quiescent/early-type galaxies may be incorrect, and the fact that very old stellar populations are present in the outermost regions of these galaxies could be rather due to the accretion of satellites with old populations (late dry mergers).  

{It should be said that the main conclusions from our study remain the same when including galaxies from the MaNGA Secondary sub-sample, i.e., those with a radial coverage that reaches 2.5 effective radii. We have not included these galaxies in order to avoid possible aperture biases, and because the mass dependence on redshift is not trivial for this sub-sample and it may introduce selection effects.} 

The fossil record method applied to IFS observations is a unique technique to infer the local and global evolution of \textit{individual} galaxies. However, we should be borne in mind that these inferences are subject to several uncertainties and degeneracies. For example, the age uncertainty in the spectrum inversion is large, specially for the contributions coming from old stellar populations. In the present work the inferred global MGHs are biased to high mass fractions assembled at the earliest epochs with respect to inferences based on  look back empirical and semi-empirical studies or even with respect to some previous archaeological inferences (see \S\S \ref{comparison}). As discussed in  \S\S \ref{caveats}, for intermediate/large LBTs, because of the low age resolution the stellar MGHs can be systematically modified, specially toward older ages. The choice  of the SPS models also affects significantly the cumulative SFHs or MGHs; for instance, according how the thermally pulsating asymptotic giant branch stars are treated, shifts of some Gyrs in the MGHs are obtained. 

More work is necessary to evaluate and control the effects of uncertainties and degeneracies in the fossil record inferences of galaxy stellar mass assembly. In defense of the general results presented here, we should say that, once the same method and analysis is applied to all galaxies and radial regions, the results related to comparisons of the means among different masses and among different radial regions should be reliable at least in a qualitative level.

\section*{Acknowledgments}
HIM received financial support through a Postdoctoral fellowship provided by CONACyT grant (Ciencia B\'asica) 180125. The authors from UNAM acknowledge CONACyT grants (Ciencia B\'asica) 180125 and 167332 for partial funding. Funding for the Sloan Digital Sky Survey IV has been provided by the Alfred P. Sloan Foundation, the U.S. Department of Energy Office of Science, and the Participating Institutions. SDSS-IV acknowledges support and resources from the Center for High-Performance Computing at the University of Utah. The SDSS web site is \url{www.sdss.org}.

SDSS-IV is managed by the Astrophysical Research Consortium for the Participating Institutions of the SDSS Collaboration including the Brazilian Participation Group, the Carnegie Institution for Science, Carnegie Mellon University, the Chilean Participation Group, the French Participation Group, Harvard-Smithsonian Center for Astrophysics, Instituto de Astrof\'isica de Canarias, The Johns Hopkins University, Kavli Institute for the Physics and Mathematics of the Universe (IPMU) / University of Tokyo, Lawrence Berkeley National Laboratory, Leibniz Institut f\"ur Astrophysik Potsdam (AIP),  Max-Planck-Institut f\"ur Astronomie (MPIA Heidelberg), Max-Planck-Institut f\"ur Astrophysik (MPA Garching), Max-Planck-Institut f\"ur Extraterrestrische Physik (MPE), National Astronomical Observatory of China, New Mexico State University, New York University, University of Notre Dame, Observat\'ario Nacional / MCTI, The Ohio State University, Pennsylvania State University, Shanghai Astronomical Observatory, United Kingdom Participation Group, Universidad Nacional Aut\'onoma de M\'exico, University of Arizona, University of Colorado Boulder, University of Oxford, University of Portsmouth, University of Utah, University of Virginia, University of Washington, University of Wisconsin, Vanderbilt University, and Yale University.

\bibliography{paper}{}

\begin{thebibliography}{}
\makeatletter
\relax
\def\mn@urlcharsother{\let\do\@makeother \do\$\do\&\do\#\do\^\do\_\do\%\do\~}
\def\mn@doi{\begingroup\mn@urlcharsother \@ifnextchar [ {\mn@doi@}
  {\mn@doi@[]}}
\def\mn@doi@[#1]#2{\def\@tempa{#1}\ifx\@tempa\@empty \href
  {http://dx.doi.org/#2} {doi:#2}\else \href {http://dx.doi.org/#2} {#1}\fi
  \endgroup}
\def\mn@eprint#1#2{\mn@eprint@#1:#2::\@nil}
\def\mn@eprint@arXiv#1{\href {http://arxiv.org/abs/#1} {{\tt arXiv:#1}}}
\def\mn@eprint@dblp#1{\href {http://dblp.uni-trier.de/rec/bibtex/#1.xml}
  {dblp:#1}}
\def\mn@eprint@#1:#2:#3:#4\@nil{\def\@tempa {#1}\def\@tempb {#2}\def\@tempc
  {#3}\ifx \@tempc \@empty \let \@tempc \@tempb \let \@tempb \@tempa \fi \ifx
  \@tempb \@empty \def\@tempb {arXiv}\fi \@ifundefined
  {mn@eprint@\@tempb}{\@tempb:\@tempc}{\expandafter \expandafter \csname
  mn@eprint@\@tempb\endcsname \expandafter{\@tempc}}}

\bibitem[\protect\citeauthoryear{{Aumer}, {White}  \& {Naab}}{{Aumer}
  et~al.}{2014}]{Aumer+2014}
{Aumer} M.,  {White} S.~D.~M.,   {Naab} T.,  2014, \mn@doi [\mnras]
  {10.1093/mnras/stu818}, \href
  {http://adsabs.harvard.edu/abs/2014MNRAS.441.3679A} {441, 3679}

\bibitem[\protect\citeauthoryear{{Avila-Reese} \& {Firmani}}{{Avila-Reese} \&
  {Firmani}}{2000}]{Avila-Reese+2000}
{Avila-Reese} V.,  {Firmani} C.,  2000, \rmxaa, \href
  {http://adsabs.harvard.edu/abs/2000RMxAA..36...23A} {36, 23}

\bibitem[\protect\citeauthoryear{{Avila-Reese}, {Col{\'{\i}}n},
  {Gonz{\'a}lez-Samaniego}, {Valenzuela}, {Firmani}, {Vel{\'a}zquez}  \&
  {Ceverino}}{{Avila-Reese} et~al.}{2011}]{Avila-Reese+2011}
{Avila-Reese} V.,  {Col{\'{\i}}n} P.,  {Gonz{\'a}lez-Samaniego} A.,
  {Valenzuela} O.,  {Firmani} C.,  {Vel{\'a}zquez} H.,   {Ceverino} D.,  2011,
  \mn@doi [\apj] {10.1088/0004-637X/736/2/134}, \href
  {http://adsabs.harvard.edu/abs/2011ApJ...736..134A} {736, 134}

\bibitem[\protect\citeauthoryear{{Bacon} et~al.,}{{Bacon}
  et~al.}{2001}]{Bacon:2001aa}
{Bacon} R.,  et~al., 2001, \mn@doi [\mnras] {10.1046/j.1365-8711.2001.04612.x},
  \href {http://adsabs.harvard.edu/abs/2001MNRAS.326...23B} {326, 23}

\bibitem[\protect\citeauthoryear{{Bauer} et~al.,}{{Bauer}
  et~al.}{2013}]{Bauer+2013}
{Bauer} A.~E.,  et~al., 2013, \mn@doi [\mnras] {10.1093/mnras/stt1011}, \href
  {http://adsabs.harvard.edu/abs/2013MNRAS.434..209B} {434, 209}

\bibitem[\protect\citeauthoryear{{Behroozi}, {Conroy}  \&
  {Wechsler}}{{Behroozi} et~al.}{2010}]{Behroozi+2010}
{Behroozi} P.~S.,  {Conroy} C.,   {Wechsler} R.~H.,  2010, \mn@doi [\apj]
  {10.1088/0004-637X/717/1/379}, \href
  {http://adsabs.harvard.edu/abs/2010ApJ...717..379B} {717, 379}

\bibitem[\protect\citeauthoryear{{Behroozi}, {Wechsler}  \&
  {Conroy}}{{Behroozi} et~al.}{2013}]{Behroozi+2013}
{Behroozi} P.~S.,  {Wechsler} R.~H.,   {Conroy} C.,  2013, \mn@doi [\apj]
  {10.1088/0004-637X/770/1/57}, \href
  {http://adsabs.harvard.edu/abs/2013ApJ...770...57B} {770, 57}

\bibitem[\protect\citeauthoryear{{Belfiore} et~al.,}{{Belfiore}
  et~al.}{2015}]{Belfiore:2015aa}
{Belfiore} F.,  et~al., 2015, \mn@doi [\mnras] {10.1093/mnras/stv296}, \href
  {http://adsabs.harvard.edu/abs/2015MNRAS.449..867B} {449, 867}

\bibitem[\protect\citeauthoryear{{Bell}, {Zheng}, {Papovich}, {Borch}, {Wolf}
  \& {Meisenheimer}}{{Bell} et~al.}{2007}]{Bell+2007}
{Bell} E.~F.,  {Zheng} X.~Z.,  {Papovich} C.,  {Borch} A.,  {Wolf} C.,
  {Meisenheimer} K.,  2007, \mn@doi [\apj] {10.1086/518594}, \href
  {http://adsabs.harvard.edu/abs/2007ApJ...663..834B} {663, 834}

\bibitem[\protect\citeauthoryear{{Ben{\'{\i}}tez} et~al.,}{{Ben{\'{\i}}tez}
  et~al.}{2009}]{Benitez:2009aa}
{Ben{\'{\i}}tez} N.,  et~al., 2009, \mn@doi [\apjl]
  {10.1088/0004-637X/692/1/L5}, \href
  {http://adsabs.harvard.edu/abs/2009ApJ...692L...5B} {692, L5}

\bibitem[\protect\citeauthoryear{{Bernard}, {Aparicio}, {Gallart},
  {Padilla-Torres}  \& {Panniello}}{{Bernard} et~al.}{2007}]{Bernard+2007}
{Bernard} E.~J.,  {Aparicio} A.,  {Gallart} C.,  {Padilla-Torres} C.~P.,
  {Panniello} M.,  2007, \mn@doi [\aj] {10.1086/520805}, \href
  {http://adsabs.harvard.edu/abs/2007AJ....134.1124B} {134, 1124}

\bibitem[\protect\citeauthoryear{{Birnboim} \& {Dekel}}{{Birnboim} \&
  {Dekel}}{2003}]{Birnboim:2003aa}
{Birnboim} Y.,  {Dekel} A.,  2003, \mn@doi [\mnras]
  {10.1046/j.1365-8711.2003.06955.x}, \href
  {http://adsabs.harvard.edu/abs/2003MNRAS.345..349B} {345, 349}

\bibitem[\protect\citeauthoryear{{Blanton} \& {Moustakas}}{{Blanton} \&
  {Moustakas}}{2009}]{Blanton+2009}
{Blanton} M.~R.,  {Moustakas} J.,  2009, \mn@doi [\araa]
  {10.1146/annurev-astro-082708-101734}, \href
  {http://adsabs.harvard.edu/abs/2009ARA%26A..47..159B} {47, 159}

\bibitem[\protect\citeauthoryear{{Bower}, {Benson}, {Malbon}, {Helly}, {Frenk},
  {Baugh}, {Cole}  \& {Lacey}}{{Bower} et~al.}{2006}]{Bower:2006aa}
{Bower} R.~G.,  {Benson} A.~J.,  {Malbon} R.,  {Helly} J.~C.,  {Frenk} C.~S.,
  {Baugh} C.~M.,  {Cole} S.,   {Lacey} C.~G.,  2006, \mn@doi [\mnras]
  {10.1111/j.1365-2966.2006.10519.x}, \href
  {http://adsabs.harvard.edu/abs/2006MNRAS.370..645B} {370, 645}

\bibitem[\protect\citeauthoryear{{Brinchmann} \& {Ellis}}{{Brinchmann} \&
  {Ellis}}{2000}]{Brinchmann:2000aa}
{Brinchmann} J.,  {Ellis} R.~S.,  2000, \mn@doi [\apjl] {10.1086/312738}, \href
  {http://adsabs.harvard.edu/abs/2000ApJ...536L..77B} {536, L77}

\bibitem[\protect\citeauthoryear{{Brook} et~al.,}{{Brook}
  et~al.}{2011}]{Brook+2011}
{Brook} C.~B.,  et~al., 2011, \mn@doi [\mnras]
  {10.1111/j.1365-2966.2011.18545.x}, \href
  {http://adsabs.harvard.edu/abs/2011MNRAS.415.1051B} {415, 1051}

\bibitem[\protect\citeauthoryear{{Brooks} \& {Christensen}}{{Brooks} \&
  {Christensen}}{2016}]{Brooks+2016}
{Brooks} A.,  {Christensen} C.,  2016, \mn@doi [Galactic Bulges]
  {10.1007/978-3-319-19378-6_12}, \href
  {http://adsabs.harvard.edu/abs/2016ASSL..418..317B} {418, 317}

\bibitem[\protect\citeauthoryear{{Bruzual} \& {Charlot}}{{Bruzual} \&
  {Charlot}}{2003}]{Bruzual:2003aa}
{Bruzual} G.,  {Charlot} S.,  2003, \mn@doi [\mnras]
  {10.1046/j.1365-8711.2003.06897.x}, \href
  {http://adsabs.harvard.edu/abs/2003MNRAS.344.1000B} {344, 1000}

\bibitem[\protect\citeauthoryear{{Bundy} et~al.,}{{Bundy}
  et~al.}{2015}]{Bundy:2015aa}
{Bundy} K.,  et~al., 2015, \mn@doi [\apj] {10.1088/0004-637X/798/1/7}, \href
  {http://adsabs.harvard.edu/abs/2015ApJ...798....7B} {798, 7}

\bibitem[\protect\citeauthoryear{{Buzzoni}}{{Buzzoni}}{1989}]{Buzzoni:1989aa}
{Buzzoni} A.,  1989, \mn@doi [\apjs] {10.1086/191399}, \href
  {http://adsabs.harvard.edu/abs/1989ApJS...71..817B} {71, 817}

\bibitem[\protect\citeauthoryear{{Cappellari} et~al.,}{{Cappellari}
  et~al.}{2011}]{Cappellari+2011}
{Cappellari} M.,  et~al., 2011, \mn@doi [\mnras]
  {10.1111/j.1365-2966.2010.18174.x}, \href
  {http://adsabs.harvard.edu/abs/2011MNRAS.413..813C} {413, 813}

\bibitem[\protect\citeauthoryear{{Cardelli}, {Clayton}  \& {Mathis}}{{Cardelli}
  et~al.}{1989}]{Cardelli:1989aa}
{Cardelli} J.~A.,  {Clayton} G.~C.,   {Mathis} J.~S.,  1989, \mn@doi [\apj]
  {10.1086/167900}, \href {http://adsabs.harvard.edu/abs/1989ApJ...345..245C}
  {345, 245}

\bibitem[\protect\citeauthoryear{{Cid Fernandes}, {Mateus}, {Sodr{\'e}},
  {Stasi{\'n}ska}  \& {Gomes}}{{Cid Fernandes}
  et~al.}{2005}]{Cid-Fernandes:2005aa}
{Cid Fernandes} R.,  {Mateus} A.,  {Sodr{\'e}} L.,  {Stasi{\'n}ska} G.,
  {Gomes} J.~M.,  2005, \mn@doi [\mnras] {10.1111/j.1365-2966.2005.08752.x},
  \href {http://adsabs.harvard.edu/abs/2005MNRAS.358..363C} {358, 363}

\bibitem[\protect\citeauthoryear{{Cid Fernandes}, {Asari}, {Sodr{\'e}},
  {Stasi{\'n}ska}, {Mateus}, {Torres-Papaqui}  \& {Schoenell}}{{Cid Fernandes}
  et~al.}{2007}]{Cid-Fernandes+2007}
{Cid Fernandes} R.,  {Asari} N.~V.,  {Sodr{\'e}} L.,  {Stasi{\'n}ska} G.,
  {Mateus} A.,  {Torres-Papaqui} J.~P.,   {Schoenell} W.,  2007, \mn@doi
  [\mnras] {10.1111/j.1745-3933.2006.00265.x}, \href
  {http://adsabs.harvard.edu/abs/2007MNRAS.375L..16C} {375, L16}

\bibitem[\protect\citeauthoryear{{Cid Fernandes} et~al.,}{{Cid Fernandes}
  et~al.}{2013}]{Cid-Fernandes:2013aa}
{Cid Fernandes} R.,  et~al., 2013, \mn@doi [\aap]
  {10.1051/0004-6361/201220616}, \href
  {http://adsabs.harvard.edu/abs/2013A%26A...557A..86C} {557, A86}

\bibitem[\protect\citeauthoryear{{Cid Fernandes} et~al.,}{{Cid Fernandes}
  et~al.}{2014}]{Cid-Fernandes:2014aa}
{Cid Fernandes} R.,  et~al., 2014, \mn@doi [\aap]
  {10.1051/0004-6361/201321692}, \href
  {http://adsabs.harvard.edu/abs/2014A%26A...561A.130C} {561, A130}

\bibitem[\protect\citeauthoryear{{Cimatti}, {Daddi}  \& {Renzini}}{{Cimatti}
  et~al.}{2006}]{Cimatti+2006}
{Cimatti} A.,  {Daddi} E.,   {Renzini} A.,  2006, \mn@doi [\aap]
  {10.1051/0004-6361:20065155}, \href
  {http://adsabs.harvard.edu/abs/2006A%26A...453L..29C} {453, L29}

\bibitem[\protect\citeauthoryear{{Conroy} \& {Wechsler}}{{Conroy} \&
  {Wechsler}}{2009}]{Conroy+2009}
{Conroy} C.,  {Wechsler} R.~H.,  2009, \mn@doi [\apj]
  {10.1088/0004-637X/696/1/620}, \href
  {http://adsabs.harvard.edu/abs/2009ApJ...696..620C} {696, 620}

\bibitem[\protect\citeauthoryear{{Cowie}, {Songaila}, {Hu}  \& {Cohen}}{{Cowie}
  et~al.}{1996}]{Cowie+1996}
{Cowie} L.~L.,  {Songaila} A.,  {Hu} E.~M.,   {Cohen} J.~G.,  1996, \mn@doi
  [\aj] {10.1086/118058}, \href
  {http://adsabs.harvard.edu/abs/1996AJ....112..839C} {112, 839}

\bibitem[\protect\citeauthoryear{{Croom} et~al.,}{{Croom}
  et~al.}{2012}]{Croom:2012aa}
{Croom} S.~M.,  et~al., 2012, \mn@doi [\mnras]
  {10.1111/j.1365-2966.2011.20365.x}, \href
  {http://adsabs.harvard.edu/abs/2012MNRAS.421..872C} {421, 872}

\bibitem[\protect\citeauthoryear{{Dale} et~al.,}{{Dale}
  et~al.}{2016}]{Dale+2016}
{Dale} D.~A.,  et~al., 2016, \mn@doi [\aj] {10.3847/0004-6256/151/1/4}, \href
  {http://adsabs.harvard.edu/abs/2016AJ....151....4D} {151, 4}

\bibitem[\protect\citeauthoryear{{Dekel} \& {Birnboim}}{{Dekel} \&
  {Birnboim}}{2006}]{Dekel:2006aa}
{Dekel} A.,  {Birnboim} Y.,  2006, \mn@doi [\mnras]
  {10.1111/j.1365-2966.2006.10145.x}, \href
  {http://adsabs.harvard.edu/abs/2006MNRAS.368....2D} {368, 2}

\bibitem[\protect\citeauthoryear{{Dekel} \& {Silk}}{{Dekel} \&
  {Silk}}{1986}]{Dekel:1986aa}
{Dekel} A.,  {Silk} J.,  1986, \mn@doi [\apj] {10.1086/164050}, \href
  {http://adsabs.harvard.edu/abs/1986ApJ...303...39D} {303, 39}

\bibitem[\protect\citeauthoryear{{Di Matteo}, {Springel}  \& {Hernquist}}{{Di
  Matteo} et~al.}{2005}]{Di-Matteo:2005aa}
{Di Matteo} T.,  {Springel} V.,   {Hernquist} L.,  2005, \mn@doi [\nat]
  {10.1038/nature03335}, \href
  {http://adsabs.harvard.edu/abs/2005Natur.433..604D} {433, 604}

\bibitem[\protect\citeauthoryear{{Di Matteo} et~al.,}{{Di Matteo}
  et~al.}{2014}]{Di-Matteo:2014aa}
{Di Matteo} P.,  et~al., 2014, \mn@doi [\aap] {10.1051/0004-6361/201322958},
  \href {http://adsabs.harvard.edu/abs/2014A%26A...567A.122D} {567, A122}

\bibitem[\protect\citeauthoryear{{Di Matteo} et~al.,}{{Di Matteo}
  et~al.}{2015}]{Di-Matteo:2015aa}
{Di Matteo} P.,  et~al., 2015, \mn@doi [\aap] {10.1051/0004-6361/201424457},
  \href {http://adsabs.harvard.edu/abs/2015A%26A...577A...1D} {577, A1}

\bibitem[\protect\citeauthoryear{{Drory} \& {Alvarez}}{{Drory} \&
  {Alvarez}}{2008}]{Drory+2008}
{Drory} N.,  {Alvarez} M.,  2008, \mn@doi [\apj] {10.1086/588006}, \href
  {http://adsabs.harvard.edu/abs/2008ApJ...680...41D} {680, 41}

\bibitem[\protect\citeauthoryear{{Drory} et~al.,}{{Drory}
  et~al.}{2015}]{Drory:2015aa}
{Drory} N.,  et~al., 2015, \mn@doi [\aj] {10.1088/0004-6256/149/2/77}, \href
  {http://adsabs.harvard.edu/abs/2015AJ....149...77D} {149, 77}

\bibitem[\protect\citeauthoryear{{Efstathiou}}{{Efstathiou}}{1992}]{Efstathiou:1992aa}
{Efstathiou} G.,  1992, \mn@doi [\mnras] {10.1093/mnras/256.1.43P}, \href
  {http://adsabs.harvard.edu/abs/1992MNRAS.256P..43E} {256, 43P}

\bibitem[\protect\citeauthoryear{{Eggen}, {Lynden-Bell}  \& {Sandage}}{{Eggen}
  et~al.}{1962}]{Eggen+1962}
{Eggen} O.~J.,  {Lynden-Bell} D.,   {Sandage} A.~R.,  1962, \mn@doi [\apj]
  {10.1086/147433}, \href {http://adsabs.harvard.edu/abs/1962ApJ...136..748E}
  {136, 748}

\bibitem[\protect\citeauthoryear{{Eisenstein} et~al.,}{{Eisenstein}
  et~al.}{2011}]{Eisenstein:2011aa}
{Eisenstein} D.~J.,  et~al., 2011, \mn@doi [\aj] {10.1088/0004-6256/142/3/72},
  \href {http://adsabs.harvard.edu/abs/2011AJ....142...72E} {142, 72}

\bibitem[\protect\citeauthoryear{{El-Badry}, {Wetzel}, {Geha}, {Hopkins},
  {Kere{\v s}}, {Chan}  \& {Faucher-Gigu{\`e}re}}{{El-Badry}
  et~al.}{2016}]{El-Badry:2016aa}
{El-Badry} K.,  {Wetzel} A.,  {Geha} M.,  {Hopkins} P.~F.,  {Kere{\v s}} D.,
  {Chan} T.~K.,   {Faucher-Gigu{\`e}re} C.-A.,  2016, \mn@doi [\apj]
  {10.3847/0004-637X/820/2/131}, \href
  {http://adsabs.harvard.edu/abs/2016ApJ...820..131E} {820, 131}

\bibitem[\protect\citeauthoryear{{Falc{\'o}n-Barroso},
  {S{\'a}nchez-Bl{\'a}zquez}, {Vazdekis}, {Ricciardelli}, {Cardiel}, {Cenarro},
  {Gorgas}  \& {Peletier}}{{Falc{\'o}n-Barroso}
  et~al.}{2011}]{Falcon-Barroso:2011aa}
{Falc{\'o}n-Barroso} J.,  {S{\'a}nchez-Bl{\'a}zquez} P.,  {Vazdekis} A.,
  {Ricciardelli} E.,  {Cardiel} N.,  {Cenarro} A.~J.,  {Gorgas} J.,
  {Peletier} R.~F.,  2011, \mn@doi [\aap] {10.1051/0004-6361/201116842}, \href
  {http://adsabs.harvard.edu/abs/2011A%26A...532A..95F} {532, A95}

\bibitem[\protect\citeauthoryear{{Farouki} \& {Shapiro}}{{Farouki} \&
  {Shapiro}}{1981}]{Farouki:1981aa}
{Farouki} R.,  {Shapiro} S.~L.,  1981, \mn@doi [\apj] {10.1086/158563}, \href
  {http://adsabs.harvard.edu/abs/1981ApJ...243...32F} {243, 32}

\bibitem[\protect\citeauthoryear{{Faucher-Gigu{\`e}re}, {Kere{\v s}}  \&
  {Ma}}{{Faucher-Gigu{\`e}re} et~al.}{2011}]{Faucher-Giguere+2011}
{Faucher-Gigu{\`e}re} C.-A.,  {Kere{\v s}} D.,   {Ma} C.-P.,  2011, \mn@doi
  [\mnras] {10.1111/j.1365-2966.2011.19457.x}, \href
  {http://adsabs.harvard.edu/abs/2011MNRAS.417.2982F} {417, 2982}

\bibitem[\protect\citeauthoryear{{Feldmann} \& {Mayer}}{{Feldmann} \&
  {Mayer}}{2015}]{Feldmann:2015aa}
{Feldmann} R.,  {Mayer} L.,  2015, \mn@doi [\mnras] {10.1093/mnras/stu2207},
  \href {http://adsabs.harvard.edu/abs/2015MNRAS.446.1939F} {446, 1939}

\bibitem[\protect\citeauthoryear{{Firmani} \& {Avila-Reese}}{{Firmani} \&
  {Avila-Reese}}{2000}]{Firmani+2000}
{Firmani} C.,  {Avila-Reese} V.,  2000, \mn@doi [\mnras]
  {10.1046/j.1365-8711.2000.03338.x}, \href
  {http://adsabs.harvard.edu/abs/2000MNRAS.315..457F} {315, 457}

\bibitem[\protect\citeauthoryear{{Firmani}, {Avila-Reese}  \&
  {Rodr{\'{\i}}guez-Puebla}}{{Firmani} et~al.}{2010}]{Firmani+2010}
{Firmani} C.,  {Avila-Reese} V.,   {Rodr{\'{\i}}guez-Puebla} A.,  2010, \mn@doi
  [\mnras] {10.1111/j.1365-2966.2010.16366.x}, \href
  {http://adsabs.harvard.edu/abs/2010MNRAS.404.1100F} {404, 1100}

\bibitem[\protect\citeauthoryear{{Fontanot}, {De Lucia}, {Monaco}, {Somerville}
   \& {Santini}}{{Fontanot} et~al.}{2009}]{Fontanot:2009aa}
{Fontanot} F.,  {De Lucia} G.,  {Monaco} P.,  {Somerville} R.~S.,   {Santini}
  P.,  2009, \mn@doi [\mnras] {10.1111/j.1365-2966.2009.15058.x}, \href
  {http://adsabs.harvard.edu/abs/2009MNRAS.397.1776F} {397, 1776}

\bibitem[\protect\citeauthoryear{{Friedli}, {Benz}  \& {Kennicutt}}{{Friedli}
  et~al.}{1994}]{Friedli+1994}
{Friedli} D.,  {Benz} W.,   {Kennicutt} R.,  1994, \mn@doi [\apjl]
  {10.1086/187449}, \href {http://adsabs.harvard.edu/abs/1994ApJ...430L.105F}
  {430, L105}

\bibitem[\protect\citeauthoryear{{Gallart}, {Stetson}, {Meschin}, {Pont}  \&
  {Hardy}}{{Gallart} et~al.}{2008}]{Gallart:2008aa}
{Gallart} C.,  {Stetson} P.~B.,  {Meschin} I.~P.,  {Pont} F.,   {Hardy} E.,
  2008, \mn@doi [\apjl] {10.1086/590552}, \href
  {http://adsabs.harvard.edu/abs/2008ApJ...682L..89G} {682, L89}

\bibitem[\protect\citeauthoryear{{Gallazzi}, {Charlot}, {Brinchmann}, {White}
  \& {Tremonti}}{{Gallazzi} et~al.}{2005}]{Gallazzi:2005aa}
{Gallazzi} A.,  {Charlot} S.,  {Brinchmann} J.,  {White} S.~D.~M.,   {Tremonti}
  C.~A.,  2005, \mn@doi [\mnras] {10.1111/j.1365-2966.2005.09321.x}, \href
  {http://adsabs.harvard.edu/abs/2005MNRAS.362...41G} {362, 41}

\bibitem[\protect\citeauthoryear{{Gallazzi}, {Brinchmann}, {Charlot}  \&
  {White}}{{Gallazzi} et~al.}{2008}]{Gallazzi+2008}
{Gallazzi} A.,  {Brinchmann} J.,  {Charlot} S.,   {White} S.~D.~M.,  2008,
  \mn@doi [\mnras] {10.1111/j.1365-2966.2007.12632.x}, \href
  {http://adsabs.harvard.edu/abs/2008MNRAS.383.1439G} {383, 1439}

\bibitem[\protect\citeauthoryear{{Gonz{\'a}lez Delgado} et~al.,}{{Gonz{\'a}lez
  Delgado} et~al.}{2014a}]{Gonzalez-Delgado:2014aa}
{Gonz{\'a}lez Delgado} R.~M.,  et~al., 2014a, \mn@doi [\aap]
  {10.1051/0004-6361/201322011}, \href
  {http://adsabs.harvard.edu/abs/2014A%26A...562A..47G} {562, A47}

\bibitem[\protect\citeauthoryear{{Gonz{\'a}lez Delgado} et~al.,}{{Gonz{\'a}lez
  Delgado} et~al.}{2014b}]{Gonzalez-Delgado+2014}
{Gonz{\'a}lez Delgado} R.~M.,  et~al., 2014b, \mn@doi [\aap]
  {10.1051/0004-6361/201322011}, \href
  {http://adsabs.harvard.edu/abs/2014A%26A...562A..47G} {562, A47}

\bibitem[\protect\citeauthoryear{{Gonz{\'a}lez Delgado} et~al.,}{{Gonz{\'a}lez
  Delgado} et~al.}{2016}]{Gonzalez-Delgado+2016}
{Gonz{\'a}lez Delgado} R.~M.,  et~al., 2016, \mn@doi [\aap]
  {10.1051/0004-6361/201628174}, \href
  {http://adsabs.harvard.edu/abs/2016A%26A...590A..44G} {590, A44}

\bibitem[\protect\citeauthoryear{{Gonz{\'a}lez-Samaniego}, {Col{\'{\i}}n},
  {Avila-Reese}, {Rodr{\'{\i}}guez-Puebla}  \&
  {Valenzuela}}{{Gonz{\'a}lez-Samaniego}
  et~al.}{2014}]{Gonzalez-Samaniego:2014aa}
{Gonz{\'a}lez-Samaniego} A.,  {Col{\'{\i}}n} P.,  {Avila-Reese} V.,
  {Rodr{\'{\i}}guez-Puebla} A.,   {Valenzuela} O.,  2014, \mn@doi [\apj]
  {10.1088/0004-637X/785/1/58}, \href
  {http://adsabs.harvard.edu/abs/2014ApJ...785...58G} {785, 58}

\bibitem[\protect\citeauthoryear{{Gonz{\'a}lez-Samaniego}, {Avila-Reese}  \&
  {Col{\'{\i}}n}}{{Gonz{\'a}lez-Samaniego}
  et~al.}{2016}]{Gonzalez-Samaniego+2016}
{Gonz{\'a}lez-Samaniego} A.,  {Avila-Reese} V.,   {Col{\'{\i}}n} P.,  2016,
  \mn@doi [\apj] {10.3847/0004-637X/819/2/101}, \href
  {http://adsabs.harvard.edu/abs/2016ApJ...819..101G} {819, 101}

\bibitem[\protect\citeauthoryear{{Governato} et~al.,}{{Governato}
  et~al.}{2015}]{Governato+2015}
{Governato} F.,  et~al., 2015, \mn@doi [\mnras] {10.1093/mnras/stu2720}, \href
  {http://adsabs.harvard.edu/abs/2015MNRAS.448..792G} {448, 792}

\bibitem[\protect\citeauthoryear{{Graham}, {Driver}, {Petrosian}, {Conselice},
  {Bershady}, {Crawford}  \& {Goto}}{{Graham} et~al.}{2005}]{Graham:2005aa}
{Graham} A.~W.,  {Driver} S.~P.,  {Petrosian} V.,  {Conselice} C.~J.,
  {Bershady} M.~A.,  {Crawford} S.~M.,   {Goto} T.,  2005, \mn@doi [\aj]
  {10.1086/444475}, \href {http://adsabs.harvard.edu/abs/2005AJ....130.1535G}
  {130, 1535}

\bibitem[\protect\citeauthoryear{{Gunn}}{{Gunn}}{1982}]{Gunn1982}
{Gunn} J.~E.,  1982, in {Brueck} H.~A.,  {Coyne} G.~V.,   {Longair} M.~S.,
  eds, Astrophysical Cosmology Proceedings. pp 233--259

\bibitem[\protect\citeauthoryear{{Gunn} \& {Gott}}{{Gunn} \&
  {Gott}}{1972}]{Gunn:1972aa}
{Gunn} J.~E.,  {Gott} III J.~R.,  1972, \mn@doi [\apj] {10.1086/151605}, \href
  {http://adsabs.harvard.edu/abs/1972ApJ...176....1G} {176, 1}

\bibitem[\protect\citeauthoryear{{Gunn} et~al.,}{{Gunn}
  et~al.}{2006}]{Gunn:2006aa}
{Gunn} J.~E.,  et~al., 2006, \mn@doi [\aj] {10.1086/500975}, \href
  {http://adsabs.harvard.edu/abs/2006AJ....131.2332G} {131, 2332}

\bibitem[\protect\citeauthoryear{{Guo}}{{Guo}}{2014}]{Guo:2014aa}
{Guo} F.,  2014, \mn@doi [\apjl] {10.1088/2041-8205/797/2/L34}, \href
  {http://adsabs.harvard.edu/abs/2014ApJ...797L..34G} {797, L34}

\bibitem[\protect\citeauthoryear{{Hidalgo}, {Mar{\'{\i}}n-Franch}  \&
  {Aparicio}}{{Hidalgo} et~al.}{2003}]{Hidalgo+2003}
{Hidalgo} S.~L.,  {Mar{\'{\i}}n-Franch} A.,   {Aparicio} A.,  2003, \mn@doi
  [\aj] {10.1086/367779}, \href
  {http://adsabs.harvard.edu/abs/2003AJ....125.1247H} {125, 1247}

\bibitem[\protect\citeauthoryear{{Hopkins}, {Lauer}, {Cox}, {Hernquist}  \&
  {Kormendy}}{{Hopkins} et~al.}{2009}]{Hopkins+2009}
{Hopkins} P.~F.,  {Lauer} T.~R.,  {Cox} T.~J.,  {Hernquist} L.,   {Kormendy}
  J.,  2009, \mn@doi [\apjs] {10.1088/0067-0049/181/2/486}, \href
  {http://adsabs.harvard.edu/abs/2009ApJS..181..486H} {181, 486}

\bibitem[\protect\citeauthoryear{{Kauffmann} et~al.,}{{Kauffmann}
  et~al.}{2003a}]{Kauffmann:2003aa}
{Kauffmann} G.,  et~al., 2003a, \mn@doi [\mnras]
  {10.1046/j.1365-8711.2003.06291.x}, \href
  {http://adsabs.harvard.edu/abs/2003MNRAS.341...33K} {341, 33}

\bibitem[\protect\citeauthoryear{{Kauffmann} et~al.,}{{Kauffmann}
  et~al.}{2003b}]{Kauffmann:2003ab}
{Kauffmann} G.,  et~al., 2003b, \mn@doi [\mnras]
  {10.1046/j.1365-8711.2003.06292.x}, \href
  {http://adsabs.harvard.edu/abs/2003MNRAS.341...54K} {341, 54}

\bibitem[\protect\citeauthoryear{{Kauffmann}, {White}, {Heckman}, {M{\'e}nard},
  {Brinchmann}, {Charlot}, {Tremonti}  \& {Brinkmann}}{{Kauffmann}
  et~al.}{2004}]{Kauffmann:2004aa}
{Kauffmann} G.,  {White} S.~D.~M.,  {Heckman} T.~M.,  {M{\'e}nard} B.,
  {Brinchmann} J.,  {Charlot} S.,  {Tremonti} C.,   {Brinkmann} J.,  2004,
  \mn@doi [\mnras] {10.1111/j.1365-2966.2004.08117.x}, \href
  {http://adsabs.harvard.edu/abs/2004MNRAS.353..713K} {353, 713}

\bibitem[\protect\citeauthoryear{{Kong} et~al.,}{{Kong}
  et~al.}{2000}]{Kong:2000aa}
{Kong} X.,  et~al., 2000, \mn@doi [\aj] {10.1086/301396}, \href
  {http://adsabs.harvard.edu/abs/2000AJ....119.2745K} {119, 2745}

\bibitem[\protect\citeauthoryear{{Kuntschner} et~al.,}{{Kuntschner}
  et~al.}{2010}]{Kuntschner:2010aa}
{Kuntschner} H.,  et~al., 2010, \mn@doi [\mnras]
  {10.1111/j.1365-2966.2010.17161.x}, \href
  {http://adsabs.harvard.edu/abs/2010MNRAS.408...97K} {408, 97}

\bibitem[\protect\citeauthoryear{{Lacerna}, {Rodr{\'{\i}}guez-Puebla},
  {Avila-Reese}  \& {Hern{\'a}ndez-Toledo}}{{Lacerna}
  et~al.}{2014}]{Lacerna+2014}
{Lacerna} I.,  {Rodr{\'{\i}}guez-Puebla} A.,  {Avila-Reese} V.,
  {Hern{\'a}ndez-Toledo} H.~M.,  2014, \mn@doi [\apj]
  {10.1088/0004-637X/788/1/29}, \href
  {http://adsabs.harvard.edu/abs/2014ApJ...788...29L} {788, 29}

\bibitem[\protect\citeauthoryear{{Larson}, {Tinsley}  \& {Caldwell}}{{Larson}
  et~al.}{1980}]{Larson:1980aa}
{Larson} R.~B.,  {Tinsley} B.~M.,   {Caldwell} C.~N.,  1980, \mn@doi [\apj]
  {10.1086/157917}, \href {http://adsabs.harvard.edu/abs/1980ApJ...237..692L}
  {237, 692}

\bibitem[\protect\citeauthoryear{{Law} et~al.,}{{Law}
  et~al.}{2015}]{Law:2015aa}
{Law} D.~R.,  et~al., 2015, \mn@doi [\aj] {10.1088/0004-6256/150/1/19}, \href
  {http://adsabs.harvard.edu/abs/2015AJ....150...19L} {150, 19}

\bibitem[\protect\citeauthoryear{{Leitner}}{{Leitner}}{2012}]{Leitner:2012aa}
{Leitner} S.~N.,  2012, \mn@doi [\apj] {10.1088/0004-637X/745/2/149}, \href
  {http://adsabs.harvard.edu/abs/2012ApJ...745..149L} {745, 149}

\bibitem[\protect\citeauthoryear{{Li} et~al.,}{{Li} et~al.}{2015}]{Li:2015aa}
{Li} C.,  et~al., 2015, \mn@doi [\apj] {10.1088/0004-637X/804/2/125}, \href
  {http://adsabs.harvard.edu/abs/2015ApJ...804..125L} {804, 125}

\bibitem[\protect\citeauthoryear{{Lin} \& {Faber}}{{Lin} \&
  {Faber}}{1983}]{Lin:1983aa}
{Lin} D.~N.~C.,  {Faber} S.~M.,  1983, \mn@doi [\apjl] {10.1086/183971}, \href
  {http://adsabs.harvard.edu/abs/1983ApJ...266L..21L} {266, L21}

\bibitem[\protect\citeauthoryear{{Lin}, {Zou}, {Kong}, {Lin}, {Mao}, {Cheng},
  {Jiang}  \& {Zhou}}{{Lin} et~al.}{2013}]{Lin:2013aa}
{Lin} L.,  {Zou} H.,  {Kong} X.,  {Lin} X.,  {Mao} Y.,  {Cheng} F.,  {Jiang}
  Z.,   {Zhou} X.,  2013, \mn@doi [\apj] {10.1088/0004-637X/769/2/127}, \href
  {http://adsabs.harvard.edu/abs/2013ApJ...769..127L} {769, 127}

\bibitem[\protect\citeauthoryear{{MacArthur}, {Gonz{\'a}lez}  \&
  {Courteau}}{{MacArthur} et~al.}{2009}]{MacArthur+2009}
{MacArthur} L.~A.,  {Gonz{\'a}lez} J.~J.,   {Courteau} S.,  2009, \mn@doi
  [\mnras] {10.1111/j.1365-2966.2009.14519.x}, \href
  {http://adsabs.harvard.edu/abs/2009MNRAS.395...28M} {395, 28}

\bibitem[\protect\citeauthoryear{{Maraston}}{{Maraston}}{2005}]{Maraston2005}
{Maraston} C.,  2005, \mn@doi [\mnras] {10.1111/j.1365-2966.2005.09270.x},
  \href {http://adsabs.harvard.edu/abs/2005MNRAS.362..799M} {362, 799}

\bibitem[\protect\citeauthoryear{{Martins}, {Gonz{\'a}lez Delgado},
  {Leitherer}, {Cervi{\~n}o}  \& {Hauschildt}}{{Martins}
  et~al.}{2005}]{Martins:2005aa}
{Martins} L.~P.,  {Gonz{\'a}lez Delgado} R.~M.,  {Leitherer} C.,  {Cervi{\~n}o}
  M.,   {Hauschildt} P.,  2005, \mn@doi [\mnras]
  {10.1111/j.1365-2966.2005.08703.x}, \href
  {http://adsabs.harvard.edu/abs/2005MNRAS.358...49M} {358, 49}

\bibitem[\protect\citeauthoryear{{McDermid} et~al.,}{{McDermid}
  et~al.}{2015}]{McDermid+2015}
{McDermid} R.~M.,  et~al., 2015, \mn@doi [\mnras] {10.1093/mnras/stv105}, \href
  {http://adsabs.harvard.edu/abs/2015MNRAS.448.3484M} {448, 3484}

\bibitem[\protect\citeauthoryear{{Mehlert}, {Thomas}, {Saglia}, {Bender}  \&
  {Wegner}}{{Mehlert} et~al.}{2003}]{Mehlert:2003aa}
{Mehlert} D.,  {Thomas} D.,  {Saglia} R.~P.,  {Bender} R.,   {Wegner} G.,
  2003, \mn@doi [\aap] {10.1051/0004-6361:20030886}, \href
  {http://adsabs.harvard.edu/abs/2003A%26A...407..423M} {407, 423}

\bibitem[\protect\citeauthoryear{{Minchev} \& {Famaey}}{{Minchev} \&
  {Famaey}}{2010}]{Minchev+2010}
{Minchev} I.,  {Famaey} B.,  2010, \mn@doi [\apj]
  {10.1088/0004-637X/722/1/112}, \href
  {http://adsabs.harvard.edu/abs/2010ApJ...722..112M} {722, 112}

\bibitem[\protect\citeauthoryear{{Minchev}, {Famaey}, {Quillen}, {Di Matteo},
  {Combes}, {Vlaji{\'c}}, {Erwin}  \& {Bland-Hawthorn}}{{Minchev}
  et~al.}{2012}]{Minchev:2012aa}
{Minchev} I.,  {Famaey} B.,  {Quillen} A.~C.,  {Di Matteo} P.,  {Combes} F.,
  {Vlaji{\'c}} M.,  {Erwin} P.,   {Bland-Hawthorn} J.,  2012, \mn@doi [\aap]
  {10.1051/0004-6361/201219198}, \href
  {http://adsabs.harvard.edu/abs/2012A%26A...548A.126M} {548, A126}

\bibitem[\protect\citeauthoryear{{Mo}, {Mao}  \& {White}}{{Mo}
  et~al.}{1998}]{Mo+1998}
{Mo} H.~J.,  {Mao} S.,   {White} S.~D.~M.,  1998, \mn@doi [\mnras]
  {10.1046/j.1365-8711.1998.01227.x}, \href
  {http://adsabs.harvard.edu/abs/1998MNRAS.295..319M} {295, 319}

\bibitem[\protect\citeauthoryear{{Mo}, {van den Bosch}  \& {White}}{{Mo}
  et~al.}{2010}]{Mo+2010}
{Mo} H.,  {van den Bosch} F.~C.,   {White} S.,  2010, {Galaxy Formation and
  Evolution}

\bibitem[\protect\citeauthoryear{{Moster}, {Somerville}, {Maulbetsch}, {van den
  Bosch}, {Macci{\`o}}, {Naab}  \& {Oser}}{{Moster} et~al.}{2010}]{Moster+2010}
{Moster} B.~P.,  {Somerville} R.~S.,  {Maulbetsch} C.,  {van den Bosch} F.~C.,
  {Macci{\`o}} A.~V.,  {Naab} T.,   {Oser} L.,  2010, \mn@doi [\apj]
  {10.1088/0004-637X/710/2/903}, \href
  {http://adsabs.harvard.edu/abs/2010ApJ...710..903M} {710, 903}

\bibitem[\protect\citeauthoryear{{Moster}, {Naab}  \& {White}}{{Moster}
  et~al.}{2013}]{Moster+2013}
{Moster} B.~P.,  {Naab} T.,   {White} S.~D.~M.,  2013, \mn@doi [\mnras]
  {10.1093/mnras/sts261}, \href
  {http://adsabs.harvard.edu/abs/2013MNRAS.428.3121M} {428, 3121}

\bibitem[\protect\citeauthoryear{{Mu{\~n}oz} \& {Peeples}}{{Mu{\~n}oz} \&
  {Peeples}}{2015}]{Munoz+2015}
{Mu{\~n}oz} J.~A.,  {Peeples} M.~S.,  2015, \mn@doi [\mnras]
  {10.1093/mnras/stv048}, \href
  {http://adsabs.harvard.edu/abs/2015MNRAS.448.1430M} {448, 1430}

\bibitem[\protect\citeauthoryear{{Ocvirk}, {Pichon}, {Lan{\c c}on}  \&
  {Thi{\'e}baut}}{{Ocvirk} et~al.}{2006}]{Ocvirk+2006}
{Ocvirk} P.,  {Pichon} C.,  {Lan{\c c}on} A.,   {Thi{\'e}baut} E.,  2006,
  \mn@doi [\mnras] {10.1111/j.1365-2966.2005.09323.x}, \href
  {http://adsabs.harvard.edu/abs/2006MNRAS.365...74O} {365, 74}

\bibitem[\protect\citeauthoryear{{Pan}, {Li}, {Lin}, {Wang}, {Fan}  \&
  {Kong}}{{Pan} et~al.}{2015}]{Pan:2015aa}
{Pan} Z.,  {Li} J.,  {Lin} W.,  {Wang} J.,  {Fan} L.,   {Kong} X.,  2015,
  \mn@doi [\apjl] {10.1088/2041-8205/804/2/L42}, \href
  {http://adsabs.harvard.edu/abs/2015ApJ...804L..42P} {804, L42}

\bibitem[\protect\citeauthoryear{{Panter}, {Jimenez}, {Heavens}  \&
  {Charlot}}{{Panter} et~al.}{2007}]{Panter+2007}
{Panter} B.,  {Jimenez} R.,  {Heavens} A.~F.,   {Charlot} S.,  2007, \mn@doi
  [\mnras] {10.1111/j.1365-2966.2007.11909.x}, \href
  {http://adsabs.harvard.edu/abs/2007MNRAS.378.1550P} {378, 1550}

\bibitem[\protect\citeauthoryear{{Patel} et~al.,}{{Patel}
  et~al.}{2013}]{Patel+2013}
{Patel} S.~G.,  et~al., 2013, \mn@doi [\apj] {10.1088/0004-637X/778/2/115},
  \href {http://adsabs.harvard.edu/abs/2013ApJ...778..115P} {778, 115}

\bibitem[\protect\citeauthoryear{{P{\'e}rez-Gonz{\'a}lez}
  et~al.,}{{P{\'e}rez-Gonz{\'a}lez} et~al.}{2008}]{Perez-Gonzalez:2008aa}
{P{\'e}rez-Gonz{\'a}lez} P.~G.,  et~al., 2008, \mn@doi [\apj] {10.1086/523690},
  \href {http://adsabs.harvard.edu/abs/2008ApJ...675..234P} {675, 234}

\bibitem[\protect\citeauthoryear{{P{\'e}rez} et~al.,}{{P{\'e}rez}
  et~al.}{2013}]{Perez+2013}
{P{\'e}rez} E.,  et~al., 2013, \mn@doi [\apjl] {10.1088/2041-8205/764/1/L1},
  \href {http://adsabs.harvard.edu/abs/2013ApJ...764L...1P} {764, L1}

\bibitem[\protect\citeauthoryear{{Petrosian}}{{Petrosian}}{1976}]{Petrosian:1976aa}
{Petrosian} V.,  1976, \mn@doi [\apjl] {10.1086/182253}, \href
  {http://adsabs.harvard.edu/abs/1976ApJ...209L...1P} {209, L1}

\bibitem[\protect\citeauthoryear{{Pezzulli}, {Fraternali}, {Boissier}  \&
  {Mu{\~n}oz-Mateos}}{{Pezzulli} et~al.}{2015}]{Pezzulli+2015}
{Pezzulli} G.,  {Fraternali} F.,  {Boissier} S.,   {Mu{\~n}oz-Mateos} J.~C.,
  2015, \mn@doi [\mnras] {10.1093/mnras/stv1077}, \href
  {http://adsabs.harvard.edu/abs/2015MNRAS.451.2324P} {451, 2324}

\bibitem[\protect\citeauthoryear{{Pozzetti} et~al.,}{{Pozzetti}
  et~al.}{2010}]{Pozzetti+2010}
{Pozzetti} L.,  et~al., 2010, \mn@doi [\aap] {10.1051/0004-6361/200913020},
  \href {http://adsabs.harvard.edu/abs/2010A%26A...523A..13P} {523, A13}

\bibitem[\protect\citeauthoryear{{Rodriguez-Gomez} et~al.,}{{Rodriguez-Gomez}
  et~al.}{2016}]{Rodriguez-Gomez+2016}
{Rodriguez-Gomez} V.,  et~al., 2016, \mn@doi [\mnras] {10.1093/mnras/stw456},
  \href {http://adsabs.harvard.edu/abs/2016MNRAS.458.2371R} {458, 2371}

\bibitem[\protect\citeauthoryear{{Rodr{\'{\i}}guez-Puebla}, {Avila-Reese},
  {Yang}, {Foucaud}, {Drory}  \& {Jing}}{{Rodr{\'{\i}}guez-Puebla}
  et~al.}{2015}]{Rodriguez-Puebla+2015}
{Rodr{\'{\i}}guez-Puebla} A.,  {Avila-Reese} V.,  {Yang} X.,  {Foucaud} S.,
  {Drory} N.,   {Jing} Y.~P.,  2015, \mn@doi [\apj]
  {10.1088/0004-637X/799/2/130}, \href
  {http://adsabs.harvard.edu/abs/2015ApJ...799..130R} {799, 130}

\bibitem[\protect\citeauthoryear{{Ro{\v s}kar}, {Debattista}, {Quinn},
  {Stinson}  \& {Wadsley}}{{Ro{\v s}kar} et~al.}{2008}]{Roskar+2008}
{Ro{\v s}kar} R.,  {Debattista} V.~P.,  {Quinn} T.~R.,  {Stinson} G.~S.,
  {Wadsley} J.,  2008, \mn@doi [\apjl] {10.1086/592231}, \href
  {http://adsabs.harvard.edu/abs/2008ApJ...684L..79R} {684, L79}

\bibitem[\protect\citeauthoryear{{Ro{\v s}kar}, {Debattista}, {Quinn}  \&
  {Wadsley}}{{Ro{\v s}kar} et~al.}{2012}]{Roskar+2012}
{Ro{\v s}kar} R.,  {Debattista} V.~P.,  {Quinn} T.~R.,   {Wadsley} J.,  2012,
  \mn@doi [\mnras] {10.1111/j.1365-2966.2012.21860.x}, \href
  {http://adsabs.harvard.edu/abs/2012MNRAS.426.2089R} {426, 2089}

\bibitem[\protect\citeauthoryear{{S{\'a}nchez-Bl{\'a}zquez}
  et~al.,}{{S{\'a}nchez-Bl{\'a}zquez} et~al.}{2006}]{Sanchez-Blazquez:2006aa}
{S{\'a}nchez-Bl{\'a}zquez} P.,  et~al., 2006, \mn@doi [\mnras]
  {10.1111/j.1365-2966.2006.10699.x}, \href
  {http://adsabs.harvard.edu/abs/2006MNRAS.371..703S} {371, 703}

\bibitem[\protect\citeauthoryear{{S{\'a}nchez-Bl{\'a}zquez}, {Forbes},
  {Strader}, {Brodie}  \& {Proctor}}{{S{\'a}nchez-Bl{\'a}zquez}
  et~al.}{2007}]{Sanchez-Blazquez:2007aa}
{S{\'a}nchez-Bl{\'a}zquez} P.,  {Forbes} D.~A.,  {Strader} J.,  {Brodie} J.,
  {Proctor} R.,  2007, \mn@doi [\mnras] {10.1111/j.1365-2966.2007.11647.x},
  \href {http://adsabs.harvard.edu/abs/2007MNRAS.377..759S} {377, 759}

\bibitem[\protect\citeauthoryear{{S{\'a}nchez-Bl{\'a}zquez}, {Courty}, {Gibson}
   \& {Brook}}{{S{\'a}nchez-Bl{\'a}zquez} et~al.}{2009}]{Sanchez-Blazquez+2009}
{S{\'a}nchez-Bl{\'a}zquez} P.,  {Courty} S.,  {Gibson} B.~K.,   {Brook} C.~B.,
  2009, \mn@doi [\mnras] {10.1111/j.1365-2966.2009.15133.x}, \href
  {http://adsabs.harvard.edu/abs/2009MNRAS.398..591S} {398, 591}

\bibitem[\protect\citeauthoryear{{S{\'a}nchez-Bl{\'a}zquez}, {Ocvirk},
  {Gibson}, {P{\'e}rez}  \& {Peletier}}{{S{\'a}nchez-Bl{\'a}zquez}
  et~al.}{2011}]{Sanchez-Blazquez+2011}
{S{\'a}nchez-Bl{\'a}zquez} P.,  {Ocvirk} P.,  {Gibson} B.~K.,  {P{\'e}rez} I.,
   {Peletier} R.~F.,  2011, \mn@doi [\mnras]
  {10.1111/j.1365-2966.2011.18749.x}, \href
  {http://adsabs.harvard.edu/abs/2011MNRAS.415..709S} {415, 709}

\bibitem[\protect\citeauthoryear{{S{\'a}nchez-Bl{\'a}zquez}
  et~al.,}{{S{\'a}nchez-Bl{\'a}zquez} et~al.}{2014}]{Sanchez-Blazquez+2014}
{S{\'a}nchez-Bl{\'a}zquez} P.,  et~al., 2014, \mn@doi [\aap]
  {10.1051/0004-6361/201423635}, \href
  {http://adsabs.harvard.edu/abs/2014A%26A...570A...6S} {570, A6}

\bibitem[\protect\citeauthoryear{{S{\'a}nchez} et~al.,}{{S{\'a}nchez}
  et~al.}{2012}]{Sanchez:2012aa}
{S{\'a}nchez} S.~F.,  et~al., 2012, \mn@doi [\aap]
  {10.1051/0004-6361/201117353}, \href
  {http://adsabs.harvard.edu/abs/2012A%26A...538A...8S} {538, A8}

\bibitem[\protect\citeauthoryear{{S{\'a}nchez} et~al.,}{{S{\'a}nchez}
  et~al.}{2014}]{Sanchez:2014aa}
{S{\'a}nchez} S.~F.,  et~al., 2014, \mn@doi [\aap]
  {10.1051/0004-6361/201322343}, \href
  {http://adsabs.harvard.edu/abs/2014A%26A...563A..49S} {563, A49}

\bibitem[\protect\citeauthoryear{{S{\'a}nchez} et~al.,}{{S{\'a}nchez}
  et~al.}{2015}]{Sanchez:2015ab}
{S{\'a}nchez} S.~F.,  et~al., 2015, \mn@doi [\aap]
  {10.1051/0004-6361/201424950}, \href
  {http://adsabs.harvard.edu/abs/2015A%26A...573A.105S} {573, A105}

\bibitem[\protect\citeauthoryear{{S{\'a}nchez} et~al.,}{{S{\'a}nchez}
  et~al.}{2016a}]{Sanchez:2016aa}
{S{\'a}nchez} S.~F.,  et~al., 2016a, preprint, \href
  {http://adsabs.harvard.edu/abs/2016arXiv160201830S} {} (\mn@eprint {arXiv}
  {1602.01830})

\bibitem[\protect\citeauthoryear{{S{\'a}nchez} et~al.,}{{S{\'a}nchez}
  et~al.}{2016b}]{Sanchez:2016ab}
{S{\'a}nchez} S.~F.,  et~al., 2016b, \rmxaa, \href
  {http://adsabs.harvard.edu/abs/2016RMxAA..52...21S} {52, 21}

\bibitem[\protect\citeauthoryear{{Sanders}, {Soifer}, {Elias}, {Madore},
  {Matthews}, {Neugebauer}  \& {Scoville}}{{Sanders}
  et~al.}{1988}]{Sanders:1988aa}
{Sanders} D.~B.,  {Soifer} B.~T.,  {Elias} J.~H.,  {Madore} B.~F.,  {Matthews}
  K.,  {Neugebauer} G.,   {Scoville} N.~Z.,  1988, \mn@doi [\apj]
  {10.1086/165983}, \href {http://adsabs.harvard.edu/abs/1988ApJ...325...74S}
  {325, 74}

\bibitem[\protect\citeauthoryear{{Sch{\"o}nrich} \& {Binney}}{{Sch{\"o}nrich}
  \& {Binney}}{2009}]{Schonrich+2009}
{Sch{\"o}nrich} R.,  {Binney} J.,  2009, \mn@doi [\mnras]
  {10.1111/j.1365-2966.2009.14750.x}, \href
  {http://adsabs.harvard.edu/abs/2009MNRAS.396..203S} {396, 203}

\bibitem[\protect\citeauthoryear{{Sellwood} \& {Binney}}{{Sellwood} \&
  {Binney}}{2002}]{Sellwood+2002}
{Sellwood} J.~A.,  {Binney} J.~J.,  2002, \mn@doi [\mnras]
  {10.1046/j.1365-8711.2002.05806.x}, \href
  {http://adsabs.harvard.edu/abs/2002MNRAS.336..785S} {336, 785}

\bibitem[\protect\citeauthoryear{{Silk}}{{Silk}}{1987}]{Silk1987}
{Silk} J.,  1987, in {Hewitt} A.,  {Burbidge} G.,   {Fang} L.~Z.,  eds,  IAU
  Symposium Vol. 124, Observational Cosmology. pp 391--411

\bibitem[\protect\citeauthoryear{{Slater} \& {Bell}}{{Slater} \&
  {Bell}}{2014}]{Slater:2014aa}
{Slater} C.~T.,  {Bell} E.~F.,  2014, \mn@doi [\apj]
  {10.1088/0004-637X/792/2/141}, \href
  {http://adsabs.harvard.edu/abs/2014ApJ...792..141S} {792, 141}

\bibitem[\protect\citeauthoryear{{Stinson}, {Dalcanton}, {Quinn}, {Kaufmann}
  \& {Wadsley}}{{Stinson} et~al.}{2007}]{Stinson+2007}
{Stinson} G.~S.,  {Dalcanton} J.~J.,  {Quinn} T.,  {Kaufmann} T.,   {Wadsley}
  J.,  2007, \mn@doi [\apj] {10.1086/520504}, \href
  {http://adsabs.harvard.edu/abs/2007ApJ...667..170S} {667, 170}

\bibitem[\protect\citeauthoryear{{Tal} et~al.,}{{Tal}
  et~al.}{2014}]{Tal:2014aa}
{Tal} T.,  et~al., 2014, \mn@doi [\apj] {10.1088/0004-637X/789/2/164}, \href
  {http://adsabs.harvard.edu/abs/2014ApJ...789..164T} {789, 164}

\bibitem[\protect\citeauthoryear{{Thomas}, {Maraston}, {Bender}  \& {Mendes de
  Oliveira}}{{Thomas} et~al.}{2005}]{Thomas:2005aa}
{Thomas} D.,  {Maraston} C.,  {Bender} R.,   {Mendes de Oliveira} C.,  2005,
  \mn@doi [\apj] {10.1086/426932}, \href
  {http://adsabs.harvard.edu/abs/2005ApJ...621..673T} {621, 673}

\bibitem[\protect\citeauthoryear{{Thomas}, {Maraston}, {Schawinski}, {Sarzi}
  \& {Silk}}{{Thomas} et~al.}{2010}]{Thomas:2010aa}
{Thomas} D.,  {Maraston} C.,  {Schawinski} K.,  {Sarzi} M.,   {Silk} J.,  2010,
  \mn@doi [\mnras] {10.1111/j.1365-2966.2010.16427.x}, \href
  {http://adsabs.harvard.edu/abs/2010MNRAS.404.1775T} {404, 1775}

\bibitem[\protect\citeauthoryear{{Tinsley}}{{Tinsley}}{1980}]{Tinsley:1980aa}
{Tinsley} B.~M.,  1980, \fcp, \href
  {http://adsabs.harvard.edu/abs/1980FCPh....5..287T} {5, 287}

\bibitem[\protect\citeauthoryear{{Tissera}, {Scannapieco}, {Beers}  \&
  {Carollo}}{{Tissera} et~al.}{2013}]{Tissera:2013aa}
{Tissera} P.~B.,  {Scannapieco} C.,  {Beers} T.~C.,   {Carollo} D.,  2013,
  \mn@doi [\mnras] {10.1093/mnras/stt691}, \href
  {http://adsabs.harvard.edu/abs/2013MNRAS.432.3391T} {432, 3391}

\bibitem[\protect\citeauthoryear{{Tojeiro}, {Heavens}, {Jimenez}  \&
  {Panter}}{{Tojeiro} et~al.}{2007}]{Tojeiro+2007}
{Tojeiro} R.,  {Heavens} A.~F.,  {Jimenez} R.,   {Panter} B.,  2007, \mn@doi
  [\mnras] {10.1111/j.1365-2966.2007.12323.x}, \href
  {http://adsabs.harvard.edu/abs/2007MNRAS.381.1252T} {381, 1252}

\bibitem[\protect\citeauthoryear{{Tojeiro}, {Wilkins}, {Heavens}, {Panter}  \&
  {Jimenez}}{{Tojeiro} et~al.}{2009}]{Tojeiro+2009}
{Tojeiro} R.,  {Wilkins} S.,  {Heavens} A.~F.,  {Panter} B.,   {Jimenez} R.,
  2009, \mn@doi [\apjs] {10.1088/0067-0049/185/1/1}, \href
  {http://adsabs.harvard.edu/abs/2009ApJS..185....1T} {185, 1}

\bibitem[\protect\citeauthoryear{{Tomczak} et~al.,}{{Tomczak}
  et~al.}{2016}]{Tomczak+2016}
{Tomczak} A.~R.,  et~al., 2016, \mn@doi [\apj] {10.3847/0004-637X/817/2/118},
  \href {http://adsabs.harvard.edu/abs/2016ApJ...817..118T} {817, 118}

\bibitem[\protect\citeauthoryear{{Vazdekis}, {S{\'a}nchez-Bl{\'a}zquez},
  {Falc{\'o}n-Barroso}, {Cenarro}, {Beasley}, {Cardiel}, {Gorgas}  \&
  {Peletier}}{{Vazdekis} et~al.}{2010}]{Vazdekis:2010aa}
{Vazdekis} A.,  {S{\'a}nchez-Bl{\'a}zquez} P.,  {Falc{\'o}n-Barroso} J.,
  {Cenarro} A.~J.,  {Beasley} M.~A.,  {Cardiel} N.,  {Gorgas} J.,   {Peletier}
  R.~F.,  2010, \mn@doi [\mnras] {10.1111/j.1365-2966.2010.16407.x}, \href
  {http://adsabs.harvard.edu/abs/2010MNRAS.404.1639V} {404, 1639}

\bibitem[\protect\citeauthoryear{{Walcher}, {Groves}, {Budav{\'a}ri}  \&
  {Dale}}{{Walcher} et~al.}{2011}]{Walcher:2011aa}
{Walcher} J.,  {Groves} B.,  {Budav{\'a}ri} T.,   {Dale} D.,  2011, \mn@doi
  [\apss] {10.1007/s10509-010-0458-z}, \href
  {http://adsabs.harvard.edu/abs/2011Ap%26SS.331....1W} {331, 1}

\bibitem[\protect\citeauthoryear{{Wang} et~al.,}{{Wang}
  et~al.}{2011}]{Wang:2011aa}
{Wang} J.,  et~al., 2011, \mn@doi [\mnras] {10.1111/j.1365-2966.2010.17962.x},
  \href {http://adsabs.harvard.edu/abs/2011MNRAS.412.1081W} {412, 1081}

\bibitem[\protect\citeauthoryear{{Wilkinson} et~al.,}{{Wilkinson}
  et~al.}{2015}]{Wilkinson:2015aa}
{Wilkinson} D.~M.,  et~al., 2015, \mn@doi [\mnras] {10.1093/mnras/stv301},
  \href {http://adsabs.harvard.edu/abs/2015MNRAS.449..328W} {449, 328}

\bibitem[\protect\citeauthoryear{{Woo} et~al.,}{{Woo}
  et~al.}{2013}]{Woo:2013aa}
{Woo} J.,  et~al., 2013, \mn@doi [\mnras] {10.1093/mnras/sts274}, \href
  {http://adsabs.harvard.edu/abs/2013MNRAS.428.3306W} {428, 3306}

\bibitem[\protect\citeauthoryear{{Yan} et~al.,}{{Yan}
  et~al.}{2016}]{Yan:2016aa}
{Yan} R.,  et~al., 2016, \mn@doi [\aj] {10.3847/0004-6256/151/1/8}, \href
  {http://adsabs.harvard.edu/abs/2016AJ....151....8Y} {151, 8}

\bibitem[\protect\citeauthoryear{{Zhang}, {Hunter}, {Elmegreen}, {Gao}  \&
  {Schruba}}{{Zhang} et~al.}{2012}]{Zhang:2012aa}
{Zhang} H.-X.,  {Hunter} D.~A.,  {Elmegreen} B.~G.,  {Gao} Y.,   {Schruba} A.,
  2012, \mn@doi [\aj] {10.1088/0004-6256/143/2/47}, \href
  {http://adsabs.harvard.edu/abs/2012AJ....143...47Z} {143, 47}

\bibitem[\protect\citeauthoryear{{van Dokkum} et~al.,}{{van Dokkum}
  et~al.}{2013}]{vanDokkum+2013}
{van Dokkum} P.~G.,  et~al., 2013, \mn@doi [\apjl]
  {10.1088/2041-8205/771/2/L35}, \href
  {http://adsabs.harvard.edu/abs/2013ApJ...771L..35V} {771, L35}

\bibitem[\protect\citeauthoryear{{van den Bosch}}{{van den
  Bosch}}{2002}]{vandenBosch2002}
{van den Bosch} F.~C.,  2002, \mn@doi [\mnras]
  {10.1046/j.1365-8711.2002.05171.x}, \href
  {http://adsabs.harvard.edu/abs/2002MNRAS.331...98V} {331, 98}

\makeatother
\end{thebibliography}
\bibliographystyle{mnras}
\end{document}